\documentclass[twocolumn]{aastex631}
\usepackage{hyperref}
\usepackage{amsmath}

\submitjournal{ApJ}
\shorttitle{{\tt SONORA} I}
\shortauthors{Marley et al.}

\begin{document}

\title{The Sonora Brown Dwarf Atmosphere and Evolution Models. I. Model Description and Application to Cloudless Atmospheres in Rainout Chemical Equilibrium.}

\author{Mark S. Marley}
\affil{Lunar \& Planetary Laboratory, University of Arizona, Tucson, AZ 85721}

\author{Didier Saumon}
\affil{Los Alamos National Laboratory, PO Box 1663, Los Alamos, NM 87545}

\author{Channon Visscher}
\affil{Dordt University, Sioux Center IA; Space Science Institute, Boulder, CO}

\author{Roxana Lupu}
\affil{BAER Institute, NASA Ames Research Center, Moffett Field, CA 94035}

\author{Richard Freedman}
\affil{SETI Institute, NASA Ames Research Center, Moffett Field, CA 94035}

\author{Caroline Morley}
\affil{Department of Astronomy, University of Texas at Austin, Austin, TX 78712, USA}

\author{Jonathan J. Fortney}
\affil{Department of Astronomy \& Astrophysics, University of California, Santa Cruz, CA 95064, USA}

\author{Christopher Seay}
\affil{Leiden Observatory, Leiden, Netherlands}
\affil{Department of Astronomy \& Astrophysics, University of California, Santa Cruz, CA 95064, USA}

\author{Adam J. R. W. Smith} 
\affil{Department of Astronomy, New Mexico State University, Las Cruces, NM 88003 USA}
\affil{Department of Astronomy \& Astrophysics, University of California, Santa Cruz, CA 95064, USA}

\author{D. J. Teal}
\affil{Department of Astronomy, University of Maryland, College Park, MD 20742, USA}
\affil{Department of Astronomy \& Astrophysics, University of California, Santa Cruz, CA 95064, USA}

\author{Ruoyan Wang}
\affil{Department of Physics and Astronomy, University of Leicester, Leicester, United Kingdom}
\affil{Department of Astronomy \& Astrophysics, University of California, Santa Cruz, CA 95064, USA}

\begin{abstract}

We present a new generation of substellar atmosphere and evolution models, appropriate for application to studies of L, T, and Y-type brown dwarfs and self-luminous extrasolar planets. The atmosphere models describe the expected temperature-pressure profiles and emergent spectra of atmospheres in radiative-convective equilibrium with effective temperatures and gravities within the ranges
$200\le T_{\rm eff}\le2400\,\rm K$ and  $2.5\le \log g \le 5.5$. These ranges 
encompass masses from about 0.5 to 85 Jupiter masses for a set of metallicities ($[{\rm M/H}] = -0.5$ to $+0.5$), C/O ratios (from 0.5 to 1.5 times that of solar), and ages. The evolution tables describe the cooling of these substellar objects through time. These models expand the diversity of model atmospheres currently available, notably to cooler effective temperatures and greater ranges in C/O. Notable improvements from past such models include updated opacities and atmospheric chemistry. Here we describe our modeling approach and present our initial tranche of models for cloudless, chemical equilibrium atmospheres. We compare the modeled spectra, photometry, and evolution to various datasets.

\end{abstract}

\section{Introduction}\label{sec:intro}

The twenty-five years following the discovery of the first indisputable brown dwarf, Gliese 229 B \citep[][]{Oppenheimer95}, have seen  a flowering of this field. Thousands of brown dwarfs have been discovered and characterized by spectroscopy and photometry \citep[][]{joergens2014}. Dynamical masses and parallaxes have been 
measured for many objects and a wealth of trends uncovered \citep[e.g.,][]{Dupuy17,best20}. In addition young, self luminous planets have been discovered and characterized \citep[e.g.,][]{Marois08}. 

Most of these objects have been characterized with the help of forward modeling in which modelers construct hundreds to thousands of `grid models'. The models, relying upon fundamental physical processes, predict spectral and evolutionary characteristics of brown dwarfs for given choices of intrinsic parameters, such as mass, age, and bulk composition. This approach typically relies upon both 
one dimensional radiative-convective equilibrium atmosphere models  and coupled interior and evolution models. The atmosphere models aim to capture the key influences on substellar atmospheres, including chemistry, dynamics, and cloud processes, in order to compute the vertical structure of  an atmosphere which conservatively transports energy  upwards from the deep interior. Evolution models apply the rate of energy loss through the atmosphere as a boundary condition, in order to compute the rate of internal energy loss through the atmosphere and consequently the evolution of radius and luminosity through time. 

Such a forward modeling approach provides a self-consistent solution for the coupled problem of understanding both atmospheric and interior physical processes.  By making predictions, models inform observers of interesting observational tests and connect observable properties, including luminosity and the spectral energy distribution, to the physical properties of the object, including mass and age. Grids of models are also essential for motivating and planning new observations.
A non-exhaustive list of forward model grids includes those of \citet[][]{Burrows97,baraffe03,Saumon08,allard14,baraffe15,Phillips20,Malik17}. 

The older models in the literature are generally out of date as our knowledge of molecular opacities, most notably water and methane which are important absorbers in substellar atmospheres, has progressed substantially over the intervening years. Several of the more recent grid models use updated opacities but are generally not coupled with self-consistent evolution calculations, and thus do not provide a self consistent evolutionary-atmospheric modeling framework, 
a notable exception being \citet[][]{Phillips20} and the ATMO2020 grid.
Here we also provide such a framework, presenting coupled atmospheric structure and evolution models for a variety of atmospheric chemical assumptions. This effort is the first in an expected series of papers, each looking at additional model complexities, including disequilibrium chemistry, clouds, and so on.
Independent modeling efforts, such as our own and \citet[][]{Phillips20}, are crucial for cross checking the importance of various physical and chemical assumptions and for overall self consistency. Thus we view the Sonora and ATMO2020 model sets to be highly complementary. Further comparisons to some of the other model sets are discussed in Section 2.6.

In the past few years 
`retrieval methods', originally developed to study planetary atmospheres, have been applied to brown dwarfs \citep[e.g.,][]{Line15,Line17,Piette20,burningham21,kitzmann20} in order to  understand the constraints on mass, luminosity, composition, radius, and other characteristics which are evident in the spectra alone, without resorting to underlying assumptions, such as solar abundance ratios, chemical equilibrium, and a radiative-convective structure. Retrievals excel at testing theoretical predictions by comparing a host of models to data, while accounting for various dataset uncertainties. 
Retrieval methods are of greatest utility when judged in the context of grid model predictions  as such comparisons test our understanding of underlying processes. By utilizing retrieval techniques \citet{Line17} and \citet{Zalesky19}, for example, unambiguously confirmed that rainout, not pure equilibrium, chemistry acts in substellar atmospheres (see further discussion in Section 2.5). 

Both types of models are needed to motivate and interpret observations. 
In order to provide a new, systematic survey of brown dwarf atmospheric structure, emergent spectra, and evolution, we have constructed a new grid of brown dwarf model atmospheres. We ultimately aim for our grid to span broad ranges of atmospheric metallicity, C/O ratios, cloud properties, atmospheric mixing, and other parameters. Spectra predicted by our modeling grid can be compared to both observations and retrieval results to aid in the interpretation and planning of future telescopic observations.

For simplicity we divide the presentation of our new models into parts. Here, in 
Paper I we present our overall modeling approach, describing our atmosphere and evolution models, as well as various model inputs, including opacities, and present results for cloudless models. In forthcoming papers in this series we will investigate disequilibrium chemistry and cloudy atmospheres. We break with previous tradition of our team by naming the models to provide clarity as to model generations. These and future models from our group are given the moniker `Sonora', after the desert spanning the southwestern United States and northern Mexico. Individual model generations (e.g., with a given set of opacities) will be denoted by names of flora and fauna of that desert. Models herein, cloudless, rainout chemical equilibrium, are thus `Sonora Bobcat'.

This paper describes our radiative-convective forward model for calculating the atmospheric structure of substellar objects and our evolution calculation for computing their trajectory through time. Section 2 describes the modeling details, Section 3 model results, and Section 4 highlights a few comparisons of model predictions to various datasets.

\section{Model Description}
We begin with a description of the atmospheric forward modeling approach. Here we term a forward model as a description of the variation of atmospheric temperature, $T$, and composition as a function of pressure, $P$, for a specified gravity, $g$, and effective temperature, $T_{\rm eff}$. In addition we specify the atmospheric metallicity and carbon to oxygen ratio. In future work we will describe additional constraints, including cloud treatments and vertical mixing.

Ultimately the selection of parameters and numerical approach employed in forward model grids depends on a series of judgment calls that balance the need for as precise as possible modeling of physical processes with numerical expediency. In this section we describe our approach to atmospheric modeling and briefly compare our choices to those of some other well known modeling schools. For a broader overview and literature survey
 of the substellar and planetary atmosphere modeling process see the review by \citet{MarlRob15}.

\subsection{Overview}
Each model case is described by a limited set of specific parameters for 1D, plane parallel atmospheres. For the initial model set presented here these are gravity, $g$ (presumed constant with height as the thickness of the atmosphere is much less than the body's radius),  effective temperature, $T_{\rm eff}$, cloud treatment, metallicity [M/H], and carbon-to-oxygen $\rm (C/O)$ ratio. 
A crucial detail is that the abundance measures refer to the bulk chemistry of the gas from which the atmosphere forms.
Various condensation processes can alter the atmosphere at any arbitrary pressure and temperature away from the bulk values \added{by the removal of elements from the gas phase. For example, the condensation of magnesium silicates can sequester up to $\sim$20\% of the atmospheric oxygen inventory (in a solar-composition gas). This may yield a C/O ratio in the observable atmosphere of an object that is greater than its bulk C/O ratio, if some oxygen has been removed by condensation processes deeper in its atmosphere (and if carbon has not likewise been removed by condensation processes of its own).}

\added{We have selected a range of model parameters, shown in Table I, to span that expected for the evolution of solar neighborhood ultracool dwarfs. Not all model combinations, particularly high-$g$, low-$T_{\rm eff}$, are meaningful as the most massive, high gravity objects will not have cooled to the lowest temperatures in the age of the universe.}
%\deleted{For the purposes of modeling...} 
%ellipsis added on advice of Laura for stupid \replaced bug
%ultracool dwarf and directly imaged giant planets, each parameter is %bounded as summarized in Table I.} 

While we account for the effect of condensation on the atmospheric composition and chemistry, here we set all condensate opacity equal to zero; cloudy models will be presented in an upcoming paper. In the future we will add additional standard parameters, such as cloud coverage fraction, or eddy mixing coefficient.  For each combination of specific parameters we iteratively compute a single radiative-convective equilibrium atmosphere model. Such a model describes the variation in atmospheric temperature $T$ as a function of pressure $P$. Given this $T(P)$ profile and the abundances of all atmospheric constituents we can post-process emergent spectra at any needed spectral resolution.

\begin{deluxetable*}{ccc}
\tablecaption{Model Parameters}
\label{params}
\tablewidth{0pt}
\tablehead{\colhead{Parameter} & \colhead{Range}  &\colhead{Step}}

\startdata\\[-8pt]
gravity & $3 \le \log g \le 5.5 $ & 0.25 \\
effective temperature & $ 200 \le  T_{\rm eff}\le 2400\,\rm K$ & 25 $(\le600)$, 50 $(\le1000)$, 100  ($> 1000\,\rm K$)\\
metallicity & $-0.5 \le [M/H]\le +0.50 $ & 0.25 \\
carbon to oxygen ratio & $0.25 \le {\rm (C/O)/(C/O)_\odot} \le 1.50 $ & 0.25 \\
\hline
\enddata
\end{deluxetable*}

\subsection{Radiative-Convective Equilibrium Model}

Because the dominant sources of atmospheric opacity, such as $\rm H_2O$, vary strongly with wavelength--particularly in cloud free models such as these--the opacity of a gas column from a given depth in the atmosphere to infinity varies strongly with wavelength. A parcel of gas of a given temperature in the deep atmosphere can first radiate to space only over a narrow wavelength range, typically in the low opacity windows within $Y$ or $J$ bands. The atmosphere begins to emit strongly if the local Planck function overlaps these opacity windows. If sufficient energy can be radiated away, the local temperature lapse rate will transition from essentially adiabatic to the local radiative lapse rate. 
However, at higher, cooler levels in the atmosphere, the Planck function shifts
 to longer wavelengths where it can again encounter a high optical depth to infinity and the radiative lapse rate steepens as a result, in some cases enough to once more trigger convection. In very cool models ($T_{\rm eff} < 500\,\rm K$) this process can repeat once more, leading to a stacked structure of up to three convective zones separated by radiative zones \citep{Marley96, Burrows97, Morley12, MarlRob15, Morley14a}. Any substellar atmosphere model must be able to capture this behavior as it alters the atmospheric temperature-pressure profile and the boundary condition for thermal evolution. 

Our radiative-convective equilibrium model solves for a \added{hydrostatic and radiative-convective equilibrium} temperature structure by starting with a first guess profile that is convective only in the greatest  depths of the atmosphere  and in radiative equilibrium elsewhere. Given this initial guess temperature profile and a radiative-convective boundary pressure, the model adjusts the temperature in the radiative regions using a straightforward Newton-Raphson scheme (see \citet{MarlRob15}) until the fractional difference between the net thermal flux and $\sigma T_{\rm eff}^4$\footnote{For irradiated models, not considered here, the net thermal flux must also carry the net incident radiation absorbed below each atmospheric level.} is everywhere less than a specified value, typically $10^{-5}$.

Convective adjustment begins once a converged radiative profile has been found for a given specification of the top of the convection zone. The local temperature gradient $\nabla=d\log T/ d\log P$ is compared to that of an adiabat, $\nabla_{\rm ad}$ \added{as tabulated employing the equation of state (\S{2.7})}. If the radiative-equilibrium lapse rate $\nabla_{\rm rad} > \nabla_{\rm ad}$ then that layer is deemed convective and $\nabla$ is set equal to $\nabla_{\rm ad}$ for that layer.
\citet{Baraffe02} have shown that for substellar, $\rm H_2$-dominated atmospheres convection is essentially adiabatic and that  mixing length theory predicts
$\nabla = \nabla_{\rm ad}$ in convective regions, regardless of the choice of the mixing length. Thus for the cases studied here setting $\nabla \equiv \nabla_{\rm ad}$ in regions where  $\nabla_{\rm rad} > \nabla_{\rm ad}$ is warranted.
To find a properly converged structure, we specifically do not attempt to adjust the entire model region where $\nabla_{\rm rad} > \nabla_{\rm ad}$ to the adiabat all at once as changes in the deep temperature profile can and do impact the thermal energy balance and temperature profile above. Instead we re-compute a radiative-equilibrium solution for the new convection zone boundary and repeat the procedure. This iterative approach follows 
\citet{Mckay89} as adapted to giant planet and brown dwarf atmospheres by \citet{Marley96,Marley99,Burrows97}. 

The model now employs 90 vertical layers which have 91 pressure--temperature boundaries, or levels. The top model pressure is typically $\sim 10^{-4}\,\rm bar$ and the bottom pressure varies with model gravity and $T_{\rm eff}$, ranging from tens of bar to 1,000 bar or more. The highest pressures are needed to capture the radiative-convective boundary in some high gravity models. The line broadening treatment, and thus the gas opacities, at such high pressures is very uncertain, which adds a source of uncertainty to the radiative-convective boundary location in such situations.

\begin{figure*}
    \centering
%    \begin{subfigure}
%    [t]{0.49\textwidth}
    \hfil
        \centering
        \includegraphics[width=0.33\textwidth]{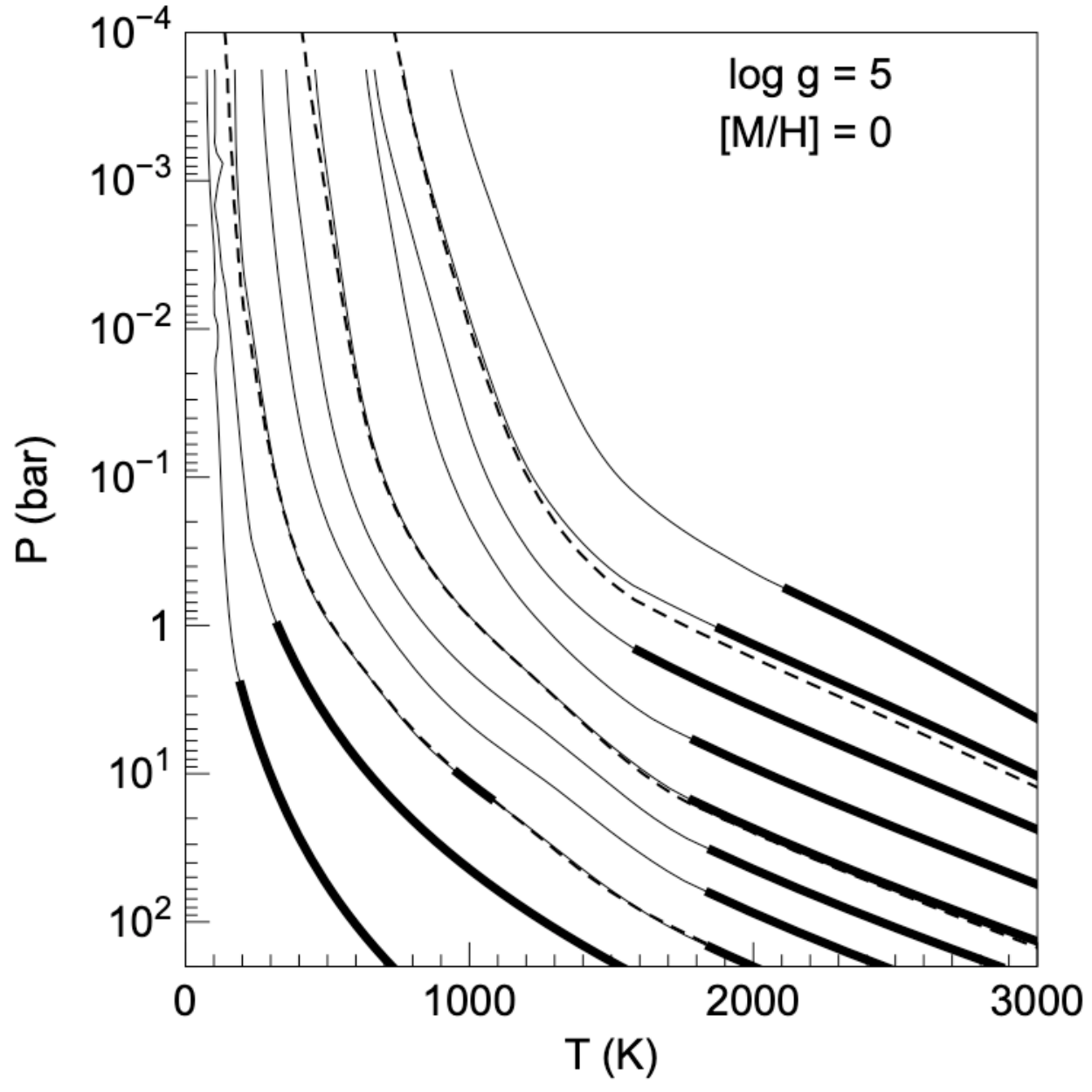} 
%    \end{subfigure}
    \hfil
%    \begin{subfigure}
%[t]{0.49\textwidth}
        \centering
        \includegraphics[width=0.33\textwidth]{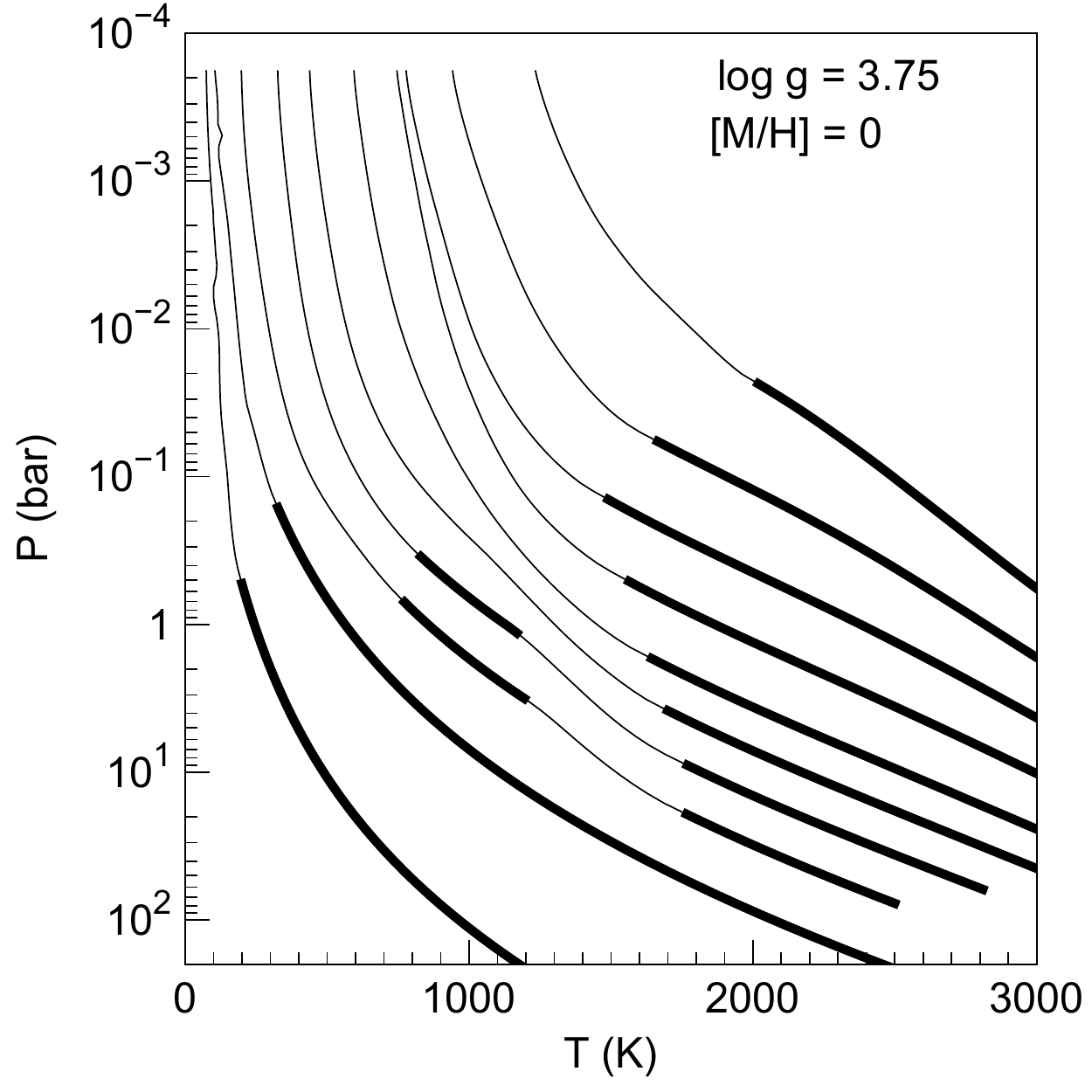} 
%    \end{subfigure}
    \hfil
    \vspace{0.5cm}
    \hfil
%    \begin{subfigure}
%    [t]{0.49\textwidth}
        \centering
        \includegraphics[width=0.33\textwidth]{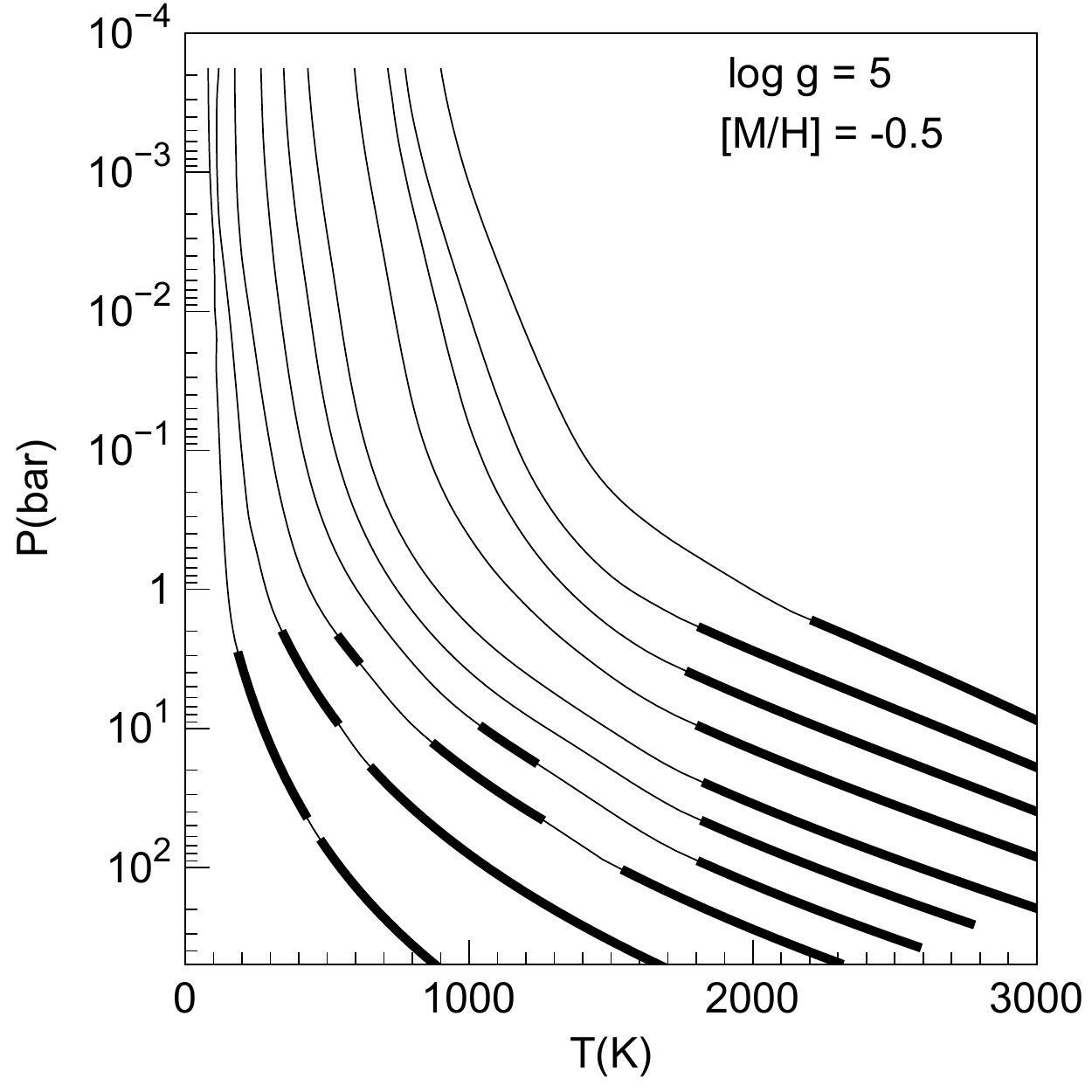} 
%    \end{subfigure}
    \hfil
%    \begin{subfigure}
%[t]{0.49\textwidth}
            \centering
        \includegraphics[width=0.33\textwidth]{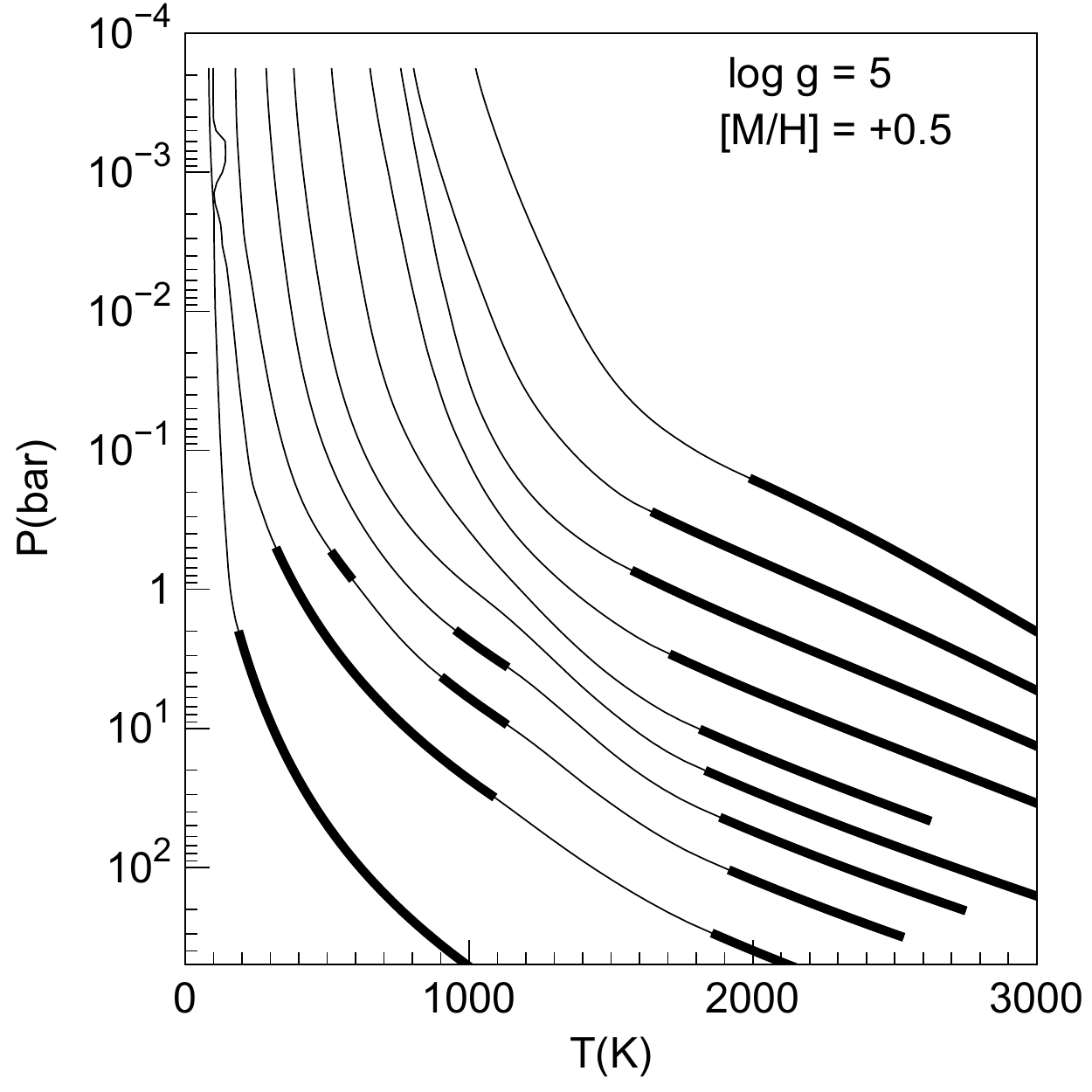}
        \hfil
%    \end{subfigure}
    \caption{Converged radiative-convective equilibrium temperature profiles computed by our modeling approach for various parameters as noted in each panel. In all panels models are shown every 200 K from $T_{\rm eff}=200$ to 1200 K and then every 300 K to 2400 K. Thick lines show convective regions of the atmospheres. Note how at higher effective temperatures the boundary lies near 2000 K, then with falling $T_{\rm eff}$, convection turns on in some portions of the radiative region, until finally these isolated zones merge and the boundary moves up to near 1 bar in the coolest models. Dashed curves show cloudless, chemical equilibrium profiles from \citet[][]{Phillips20} for $\log g =5$ and $T_{\rm eff}$ 600, 1200, and 2100 K.
    }
\label{fig:profiles4}
\end{figure*}

Once there is a converged solution for the top of the deepest convection zone the radiative equilibrium profile in the remainder of the atmosphere is compared to the local adiabat. Convective layers are inserted, one at a time, as necessary. Examples of converged radiative-convective equilbrium models with one, two and three convection zones are shown in Figure \ref{fig:profiles4} for a selection of model gravities and metallicities. 

For most cases the lowermost radiative-convective boundary falls around 2000 K at which temperature the peak of the Planck function falls near the $H$ band spectral window in water opacity, permitting cooling to space. 
%{\bf (Deleted:} As expected the radiative-convective boundary for higher metallicity models is located at lower pressures in the atmosphere than low metallicity models for the same effective temperature and gravity.}
\added{As expected from stellar atmosphere theory, the higher overall opacity of higher metallicity models shifts the entire structure to lower pressures for the same effective temperature and gravity.}

%\begin{figure*}[htbp]
%    \centering
%    \includegraphics[width=0.95\textwidth]{figures/T_profiles.pdf}
%    \caption{Set of converged radiative-convective equilibrium models for atmopsheres computed by our modeling approach for two different gravities noted in upper right. In both panels models are shown in steps of 200 K from $T_{\rm eff}=200$ to 1200 K and then every 300 K up to 2400 K. Thick lines show convective zones in the atmosphere. Note how the convection zone pulls back from the $\sim 1\,\rm bar$ region, breaks up, then further retreats as $T_{\rm eff}$ increases.\textcolor{red}{DS: Fig2 panels are too small}
%    \label{fig:profiles}
%\end{figure*}

\subsection{Opacities}

\begin{deluxetable*}{lll}
\tablecaption{Molecular Opacity Sources}
\label{tab:opacity}
\tablewidth{0pt}
\tablehead{\colhead{Molecule} & \colhead{Opacity Source(s)  \tablenotemark{a}}&\colhead{Line widths\tablenotemark{b}}
}

\startdata\\[-8pt]
C$_2$H$_2$ & H12 & W16   \\
 C$_2$H$_4$ & H12 & air  \\
 C$_2$H$_6$ & H12 & air  \\
 CH$_4$ &  \citet{Yurchenko13, Yurchenko14}\tablenotemark{d}; $\rm ^{13}CH_4$ STDS & P92\\
 CO & HT10; isotopologues \citet{Li15}&L15 \\
 CO$_2$ &  \citet{Huang14}& scale \\
 CrH &  \citet{Burrows02}& lin.  \\
 FeH &  \citet{Dulick03}; \citet{Hargreaves10} &lin. \\
 H$_2$O & \citet{Tennyson18};  isotopologues (HDO not included) \citet{Barber06} & UCL \\
 H$_2$S & ExoMol\tablenotemark{d}; \citet{Azzam15};  isotopologues H12 & K02   \\
 HCN & \citet{Harris06}; \citet{Harris08};  isotopologues GEISA\tablenotemark{f} & lin.\\
 LiCl & \citet{Weck04}\tablenotemark{c}& lin. \\
 MgH & \citet{Weck03b}\tablenotemark{c}& lin. \\
 N$_2$ &  H12 & air \\
 NH$_3$ & \citet{Yurchenko11}& W16  \\
 OCS & H12 & W16  \\
 PH$_3$ &  \citet{SousaSilva15}\tablenotemark{d} & S04 \\
 SiO & \citet{Barton13}; \citet{Kurucz11}\tablenotemark{f} & lin. \\
 TiO & \citet{Schwenke98}; \citet{Allard00} & lin. \\
 VO &  \citet{McKemmish16}; ExoMol & lin. \\
\enddata
\tablenotetext{a}{H12 = HITRAN 2012 \citep{Rothman13}; {\href{http://www.cfa.harvard.edu/hitran/updates.html}{http://www.cfa.harvard.edu/hitran/updates.html}};   HT10 = HITEMP 2010 \citep{Rothman10}; {\href{http://www.cfa.harvard.edu/hitran/HITEMP.html}{http://www.cfa.harvard.edu/hitran/HITEMP.html}}}
\tablenotetext{b}{ lin. = linear estimate for $\gamma$, see text; air = air widths from H12; scale = $1.85\times$ self broadening; K02 = \citet{Kissel02}; L15 = \citet{Li15}; P92 = \citet{Pine92}; S04 = \citet{Salem04}; W16 = \citet{Wilzewski16}; UCL = ExoMol web page$^{\rm d}$}
\tablenotetext{c}{\href{http://www.physast.uga.edu/ugamop/}{http://www.physast.uga.edu/ugamop/} }
\tablenotetext{d}{{\href{http://www.exomol.com}{http://www.exomol.com}} \citep{Tennyson12}}
\tablenotetext{e}{\href
{http://ether.ipsl.jussieu.fr/etherTypo/?id=950}{http://ether.ipsl.jussieu.fr/etherTypo/?id=950}}
\tablenotetext{f}{\href{http://kurucz.harvard.edu/molecules.html}{http://kurucz.harvard.edu/molecules.html}}
\end{deluxetable*}

We consider the opacity of 20 molecules and five atoms  (see Table II). Details on how line widths are applied to a given line list and the opacity calculation in general are presented in \citet{Freedman08,Freedman14,Lupu14}.  Since those publications we have updated several notable sources of opacity including $\rm H_2O$, $\rm CH_4$, the alkali metals, and FeH. Below we discuss our opacity sources for these species as well as our construction of opacity tables for use in the radiative-convective
model and for the calculation of high spectral resolution spectra. We note that opacity line lists are constantly being updated and it is necessary to freeze the choice of line lists in order to produce a model set. Future versions of the Sonora models will include updated opacities as warranted.

\subsubsection{Neutral Alkali Metals and Atoms}
We  use a new calculation of atomic line absorption from the neutral alkali metals (Li, Na, K, Rb and Cs). These are now included using the VALD3 line list\footnote{\url{http://vald.astro.univie.ac.at/~vald3/php/vald.php)}} \citep{Ryabchikova15}.
Atomic line profiles, with the exception of the \ion{Na}{1} and \ion{K}{1} D lines, are  assumed to be Voigt profiles without applying a line cutoff in 
strength or frequency.  The line width is calculated from the 
Van der Waals broadening theory for collisions with H$_2$ molecules using the coefficient
tabulated in the VALD3 data base when available or from the codes of P.~Barklem {\href{https://github.com/barklem}{(https://github.com/barklem)} otherwise. In all cases the classical Uns\"old width \citep[e.g.,][]{Kurucz1981} is corrected for a background gas of $\rm H_2$ and He rather than atomic H, accounting for the differences in polarization and reduced mass.\added{The choice of line cutoff for the atomic species, with the exception of the D lines, has no material effect on the models.}

The D resonance doublets of ${\rm Na\, \scriptstyle I}$ ($\sim 0.59\,\mu$m) and 
${\rm K\, \scriptstyle I}$ ($\sim 0.77\,\mu$m) can become extremely strong in
brown dwarf spectra and their line profiles can be detected as far as $\sim 3000\,\rm cm^{-1}$
from the line center in T dwarfs \citep{Burrows00b, Liebert00, Marley02}.  
Under these circumstances, a Lorentzian
line profile becomes grossly inadequate in the line wings and a detailed calculation 
is required.  

For these two doublets, we have implemented line wing profiles based on the unified line shape theory \citep{Allard07, Allard07a, Rossi85}. The tabulated profiles (Allard N., private communication) are calculated for the D1 and D2 lines of
${\rm Na\, \scriptstyle I}$ and ${\rm K\, \scriptstyle I}$ broadened by collisions with H$_2$ and He, for temperatures ranging between 500 and 3000\,K and at a reference
perturber (H$_2$ or He) number density of $n_{\rm pert}=10^{19}$ (K--H$_2$ profiles) or $10^{20}\,$cm$^{-3}$ (Na--H$_2$, Na--He and K--He profiles). 
Two collisional geometries are considered for broadening by H$_2$ and averaged to obtain the final profile. 
The line core is described with a symmetric Lorentz profile with a width calculated from
the same theory, with  coefficients given in  \citet{Allard07a}. 

The line profiles are provided as a set of coefficients in a density expansion that allows their evaluation at a range of densities
other than the reference density of the tabulation. The third order expansion is considered suitable up to perturber densities of $10^{20}\,$cm$^{-3}$
(Allard, N., priv. communication) and the Lorentzian line width is linear in perturber density \citep{Allard16}. Using those expressions and
by interpolating in temperature, we produce a set of profiles covering 500 - 3000$\,$K and $15 \le \log n_{\rm pert} \le 20$ on a uniform grid. In 
atmosphere models and spectra 
that may exceed these limits, we refrain from extrapolating. This is generally acceptable since the ${\rm Na\, \scriptstyle I}$ and ${\rm K\, \scriptstyle I}$ resonance doublets 
play a lesser role in the total opacity outside of these boundaries. Nonetheless, calculations that reach higher densities are valuable.
After the models presented here were computed, Allard and collaborators presented new tables for K--H$_2$ \citep{Allard16}
and Na--H$_2$ \citep{Allard19} valid to higher pressure. We will use those tables in future updates to the model grid. Tests show that while the new treatment does impact the model spectral slope near $1\,\rm \mu m$, it does not appreciably alter the temperature profiles from the present model generation.

\subsubsection{FeH}

For FeH we use the line list of \citet{Dulick03} which was our
opacity source in previous models. However this list did not include the E-A band at $1.6\,\rm \mu m$ which
is prominent in M and L dwarfs. Here we  include this rovibrational $E\,^4\Pi_i-A\,^4\Pi_i$ electronic
transition, employing a line list from \citet{Hargreaves10}. This list was constructed by fitting to empirical spectra of cool M dwarfs and to laboratory measurements. 
As a consequence there are multiple uncertainties in the line list, including the lower energy level of many of the lines, which are set to a constant value. \citet{Hargreaves10} ultimately applied an enhancement factor of 2.5 to their initial line strengths in stellar atmosphere models in order to match observations.
In our own initial test models employing this line list we found that the predicted band strength as seen in our models was far in excess of that observed in
early L dwarfs. After discussions with Hargreaves (priv. comm.) on various sources of uncertainty we decided to reduce the line strengths, for this band only, by a factor of 1/3, slightly over correcting to remove the correction factor of 2.5 to better match observations. This is the only absorption band in our entire model set for which we have applied such an empirical correction.

\subsubsection{Line Profiles}
 The proper choice of line widths to apply to each individual molecular line is a difficult problem as there is often little to no theoretical or laboratory guidance, particularly for higher quantum number transitions that are important at higher temperatures. For each molecule we have applied the best information available at the time our opacity database was constructed although in many cases we have had to estimate the broadening  with limited information.
 Table II summarizes our choices for line widths, including cases in which we used literature values or widths appropriate to air, rather than $\rm H_2 - He$ mixtures.
 
 In most molecular line lists, the maximum $J$ quantum number is above 100. However, Lorentz broadening coefficients are typically only available up to  $J\sim 50$, which we term $J_{\rm low}$. 
 In such cases where data is lacking, we extrapolate the broadening parameter $\gamma$ by assuming a value, $\sim 0.075\,\rm cm^{-1}\,atm^{-1}$,  for the lowest $J$ value in any set then adjusting the broadening at any $J$ by a linear expression in $J- J_{\rm low}$ up to some maximum $J$. Above that $J$, $\gamma$ is held fixed at $\sim 0.04\,\rm cm^{-1}\,atm^{-1}$. A similar approach has been followed by the UCL group (Tennyson, pers. comm.).
 
% = J_{\rm low} - fJ$ where the scale factor $fJ$
%is adjusted to give a low value of the broadening ($\sim .04$ or less
%at a cutoff value of $J$ typically $\sim 50$). $J_{\rm low}$ is typically estimated as  $\sim 0.075\,\rm cm^{-1}/atm$. Above that $J$, $\gamma$ is held fixed. A similar approach has been followed by the UCL group (Tennyson, pers. comm.).\textcolor{red}{DS: This is not very clear}
 
%In most molecular line lists, the maximum $J$ quantum number is above 100. However, Lorentz broadening coefficients are only available up to $J_{\rm low}\sim 50$ and are estimated as
%$\sim 0.075\,\rm cm^{-1}/atm$ for $J=50$.
%In these cases, we use a small constant number for Lorentz coefficients for $J < J_{\rm low}$ $(\sim 0.04\,\rm cm^{-1}/atm$ or less). A similar approach has been followed by the UCL group (Tennyson, pers. comm.).

\subsubsection{Opacity Tables}
Opacities are computed at each of 1060 distinct pressure-temperature points covering the range $75 \le T \le 4000\,\rm K$ and $10^{-6} \le P \le 300 \,\rm bar$ 
 on a wavenumber grid constructed such that there are about 3 points per Doppler width for the $\rm H_2O$ molecule. This typically amounts to roughly $2\times 10^6$ individual points for intermediate $T$ and $P$ with up to $10^7$ points at the lowest pressures 
 at which we compute the opacity from up to $10^{10}$ individual spectral lines (for example in the case of $\rm CH_4$).  This tabulated opacity is used both in post-processing to construct high resolution spectra for individual models and also to compute {\it k}-coefficients which are used in the atmospheric structure calculation. 

For the {\it k}-coefficient calculation we sum the contribution of every molecule to the total molecular opacity by weighting by their relative equilibrium abundances (see next section). Within each of 190 spectral bins covering the wavelength range 0.4 to $320\,\rm \mu m$, we then compute the {\it k}-coefficients using the summed opacity. We note that this is more accurate than later combining {\it k}-coefficients computed for individuals gasses \citep[e.g.,][]{Lacis91,Amundsen17}}. This approach, however, removes flexibility in adjusting local gaseous mixing ratios, for example to account for disequilibrium chemistry effects  (see \citet{Amundsen17} for a recent discussion). 
We use 8 Gauss points for the {\it k}-coefficients, following a double Gaussian scheme in which 4 points cover the range 0 to 0.95 of the cumulative distribution and 4 additional points cover the range 0.95 to 1.00. This permits more precise resolution of the strongest few percent of the molecular lines within any given spectral bin. Tests (M. Line, pers. comm.) have shown that this double-Gauss approach yields essentially identical thermal profiles computed with 20 Gauss points covering the full distribution range of 0 to 1.0 and the same opacities. The {\it k}-coefficients for all of the model cases presented here are available for download  \dataset[here]{https://zenodo.org/record/4755012}. 

In addition to these opacity sources we also account for several other continuum opacity sources. Pressure-induced opacity from $\rm H_2$--$\rm H_2$, $\rm H_2$--$\rm H$, $\rm H_2$--$\rm He$, $\rm H_2$--$\rm N_2$, and $\rm H_2$--$\rm CH_4$ is accounted for as described in \citet{Saumon12}. We also include bound-free and free-free opacity from $\rm H^-$ and $\rm H_2^+$ and free-free opacity from $\rm H_2^-$ and $\rm He^-$ as well as electron scattering. Rayleigh scattering from $\rm H_2$, H, and He is also included. We note in passing that, despite the recent 'discovery' of the importance of $\rm H^-$ opacity in transiting planet atmospheres\footnote{e.g., \citet{Arcangeli18}  proposed $\rm H^-$ and $\rm H_2O$ dissociation as the natural, expected, solution to explain Ultra-Hot Jupiter spectra, as earlier  studies neglected to include this opacity source.}, H$^-$ has long been recognized as an important source of continuum opacity in cool stars \citep{chandra46}. Our models  have always included this opacity source, since \citet{Marley96}. The models in this paper do not include cloud opacity, our method for accounting for such opacity will be presented in a future paper in the series.

\subsection{Radiative Transfer}
In the radiative-convective model we compute radiative fluxes through each model layer using the `two-stream source function method' outlined in \citet{Toon89}. This scheme first computes a two stream (up and down) solution to the flux and then uses the two stream
 solution as the source function for scattering in the second calculation step that computes the flux in six discrete beams. 
 Following Toon et al., within each
 model layer $n$, with optical depth ranging from $\tau=0$ at the top to $\tau=\tau_{\rm bot}$ at the bottom, the source function is linearized as $B_n(\tau)=B_{0n}+B_{1n}\tau$ where $B_{0n}$ is the Planck function at the temperature of the top of the layer and $B_{1n}=[B(T(\tau_{\rm bot}))+B_{0n}]/\tau_n$.
 The final upwards and downwards fluxes are computed by integration over the multiple streams. The method is exact for pure absorption and provides an acceptable balance between accuracy and speed for cases with particle scattering. \citet{Toon89} provide tables of the size of the error in various cases. The lower and upper boundary conditions are those commonly used in the stellar atmospheres problem for semi-infinite atmospheres \citep{Mihalas14}. Recently
 \citet{Heng17} have extended the Toon et al.\ method for greater accuracy in strongly scattering atmospheres, a limit we do not reach in the cloudless models
 presented here.

Once we have a converged $T(P)$ model we compute high resolution spectra
by solving the radiative transfer equation for nearly 362,000 monochromatic frequency
points between 0.4 and 50$\,\mu$m.  The monochromatic opacities are calculated from the same opacity data base, line broadening, and chemistry
tables as the {\it k}-coefficients and are pre-tabulated on the same $(T,P)$ grid. In these cloudless models, the radiative transfer
equation is solved with the Rybicky solution \citep{Mihalas14} assuming that Rayleigh scattering is isotropic.
The resulting spectral energy distributions are in excellent agreement with those obtained from the lower resolution
{\it k}-coefficients method with their respective integrated fluxes differing by less than 2\% in nearly all cases.

\subsection{Chemical Equilibrium}

Chemical equilibrium abundances at the grid $(P,T)$ points were calculated using a modified version of the NASA CEA Gibbs minimization code 
\cite[see][]{Gordon94}, based upon prior thermochemical models of substellar atmospheres \citep{Fegley94,Fegley96,Lodders99,Lodders02,Lodders02b,Visscher06,Visscher10,Visscher12} and recently used to explore gas and  condensate chemistry over a range of atmospheric conditions \citep{Morley12,Morley13,Moses13,Kataria16,Skemer16,Wakeford17,Burningham17}.
 
 Equilibrium abundances (with a focus on key constituents that are included in the opacity calculations: 
$\rm H_2$, H,  H$^+$, H$^-$, $\rm H_2^-,\ H_2^+,\ H_3^+$, He, $\rm H_2O,\ CH_4,$ CO, $\rm CO_2$, OCS, HCN, $\rm C_2H_2,\ C_2H_4,\ C_2H_6,$
$\rm NH_3,\ N_2,\ PH_3,\ H_2S,$ SiO, TiO, VO, Fe, FeH, MgH, CrH, Na, K, Rb, Cs, Li, LiOH, LiH, LiCl, and $\rm e^-$) were calculated over a wide range of atmospheric pressures ($1\,\rm \mu bar$ to 300 bar) and temperatures (75 to 4000 K), and over a range of metallicities (-1.0 dex to +2.0 dex relative to solar abundances, assuming uniform heavy element enrichment), and C/O element abundance ratios (0.25 to 2.5 times the solar C/O abundance ratio of $\rm C/O = 0.458$) using protosolar abundances from \citet{Lodders10} \added{which represent the bulk solar system composition and provides continuity with earlier iterations of the chemical models. Other C/O ratios can be adopted (e.g., 1.25$\times$ C/O, corresponding to C/O=0.57) for consistency with more recent determinations of the photospheric C/O ratio \citep[e.g.,][]{Asplund09,Caffau11,Lodders20,Asplund21}}. 

For a given metallicity, variations in the C/O ratio were computed while holding the C+O abundance constant, so that the total heavy element abundance relative to hydrogen ($Z/X$), characterized by $[M/H]$, remains constant. \added{For example, to achieve C/O ratios greater than the solar value (e.g., $1.25\times$ or $1.5\times$ the adopted solar ratio of $\rm C/O=0.458$), the oxygen abundance is slightly diminished while the carbon abundance is slightly enhanced.}
For most species, we utilized the thermochemical data of \citet{Chase98} with additional thermochemical data from \citet{Gurvich89,Gurvich91,Gurvich94,Burcat05} and \citet{Robie95} for several mineral phases.  

Condensation from the gas phase is included with the ``rainout'' approach of Lodders \& Fegley, wherein condensates are removed from the gas mixture lying above the condensation level. This prevents further gas-condensate chemical reactions from occurring higher up in the atmosphere, above the condensation point. Rainout has been validated for alkali species by the sequence of the disappearance of Na and K 
spectral features \citep{Marley02,Line17,Zalesky19}.   

As in previous thermochemical models, the equilibrium abundance of any condensate-forming species at higher altitudes is determined by its vapor pressure above the condensate cloud \cite[e.g., see][]{Visscher10}.  The detailed equilibria models of \citet{Lodders02b} show that several elements (such as Ca, Al, Ti) may be distributed over a number of different condensed phases depending upon pressure and temperature conditions.  For simplicity, here we consider the vapor pressure behavior of TiO and VO in equilibrium with $\rm Ti_2O_3$, $\rm Ti_3O_5$, or $\rm Ti_4O_7$, 
and  $\rm V_2O_3$, $\rm V_2O_4$, and $\rm V_2O_5$ respectively as we are primarily interested in the behavior of Ti and V  above the cloud deck and the current grid lacks the resolution for detailed Ca-Ti equilibria as a function of temperature.  The behavior of Mg-, Si-, and Fe-bearing gases was calculated following the approach of \citet{Visscher10} and included the equilibrium condensation of forsterite ($\rm Mg_2SiO_4$), enstatite ($\rm MgSiO_3$), and Fe metal.  As in \citet{Lodders99}, major condensates for the alkali elements included LiCl, LiF, $\rm Na_2S$ \cite[see also][]{Visscher06}, KCl \cite[see also][]{Morley12} RbCl, and CsCl.

\begin{figure*}[htbp]
    \centering
    \includegraphics[width=0.95\textwidth]{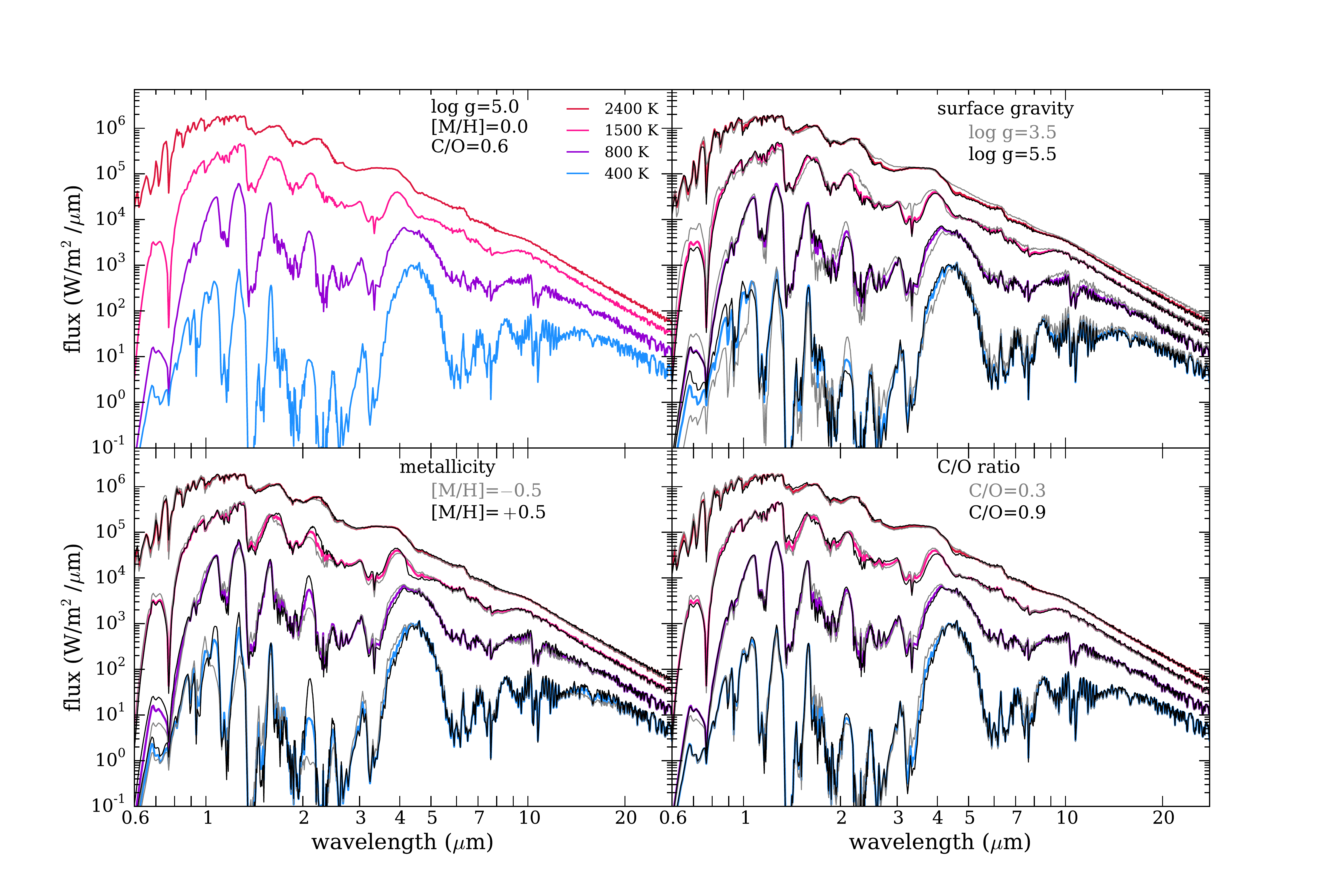}
    \caption{Overview of complete model spectra for various combinations of surface gravity, metallicity, and C/O ratio. The line color labels the  effective temperature as shown in the upper left panel. Spectral resolution $R$ varies across the figure for clarity. Effective temperatures as labeled in top left panel. Models shown in top left panel are repeated in the other three panels along with two variations (in gravity, metallicity, and C/O ratio) in black and grey, as labeled in each sub-panel. Of particular interest is the effect of metallicity on {\em K}-band emission \citep[][]{Fortney08b}}.
    \label{fig:quad_comp}
\end{figure*}

\begin{figure*}[htbp]
    \centering
    \includegraphics[width=0.95\textwidth]{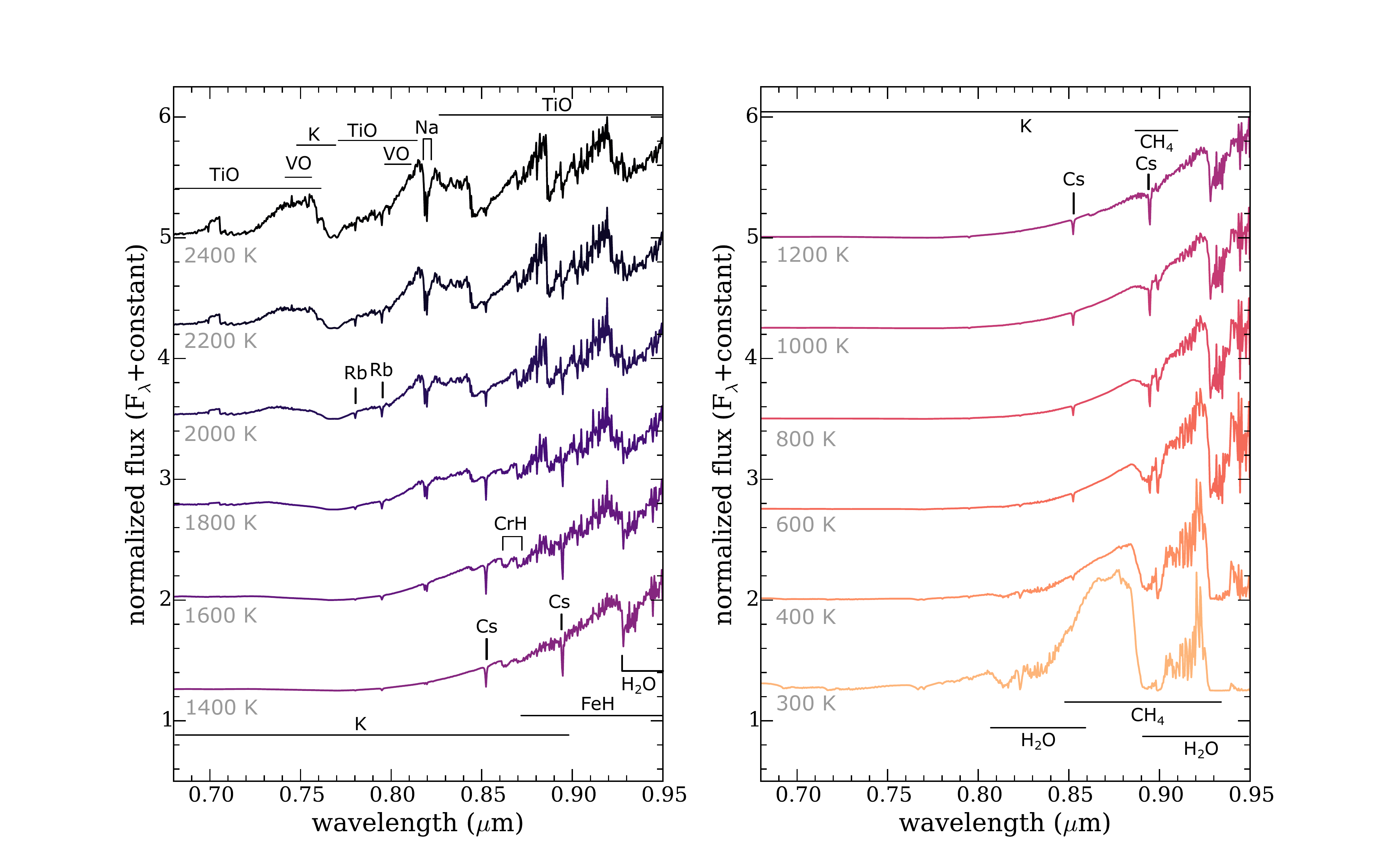}
    \caption{Red optical spectral sequence ($R\sim 2,000$) of models for solar metallicity and C/O ratio at $\log g=5$, shifted vertically for clarity. Notable absorption features are labeled. Models are shifted for clarity.}
    \label{fig:sequence0}
\end{figure*}
\begin{figure*}[htbp]
    \centering
    \includegraphics[width=0.95\textwidth]{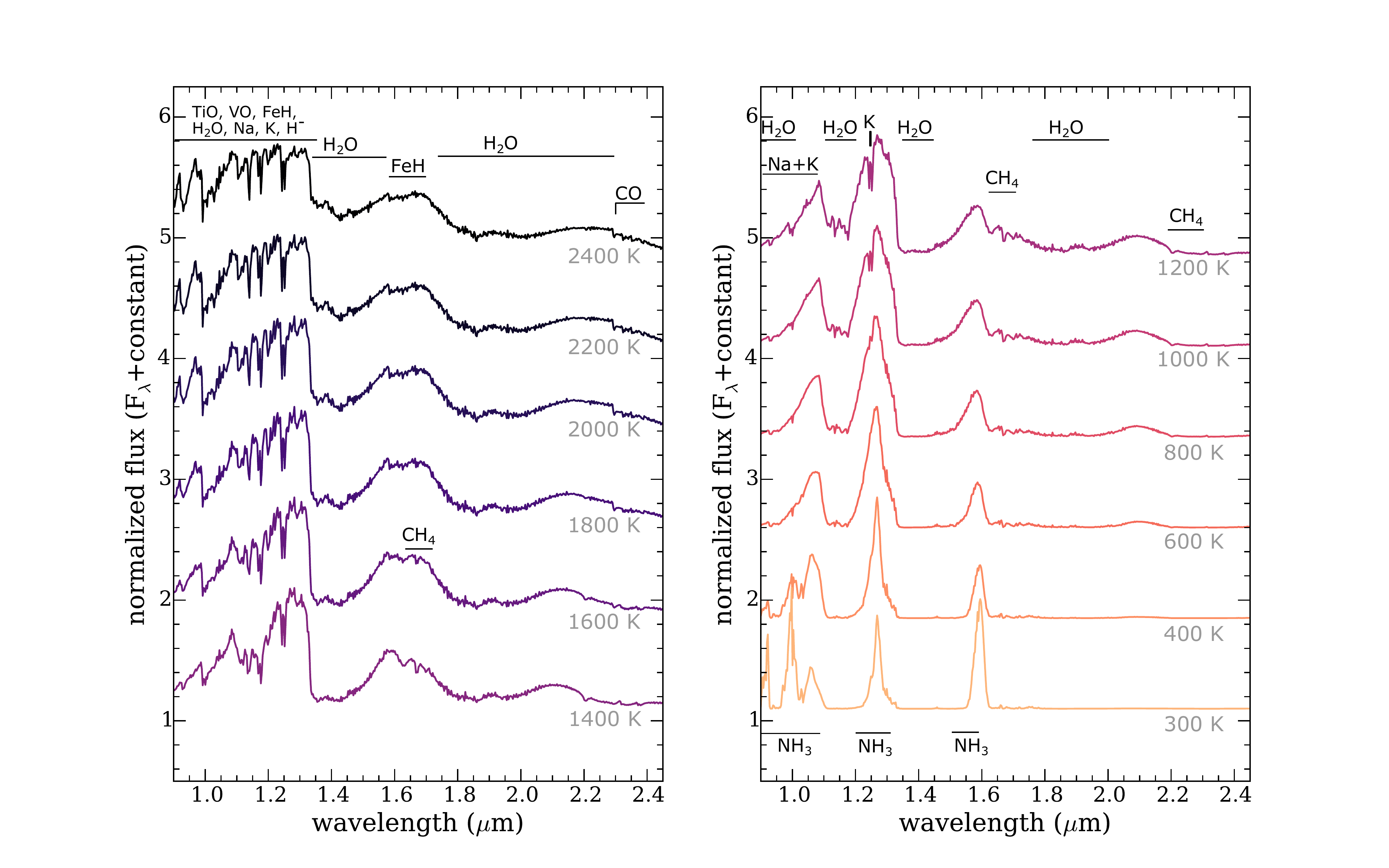}
    \caption{Same as Figure \ref{fig:sequence0} but for far red through near-IR spectral range with $R\sim600$. }
    \label{fig:sequence1}
\end{figure*}
\begin{figure*}[htbp]
    \centering
    \includegraphics[width=0.95\textwidth]{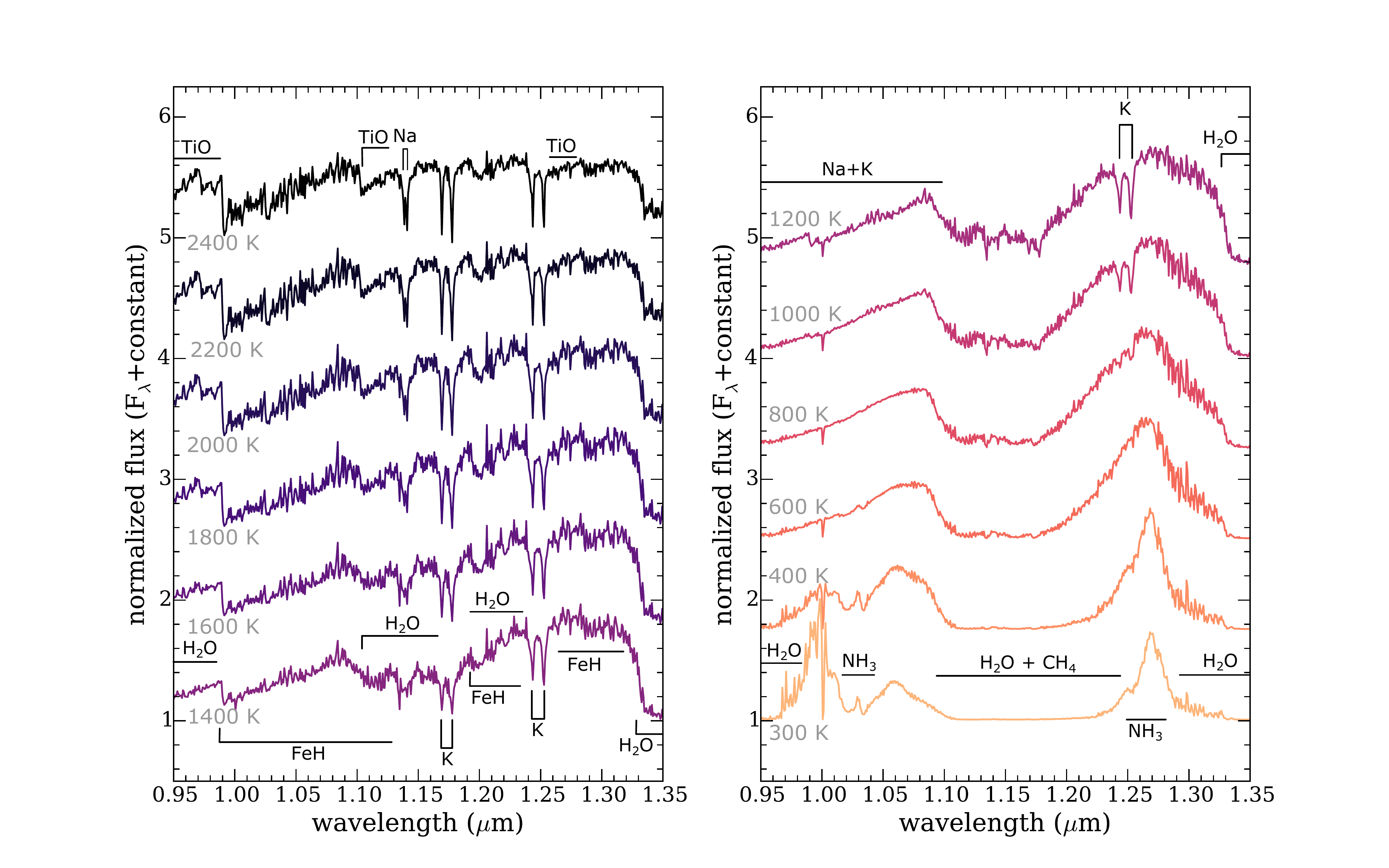}
    \caption{Same as Figure \ref{fig:sequence0} but highlighting the spectral region near $1\,\rm \mu m$ with $R\sim 1,500$.  In the coldest models the loss of almost all major gaseous absorbers, particularly the alkali metals, but excepting $\rm H_2O$, $\rm NH_3$, and $\rm CH_4$ transforms the appearance of this spectral region.}
    \label{fig:zoom}
\end{figure*}
\subsection{Comparison to Other Approaches}

As mentioned above, the construction of any forward model involves numerous choices and trade offs. When comparing models from different groups, it is worth keeping in mind the various approximations and assumptions behind the models. Here we highlight a few such differences.

Our modeling scheme, with roots in solar system atmospheres, was first applied to a brown dwarf (Gl 229 B) in \citet{Marley96}. That same year \citet{Allard96} likewise applied the {\tt PHOENIX} modeling scheme, with roots in stellar atmospheres, to the same object. Both models developed over time with
the  {\tt PHOENIX} model ultimately producing the widely cited {\tt COND} and {\tt DUSTY} model sets \citep{Allard01}. In contrast to our approach of using pre-tabulated chemistry and opacities, the {\tt PHOENIX} model computes chemistry and opacities on the fly.  {\tt COND} also employs full equilibrium chemistry rather than rainout chemistry to compute molecular abundances. As we note above, recent retrieval studies have validated rainout chemistry in the context of the T dwarfs. 
To compute fluxes {\tt PHOENIX} uses the sampling method in which the radiative transfer is only computed at a finite number ($\sim10^4$, T. Barman, pers. comm.) of wavelength points. 
Our method accounts for the opacity and uses the information at many more ($10^6$ to $10^7$) points, but only in a statistical sense as the opacity distribution within wavelength bins is 
described by {\it k}-coefficients. 

A more recent version of the {\tt PHOENIX} models is known at BT-Settl \citep[][]{allard14}, where the BT denotes the source of the water opacity employed \citep[][]{Barber06} and the Settl denotes the handling of condensate opacity. These models have not been as thoroughly described and a more detailed discussion is not yet possible. 

Recent studies of radiative transfer for transiting planets have found that the {\it k}-coefficient method more closely reproduces calculations performed at very high spectral resolution that the sampling method, since the entire range of both low and high opacity wavelengths is considered \citep[][]{garland19}.  In practice for most of the atmospheres considered here this is unlikely to be a major
difference. However our own tests of fluxes computed with opacity sampling find that at low temperatures, where
there are relatively few opacity sources and the opacity can vary wildly with wavelength, sampling can produce large errors unless many more points are employed than typically used. A systematic comparison between the approaches would be  informative. Differences between the approaches in the context of cloud opacity will be discussed in a future paper.

Another widely cited model set is that of the Adam Burrows group which consists of two different approaches. 
The models presented in \citet{Burrows97, Burrows01} were computed using the
\citet {Marley96, Marley99} radiative-convective model described above. After 2001  the Burrows group transitioned to a modeling framework based on a widely used stellar atmospheres code {\tt TLUSTY} \citep{Hubeny88, Hubeny95} adapted for use in brown dwarfs \citep[e.g.,][]{Burrows02, Burrows04,Burrows06}.
This work uses the sampling method to handle the opacities, like {\tt PHOENIX}, but with the ability to handle rainout and quenching \citep{Hubeny07}
rather than relying on pure equilibrium chemistry. There are also differences
in the treatment of radiative transfer and the global numerical method used to solve the set of structural equations. \citet{Hubeny17} provides a thorough, critical comparison of various approaches employed in substellar models.

Recently  \citet{Malik17} have released the open-source atmosphere radiative-equilibrium modeling framework {\tt HELIOS}. As of this date the model  structures are strictly in radiative equilibrium. {\tt HELIOS} computes atmospheric structure assuming true chemical equilibrium, not rainout chemistry, using a fast analytic approximation of the C, N, O chemistry \citep{Heng16}. The model uses {\it k}-coefficients of individual molecules which they combine on the fly using `random overlap' approximation rather
than the pre-mixed opacity tables employed here or the `re-bin and re-sort' method usually used to combine {\it k}-coefficients (see \citet{Amundsen17}). While fast, this approximation can produce large flux errors in certain cases because in fact
the opacities are not random but are vertically correlated through the atmosphere \citep{Amundsen17}.  There has not yet been a systematic comparison
of model atmospheres computed by {\tt HELIOS} with other models using more traditional methods for computing non-irradiated substellar atmospheres.

\added{As noted in the introduction, the model set most similar to our own is \citet[][]{Phillips20}. Those authors made generally similar modeling choices, including rainout equilibrium, although for the opacities they mix single gas k-coefficients following \citet[][]{Amundsen17} rather than pre-computing them for the mixture, as we do here. They do not discuss whether their models produce multiple layer convection zones, although it appears from their posted structures that they in fact do, with generally similar structure as our Figure 4. Their models are only for solar metallicity and those authors advise users to only trust models below $T_{\rm eff}=2000\,\rm K$, despite the grid going to higher temperatures as they neglect some important opacity sources. Where they can be compared, the Phillips $T(P)$ profiles are very similar to our own, as seen in Figure 1, except above 2000 K where they are slightly cooler, as expected given their neglect of some opacities. }

When comparing atmospheric abundances computed by our approach and other methods it is crucial to consider the impact of our rainout chemistry assumption \citep[see, e.g., ][]{Lodders06}. Because of rainout, certain species which can potentially be significant sinks of atoms of interest,
do not form. For example under rainout conditions $\rm Fe_2O_3$,
does not form as the Fe condensate is removed from the atmosphere. Rainout atmospheres cooler than the iron oxide condensation temperature will appear to have larger O abundances than other treatments with the same initial O abundance. Likewise, rainout causes the removal of condensed aluminum oxide $\rm Al_2O_3$ from the atmosphere, preventing the formation of albite (NaAlSi$_3$O$_8$) which would otherwise remove atomic sodium from the atmosphere at about 1000 K. Retrieval studies have shown  \citep{Zalesky19} that the rainout chemistry prediction is most consistent with the observed spectra of T dwarfs.

The rainout chemical equilibrium abundance tables used to compute the models presented here are available along with the models at our model page\footnote{ \dataset[https://doi.org/10.5281/zenodo.5063476]{https://doi.org/10.5281/zenodo.5063476}}.

\subsection{Evolution Model}
\label{sec:evoln_model}

The model for the interior and the evolution is discussed in detail in \cite{Saumon08} (hereafter, SM08). Briefly, the interior is modeled as fully 
convective and adiabatic and uses the atmosphere models described herein as the surface boundary condition.  We generate sequences
for the three metallicities of the atmosphere models ([M/H]= $-0.5$, 0 and 0.5).  The models are 
started with a large initial entropy (``hot start'') and include fusion of the initial deuterium content. 

We incorporate three
significant improvements over the models of SM08. First, we now account for metals in the equation of state by 
using an effective helium mass fraction \citep{cb97}
\begin{equation}
  Y^\prime = Y + Z
  \label{Yeff}
\end{equation}
where $Y=0.2735$ is the primordial He mass fraction and $Z$ the mass fraction of metals ($Z= 0.00484$, 0.0153 and 0.0484, \added{corresponding to $[M/H]=-0.5$, 0, and +0.5, respectively). The hydrogen and helium equations of state are from the tables of \cite{SCVH}, as in SM08.} Second,
we use the improved nuclear reaction screening factors of \cite{pc12}.
Third, the new surface boundary condition is provided by the  atmospheres presented here which are defined over a finer ($T_{\rm eff}$, $\log g$) mesh and extend to 
lower gravity ($\log g=3$) and lower $T_{\rm eff} = 200\,$K. This allows modeling of lower mass objects, down to 0.5\,$M_{\rm J}$.

\subsection{Previous Applications}
Our group has applied the same basic modeling approach described here to a number of topics related to ultracool dwarfs and extrasolar giant planets. This work is generally summarized in \citet{MarlRob15}. Notable applications have included computation of the evolution tracks presented in \citet{Burrows97}, characterization of L and T dwarfs from their near-infrared spectra in \citet{Cushing08, Stephens09}, calculation of atmospheric thermal profiles of irradiated giant planets \citep{Fortney05}, and modeling of young directly imaged planets in \citep{Marley12}. Our prediction that the clearing of clouds at the L/T transition would result in an excess of transition brown dwarfs \citep{Saumon08} was recently validated with the 20$\,$pc brown dwarf census of   \cite{kirkpatrick20}. 

In addition \citet{Fortney08b} investigated the atmospheric structure and spectra of cloud-free gas giants from 1--10 Jupiter masses, for models with $T_{\rm eff}<1400$K.  The role of enhanced atmospheric metallicity, an outcome expected from the core-accretion model of planet formation \citep[e.g.,][]{Fortney13}, was investigated as a way to observationally distinguish planets from stellar-composition brown dwarfs.  These atmosphere models were coupled to the ``cold start" and ``hot start'' evolutionary tracks of \citet{Marley07} to better understand the magnitudes and detectability of young giant planets.  Later, \citet{Fortney11} modeled the  evolution of the atmospheres of the solar system's giant planets to investigate how metal-enrichment and the time-varying (but modest) insolation effect our understanding of the cooling history of these planets.

While earlier models included the iron and silicate clouds that form in L dwarf atmospheres, \citet{Morley12} considered the salt and sulfide clouds \citep[][]{Visscher06} that likely form in cooler T dwarf atmospheres, finding that models that included the formation of these additional species could better match the colors of the T dwarf population. \citet{Morley14a} focused on even colder objects, the Y dwarfs, many of which are cold enough to condense volatiles like water into ice clouds. \citet{Morley14b} considered how either clouds or hot spots could cause wavelength-dependent variability in T and Y dwarf spectra.

\section{Model Results}

\subsection{Thermal and Composition Profiles}
Radiative-convective equilibrium thermal profiles are shown for a selection of model parameters in Figure \ref{fig:profiles4} for two gravities at solar metallicity and two metallicities at fixed gravity. The general behavior of the extent of the convection zones apparent in previous work is also seen here. At the highest effective temperatures
the single radiative-convective boundary lies near 2000 K and at pressures in the range of 1 to 0.01 bar, depending on gravity. As the effective temperature falls the top of this  convection zone stays near 2000 K but moves to progressively higher pressures. Finally, for effective temperatures below 1000K, the denser regions of the atmosphere begin to be cooler than about 1000 K, resulting in  the peak of the local Planck function lying not in the relatively clear, low opacity regions from 1 to $2\,\rm \mu m$ but rather in the higher gas opacity region between 2 and $4\,\rm \mu m$. Consequently a second detached convection zone forms around 1000 K.  

The second transition from convective to radiative transport occurs when sufficient flux can emerge through the spectral window at 4 to $5\,\rm \mu m$. In some cases a third small convection zone briefly develops higher in the atmosphere. Finally, with falling effective temperature, these detached zones merge and the atmosphere is finally fully, or almost fully, convective all the way up to about 0.1 to 1 bar, depending on gravity. The universality of $\sim 0.1\,\rm bar$ radiative-convective boundaries at low $T_{\rm eff}$ and $g$ has been  explored by \citet[][]{robinson14c}.

Correctly mapping the detached convection zones is important for the evolution calculation as they change the atmospheric temperature gradient away from that of pure radiative equilibrium. This in turn alters the temperature and pressure--and thus atmospheric entropy--of the deep radiative convective boundary, which ultimately controls the thermal evolution of the entire object. 

Likewise this behavior of the radiative regions rapidly merging, raising the radiative convective boundary to pressures near 1 bar at cooler effective temperatures is likely the explanation for the rapid rise in the eddy mixing coefficient, as inferred from observed disequilibrium chemistry below 400 K \citep[][]{miles20}.

\begin{figure*}[htbp]
    \centering
    \includegraphics[width=0.95\textwidth]{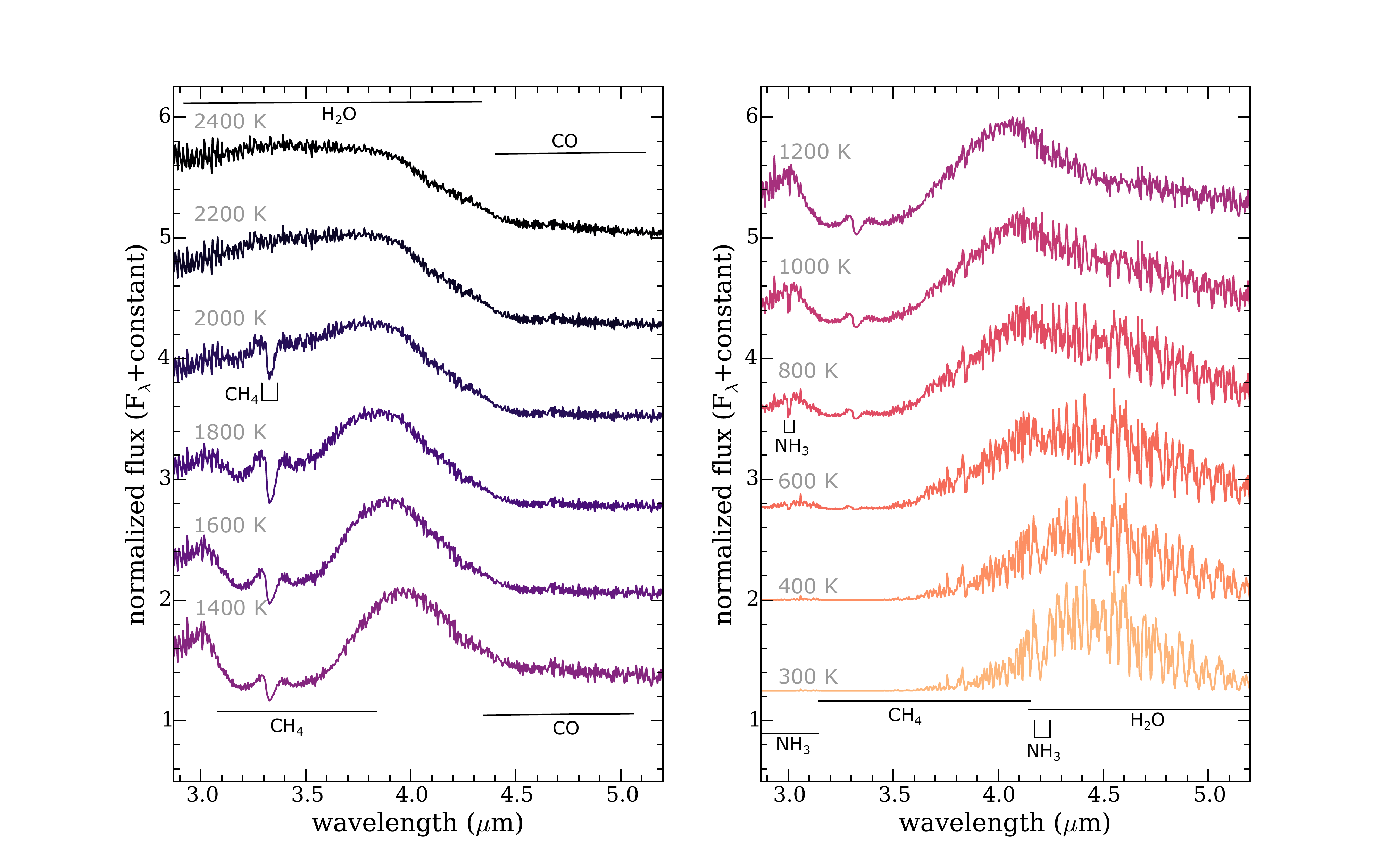}
    \caption{Same as Figure \ref{fig:sequence0} but for 3 to $5\,\mathrm{\mu m}$ spectral region and $R\sim 1,000$. Note how the peak in emission shifts from $\sim 4$ to $\sim 4.5\,\mathrm{\mu m}$ through this sequence.}
    \label{fig:sequence2}
\end{figure*}

\begin{figure*}[htbp]
    \centering
    \includegraphics[width=0.98\textwidth]{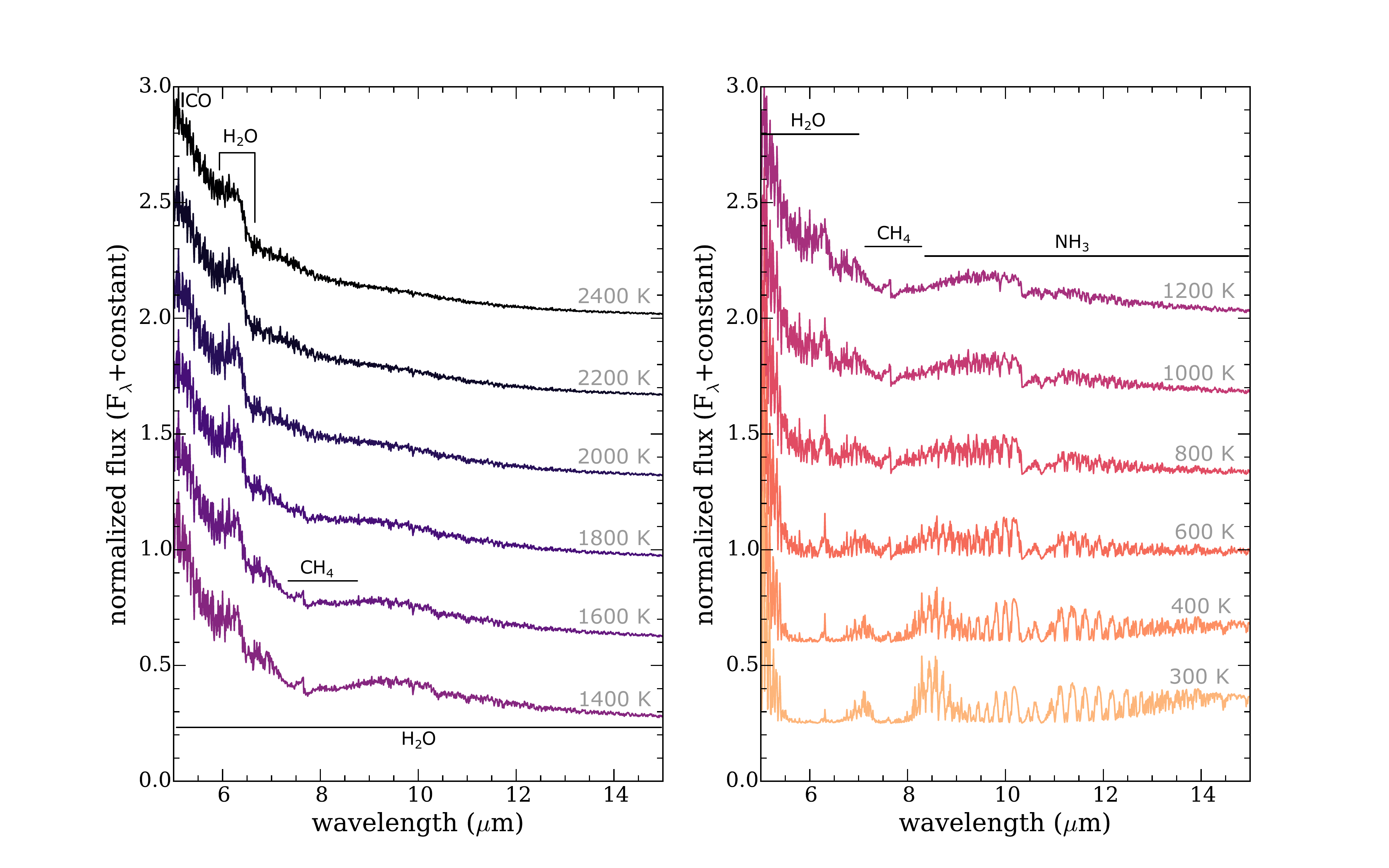}
    \caption{Same as Figure \ref{fig:sequence0} but for mid-IR spectral range with $R\sim 1,000$. Note how the arrival of $\rm NH_3$ dramatically transforms this region below about 1200 K in these equilibrium chemistry models.}
    \label{fig:sequence3}
\end{figure*}
\subsection{Model Spectra}

All of the model spectra described in this paper are available online\footnote{ \dataset[https://doi.org/10.5281/zenodo.5063476]{https://doi.org/10.5281/zenodo.5063476}} Here we briefly describe the characteristics of the model set.

Spectra for a selection of model cases and wavelength ranges are shown in Figures \ref{fig:quad_comp} through \ref{fig:sequence3}. While these cloudless models are of course less relevant to observed spectra through much of the L dwarf regime, they are nevertheless instructive for illustrating how the atmospheric chemistry evolves as effective temperature falls through the M, L,  T, and  Y spectral types.

As the atmosphere cools, molecular features become more prominent and the roughly blackbody spectra apparent at 2400 K is nearly unrecognizable by 400 K (Figure \ref{fig:quad_comp}). The universality of the M-band flux excess, first noted in \citet{Marley96} is apparent as are the familiar excesses in Y, J, H, and K bands. All of these arise from opacity windows allowing flux to escape from deep seated regions of the atmospheres where the local temperature often exceeds $T_{\rm eff}$. The folly, for most purposes, of attempting to describe the thermal emission of brown dwarfs or self-luminous planets as being blackbody-like is evident from casual inspection of Figure \ref{fig:quad_comp}.

Spectral sequences for limited spectral ranges are shown in Figures \ref{fig:sequence0} through \ref{fig:sequence3}. The red optical (Figure \ref{fig:sequence0}) is primarily sculpted by the \ion{K}{1} resonance doublet centered around $0.77\,\rm \mu m$\citep[][]{BMS}. Because this spectral region otherwise has low gas opacity, the influence of the K absorption lingers in these cloudless models well below the $T_{\rm eff}$ at which potassium condensates form (around 700 K). Including the effects of the Na and K condensates changes the opacity as shown by \citet[][]{Morley12}, but in the present cloudless models deep seated K still influences the spectra down to $T_{\rm eff}\sim 400\,\rm K$. Finally at the lowest effective temperatures the signatures of water and methane appear, strikingly altering the predicted optical spectra redward of $0.80\,\rm \mu m$.

Figure \ref{fig:sequence1} depicts the crucial near-infrared region. With falling $T_{\rm eff}$ the familiar  gradual departure of the refratory species and the alkalis, as well as the appearance of methane are apparent. The cloudless models are cooler than models that account for condensate opacity, thus methane appears around $T_{\rm eff} = 1600\rm \, K$ in {\em H} band, which is warmer than it is seen to do so in nature \citep[][]{Kirkpatrick05}. The progressive loss of flux in {\em K} band, attributable to the increasing influence of pressure-induced opacity of molecular hydrogen, continues through the entire sequence. The flux peaks are progressively squeezed between stronger and stronger molecular absorption until, by 300 K, they become sharp and well separated. As in the previous figure, the coldest model shown, at 300 K, has a notably different morphology as ammonia features are apparent near $1\,\rm \mu m$. 
This interesting small spectral region is further expanded in Figure \ref{fig:zoom}. 

The wavelength range to be explored in fresh detail by {\em James Webb Space Telescope}, beyond $3\,\rm\mu m$ is explored in Figures \ref{fig:sequence2} and \ref{fig:sequence3}. This is a region broadly sculpted by water opacity with important contributions from methane and ammonia at lower effective temperature. The five micron spectral window allows deep seated flux to emerge, as in familiar images of Jupiter \citep[e.g.,][]{Orton96}. The long pathlengths through the atmosphere, as in the near-infrared permit rare molecules to be detected \citep[][]{Morley19}. In an upcoming paper (Gharib-Nezhad et al.\ in prep.) we will look at the detectability of lithium-bearing molecules in this region.

\begin{figure*}[htbp]
    \centering
    \includegraphics[width=0.75\textwidth]{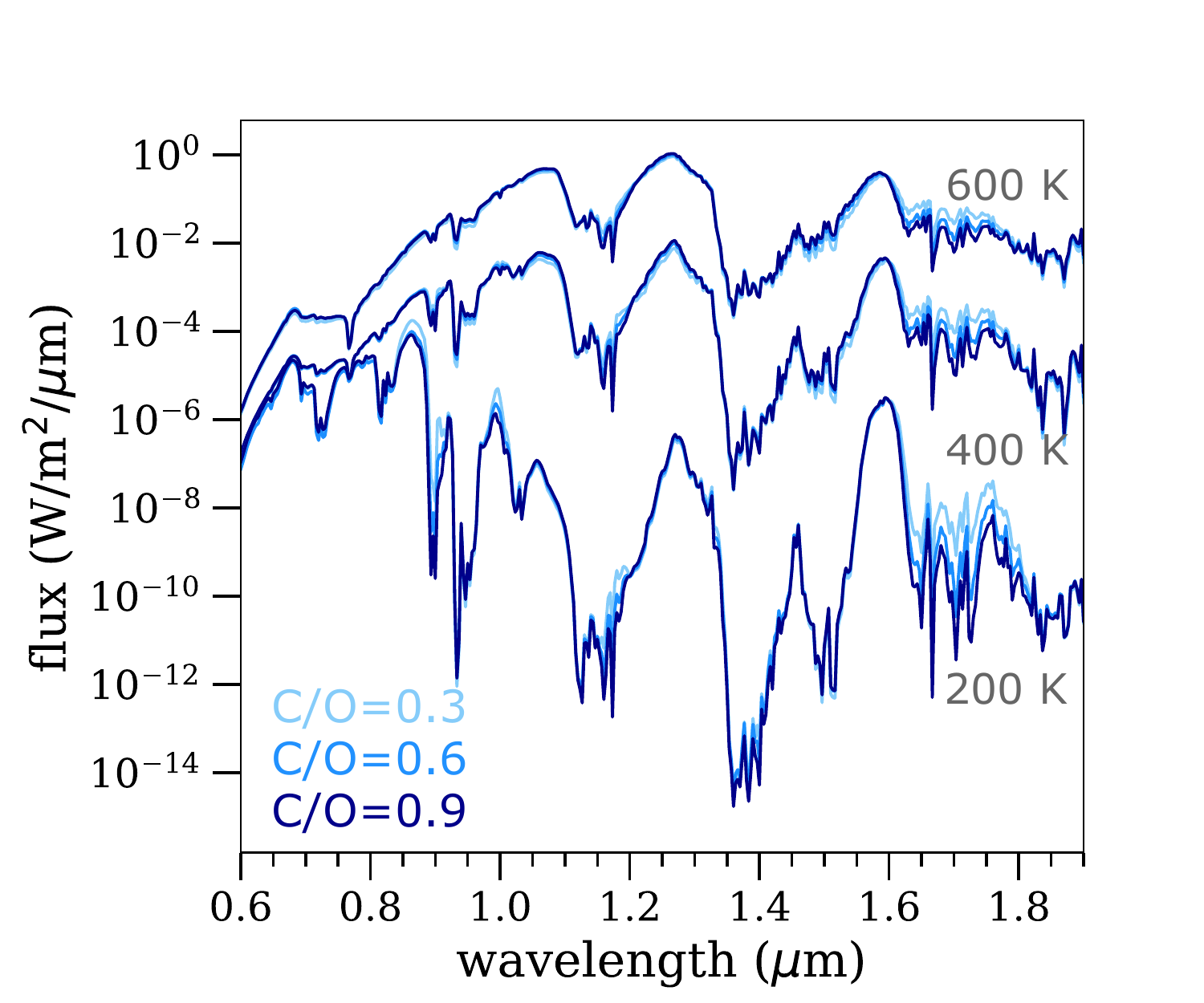}
    \caption{Comparison of model spectra ($R\sim 300$) for $\log g = 5$ at three $T_{\rm eff}$ and three C/O ratios as indicated. Note that the 400 and 200 K models have been scaled down by $10^2$ and $10^4$, respectively, for clarity in plotting. Note that the indicated C/O ratios are for the bulk gas composition and not the gas composition in the photosphere probed here, as about 20\% of the O atoms are lost to silicate grain formation deeper in the atmosphere.}
    \label{fig:CO}
\end{figure*}

Figure \ref{fig:CO} explores the influence of the C/O ratio in the near-infrared among the coolest models where differences in methane abundance are most notable. Although these are pure chemical equilibrium models, the influence of disequilibrium chemistry is less notable in this spectral region at these cool effective temperatures, thus providing better tracers of the atmospheric C/O ratio, with the usual caveat of accounting for loss of O into condensates at higher temperatures.

\begin{figure*}[htbp]
    \centering
    \includegraphics[width=0.9\textwidth]{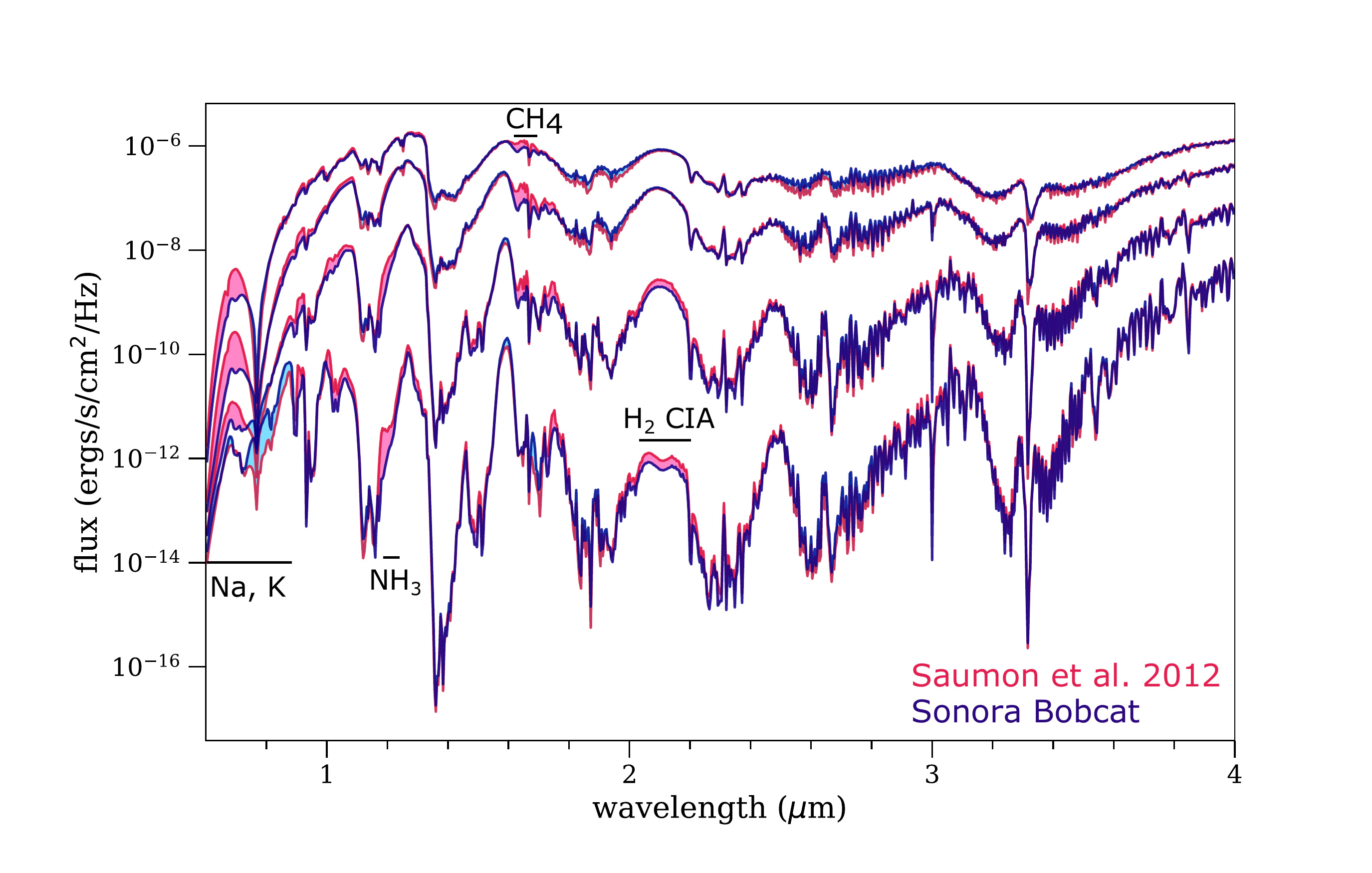}
    \caption{Comparison of the present Sonora Bobcat model spectra ($R\sim 1,000$) with those of \citet[][]{Saumon12} at $T_{\rm eff}=1300$, 900, 500, and $300\,\rm K$ (top to bottom). Absorbers responsible for notable differences are labeled. }
    \label{fig:comparison}
\end{figure*}

Finally \ref{fig:comparison} compares the model spectra presented here to our earlier generation of models \citep[][]{Saumon12} through the effective temperature range of the T and Y dwarfs. Most notable differences arise from changes in the alkali D line treatment and updated ammonia and methane opacity (\S2.3). We have found that more recent updates to the molecular line lists generally are not similarly apparent at the illustrated spectral resolution $R\sim 1,000$, but do appear at much higher spectral resolution $R>10,000$. 

\subsection{Evolution}

\begin{figure*}[htbp]
 \centering
 \includegraphics[width=0.65\textwidth]{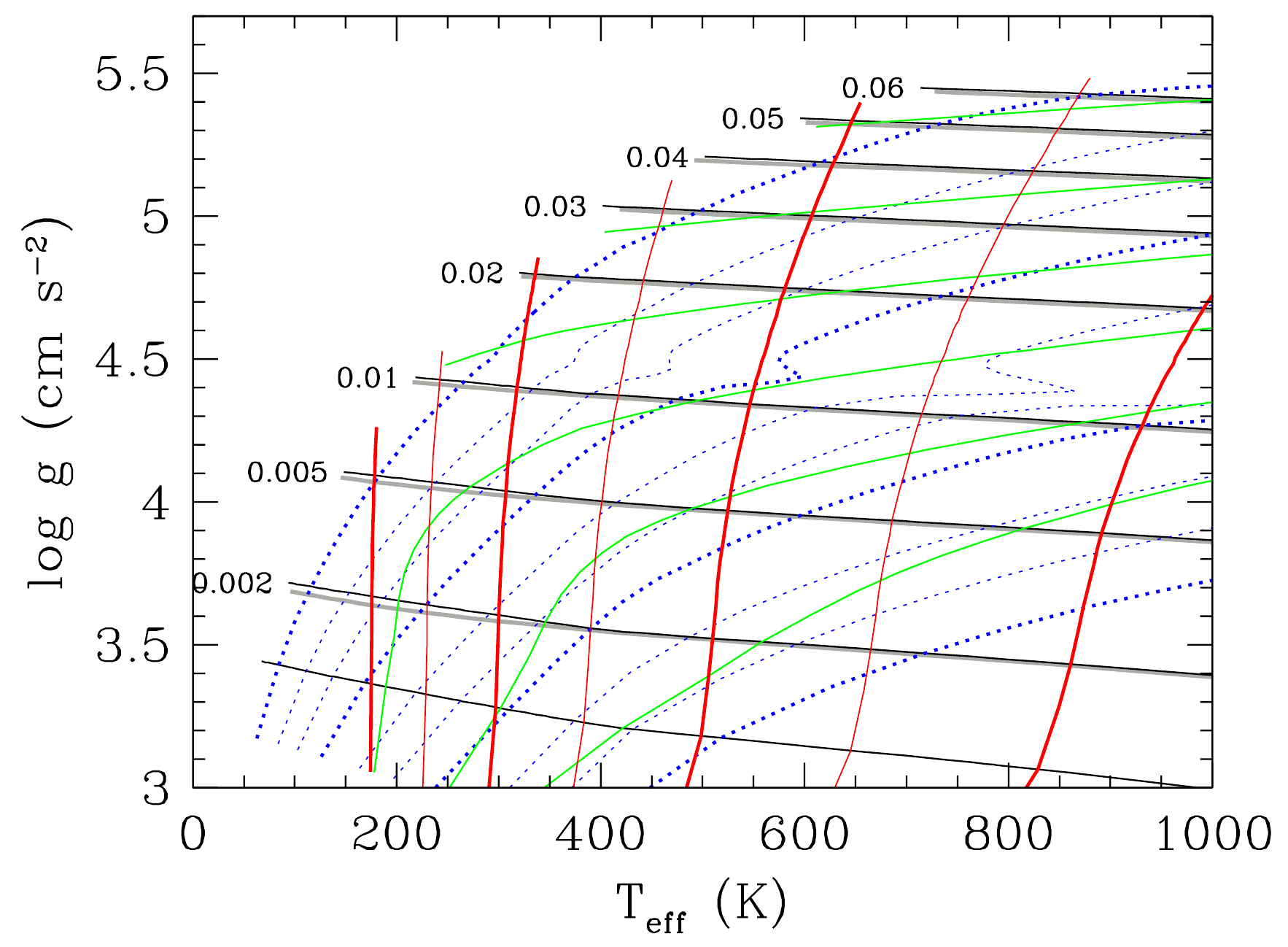}
   \caption{Low temperature end of the cooling tracks of brown dwarfs in $T_{\rm eff}$ and gravity for the sequence based on cloudless atmospheres of solar metallicity. The evolution proceeds 
 	   from right to left along the heavy black lines, which are labeled with the mass in M$_\odot$ (the unlabeled track at the bottom has $M=0.001\,M_\odot$). 
 	   Light gray cooling tracks are from \cite{Saumon08}.
 	    Isochrones are shown by
             the blue dotted lines: (from right to left) {\bf 0.01}, 0.02, 0.04, {\bf 0.1}, 0.2, 0.4, {\bf 1}, 2, 4, and {\bf 10}$\,$Gyr.
             The nearly vertical red lines are curves of constant luminosity: (from left to right): $\log L/L_\odot= -8$, to $-5$\ in
             steps of 0.5. Curves of constant radius are shown in green: (from top to bottom) 0.08 to 0.13$\,R_\odot$  in steps of 0.01.
             The phase of deuterium burning is revealed by the kink in the isochrones for objects with masses between 0.01 and 0.015$\,M_\odot$.
%           [{\it See the electronic edition of the Journal for a color version of this figure.}]
}
 \label{fig:cooling_tracks}
 \end{figure*}

\begin{figure}[htbp]
\centering
\includegraphics[angle=-90,width=0.55\textwidth]{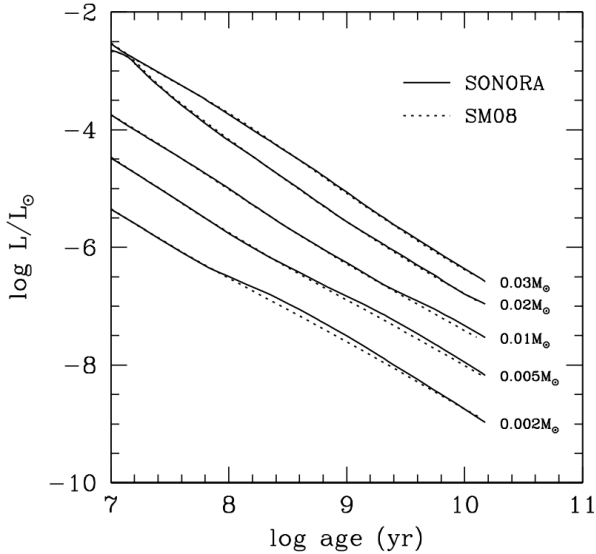}
\caption{Late evolution of the luminosity of low-mass ultracool dwarfs. The present Sonora models (solid lines), include water condensation and remain more luminous 
	   (cool more slowly) below $T_{\rm eff} \sim 400\,$K than the \cite{Saumon08} models (SM08, dotted lines). Each curve is labeled by the mass. The more massive dwarfs
	   are not affected as they remain hotter than 400$\,$K over the age of the Galaxy (See Figure \ref{fig:cooling_tracks}).
%           [{\it See the electronic edition of the Journal for a color version of this figure.}]
}
      \label{fig:Lbol}
\end{figure}

The newly computed evolution tracks are shown in Figures \ref{fig:cooling_tracks} and \ref{fig:Lbol}. 
The extension of the surface boundary condition to lower $T_{\rm eff}$  illuminates an interesting feature of Y dwarf evolution. The SM08 models were based on atmospheres that 
extended only to 500$\,$K and the corresponding boundary condition was extended to lower $T_{\rm eff}$ with a plausible, constrained  extrapolation. 
That extrapolation was inaccurate because it did not account for the condensation of water, which the new low temperature atmosphere models
include. As water is the dominant absorber in low $T_{\rm eff}$ Y dwarfs, its disappearance below $T_{\rm eff} < 400\,$K results in more 
transparent atmospheres, which affects the evolution.  

The impact of this accounting for water condensation can be seen in Figure \ref{fig:cooling_tracks} as a divergence in the respective cooling tracks
of low-mass objects ($<0.01\,M_\odot$) below $T_{\rm eff} \sim 400\,$K. The new models have slightly higher gravity and luminosity (Figure \ref{fig:Lbol}).
Qualitatively, this is similar to the hybrid model of SM08 for the  L/T transition where the disappearance of cloud opacity causes a pile up of brown dwarfs at the 
transition \citep{kirkpatrick20},
although the effect appears to be weaker. 

This result should not be taken at face value, however. The present cloudless model atmospheres include condensation of water but not 
the resulting cloud opacity. Unlike magnesium-silicate clouds, that have considerable opacity but whose gas phase precursors (e.g.,\,MgH) have barely been detected in spectra \citep[see, e.g.,][]{Kirkpatrick05}
 and thus have negligible opacity,
water is the dominant gas phase absorber throughout the LTY sequence. Thus, water condensation transforms the opacity from a series of strong molecular bands throughout the near- and mid-IR to
a continuum of condensate opacity. Until Y dwarf models with a reliable description of water clouds are produced, the net effect of water condensation on very cool brown dwarf evolution
will remain uncertain.

\begin{figure}[hbtp]
      \centering
 \includegraphics[width=0.45\textwidth]{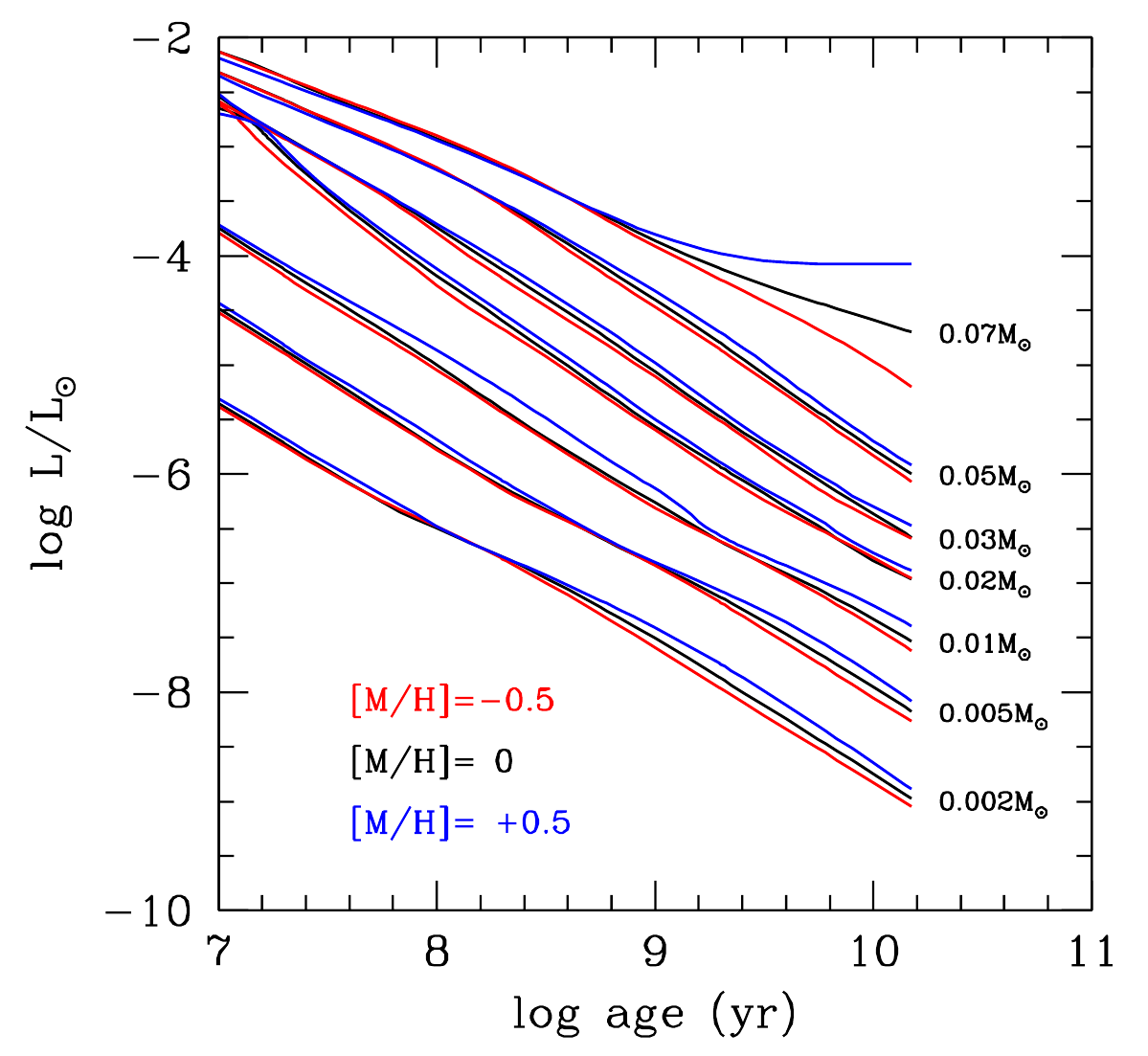}
   \caption{Comparison of brown dwarfs luminosity cooling tracks for three different metallicities. Each triplet of tracks is 
	    labeled with the mass in solar masses. For a given mass and age, a higher metallicity results in a slightly higher luminosity.
	    The $0.07\,M_\odot$ models are just below the hydrogen burning minimum mass and their fates diverge after 2$\,$Gyr, 
	    depending on the metallicity. The [M/H]=+0.5 track approaches a stable equilibrium on the main sequence while the lower 
	    metallicity tracks for that mass fail to do so.}
      \label{fig:lbol_cooling_tracks}
\end{figure}

\begin{figure}[hbtp]
      \centering
 \includegraphics[width=0.45\textwidth]{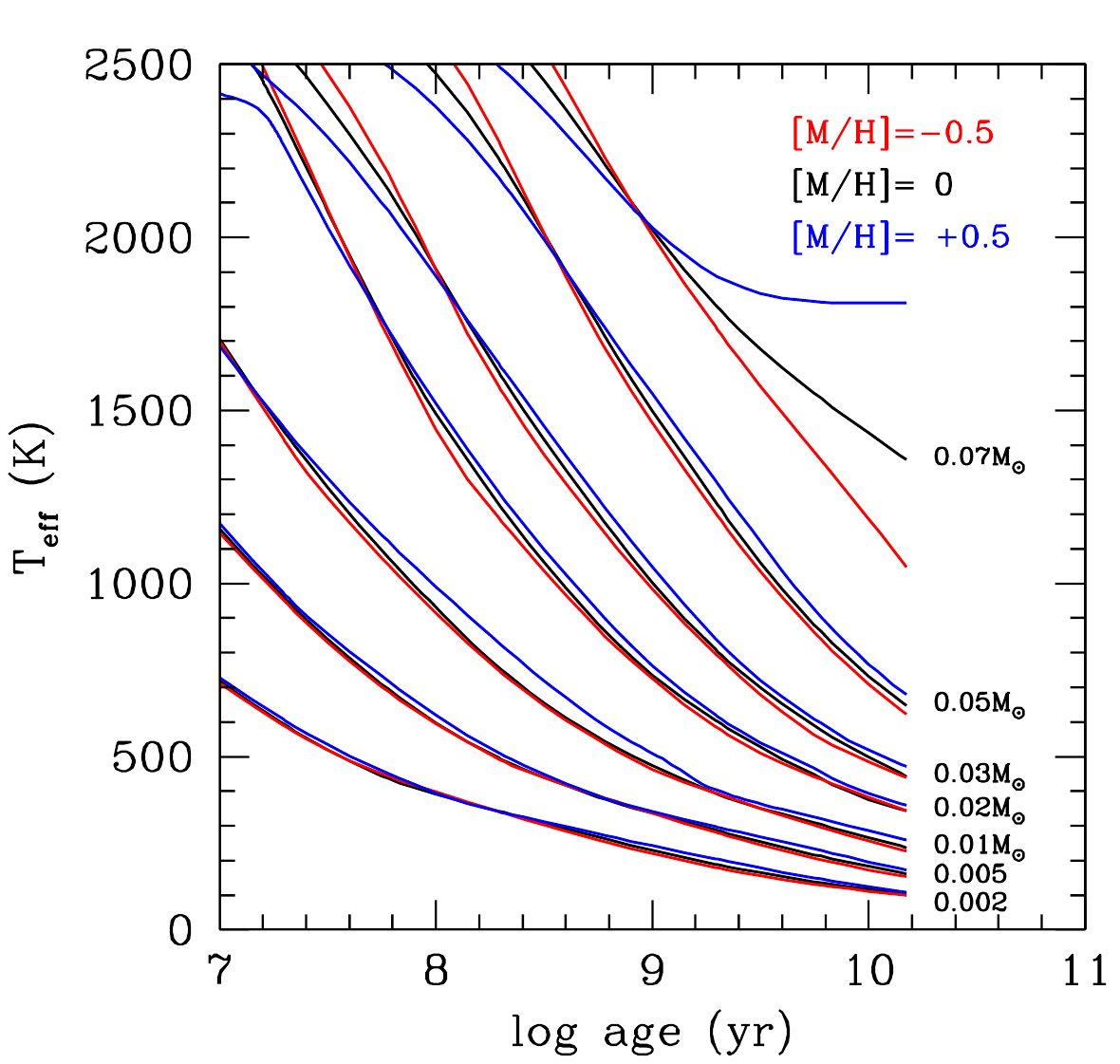}
 \caption{Same as Figure \ref{fig:lbol_cooling_tracks} but for the effective temperature.}
      \label{fig:teff_cooling_tracks}
\end{figure}

The effect of metallicity on the coolings tracks is shown in Figure \ref{fig:lbol_cooling_tracks} for the luminosity and 
Figure \ref{fig:teff_cooling_tracks} for $T_{\rm eff}$. Generally, the higher metallicity models are slightly more luminous
at a given age because the higher opacity of the atmosphere slows their cooling. All masses show the same trend with 
$\Delta \log L /\Delta {\rm [M/H]} \lesssim 0.15$. Deuterium burning causes the apparent anomalies in the 0.01 and 0.02$\,M_\odot$
tracks. The 0.07$\,M_\odot$ tracks are just below the hydrogen burning minimum mass (HBMM). Since the HBMM decreases with increasing metallicity,
the [M/H]=+0.5 model approaches a main sequence equilibrium state while the other do not produce enough nuclear energy to prevent 
further cooling.  The evolution of $T_{\rm eff}$ is very similar to that of $L_{\rm bol}$.

Because of the systematic differences that persist between forward model spectra and data, determinations of $T_{\rm eff}$  and especially 
the gravity by fitting observed spectra remain rather uncertain. Atmospheric parameters determined from models 
from different groups often disagree \citep[e.g.,][]{patience12}.  On the other hand, the modeled bolometric luminosity,
because it integrates over all wavelengths, is much less sensitive to such model errors and can also be determined fairly reliably from observations.
Benchmark brown dwarfs, either bound in a binary system with a more easily characterized primary star or members of a moving 
group also have well determined metallicities 
and fairly well constrained ages.  Equipped with $L_{\rm bol}$, [M/H] and the age, the mass (e.g. Figure \ref{fig:lbol_cooling_tracks}), 
$T_{\rm eff}$, radius, and gravity follow from evolution sequences with a good degree of confidence.

The evolution sequences are available for all three metallicities\footnote{ \dataset[https://doi.org/10.5281/zenodo.5063476]{https://doi.org/10.5281/zenodo.5063476}}. The tables provide
mass, age, radius, luminosity, gravity and $T_{\rm eff}$ along cooling tracks at constant mass and, for convenience, along isochrones,
along constant luminosity curves and for a given pair of ($T_{\rm eff}$, $\log g$).
The moment of inertia $I$ is also provided for each mass as a function of time.
For a spherically symmetric body of radius $R$,
$$ I = \frac{8\pi}{3} \int_0^R \rho(r)r^4\,dr $$
which is a useful quantity for studies of the angular momentum of brown dwarfs and their deformation under rotation
\citep{Barnes03,Sengupta2010,Clem20}.

  \begin{figure}[htbp]
      \centering
      \includegraphics[width=0.46\textwidth]{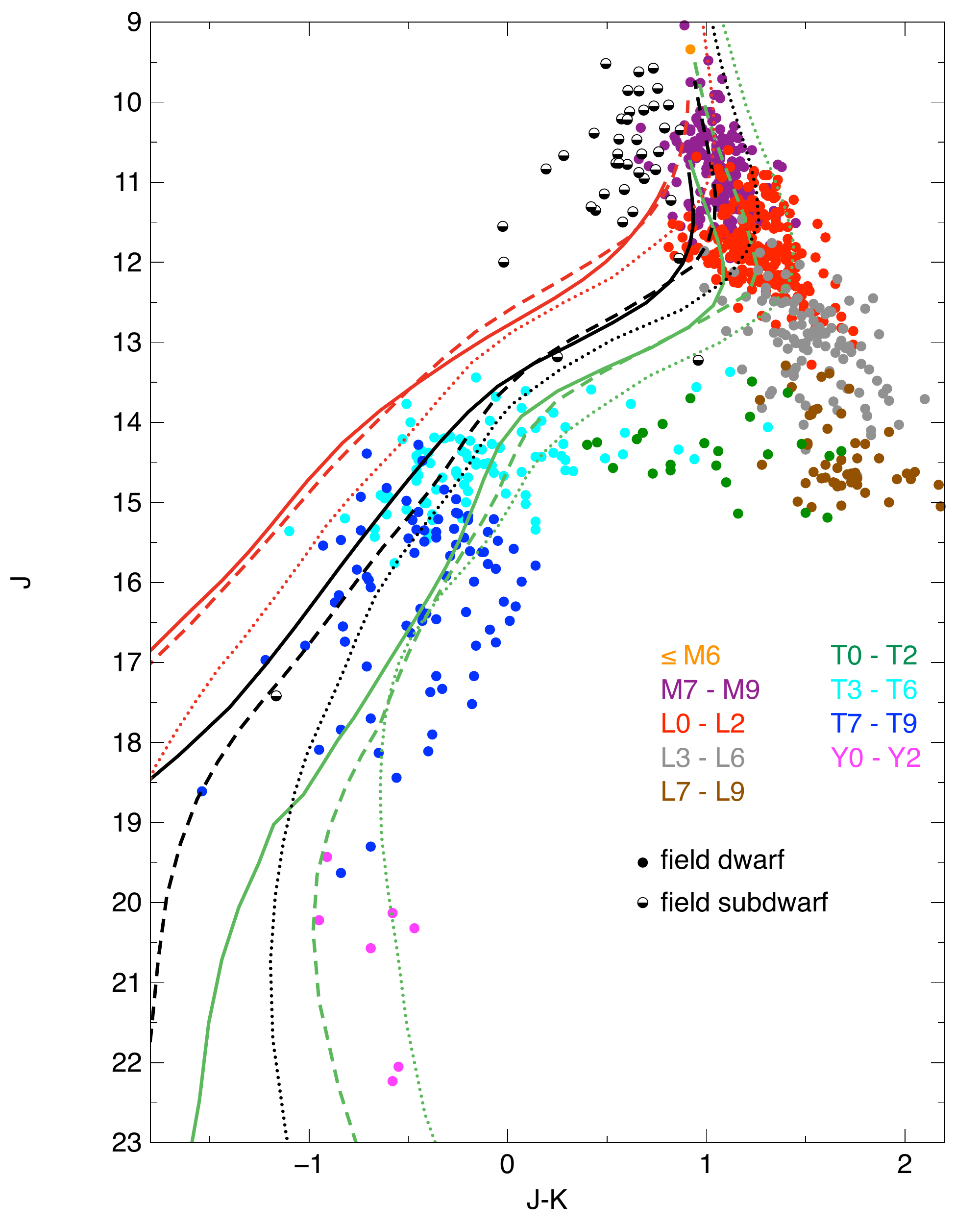}
      \caption{Cloudless Sonora Bobcat cloudless colors for different constant values of gravity and metallicity as compared to Mauna Kea Observatory {\em J-} and {\em K}-band photometry of nearby M, L, T, and Y dwarfs from \citep[][]{ultracoolsheet}. The curves are coded by their line types (solid, dashed, dotted) for $\log g({\rm cm\,s^{-2}}) =5$, 4, and 3 respectively and line colors, from top to bottom groups $[{\rm M/H}]=-0.5$, 0.0, and +0.5 for red, black, and green. The near-infrared spectral types are denoted by the color of the dot. Half-filled circles are subdwarfs.}
      \label{fig:bobcat}
  \end{figure}

  \begin{figure*}[htbp]
      \centering
      \includegraphics[width=0.46\textwidth]{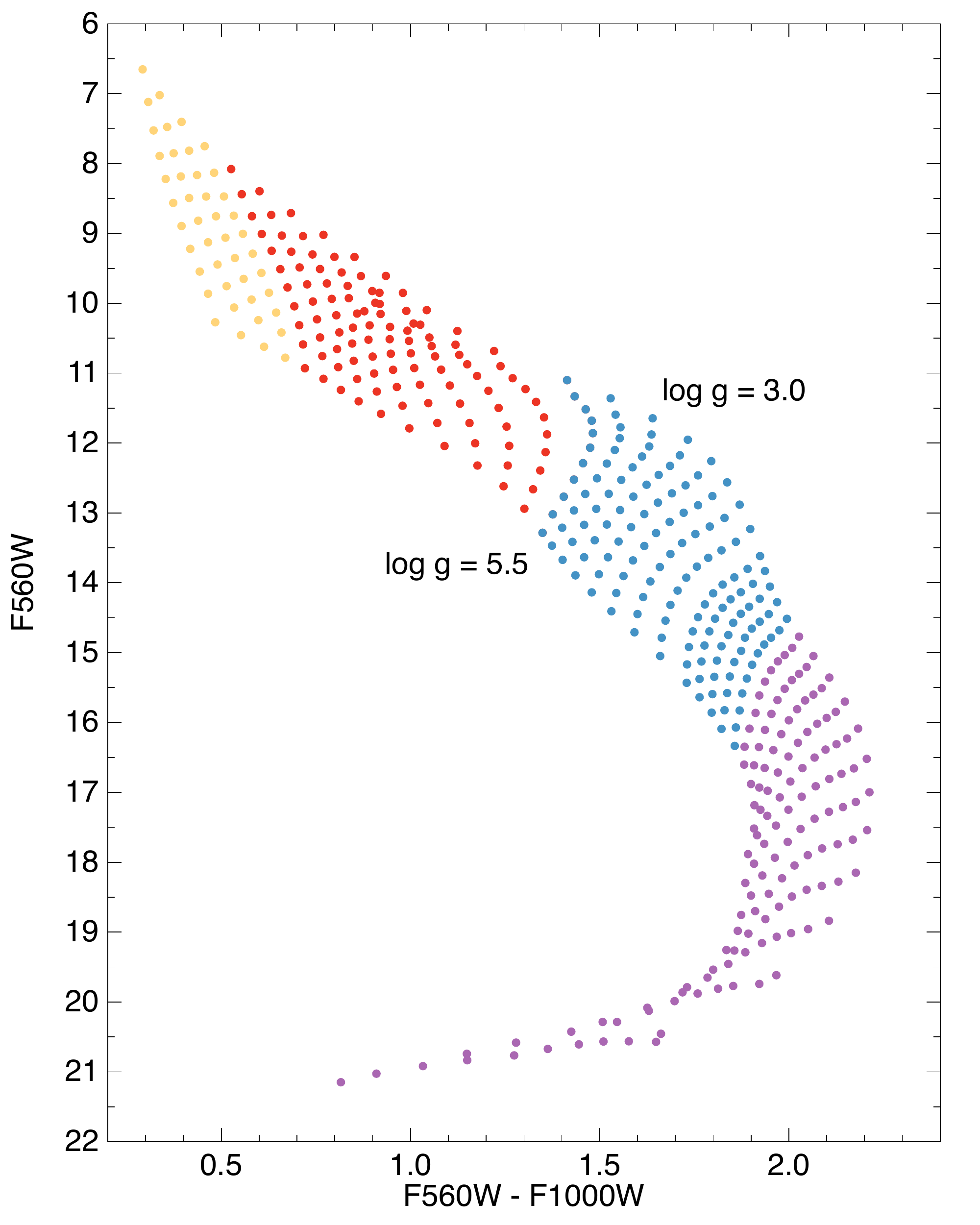}
            \hfill
     \includegraphics[width=0.46\textwidth]{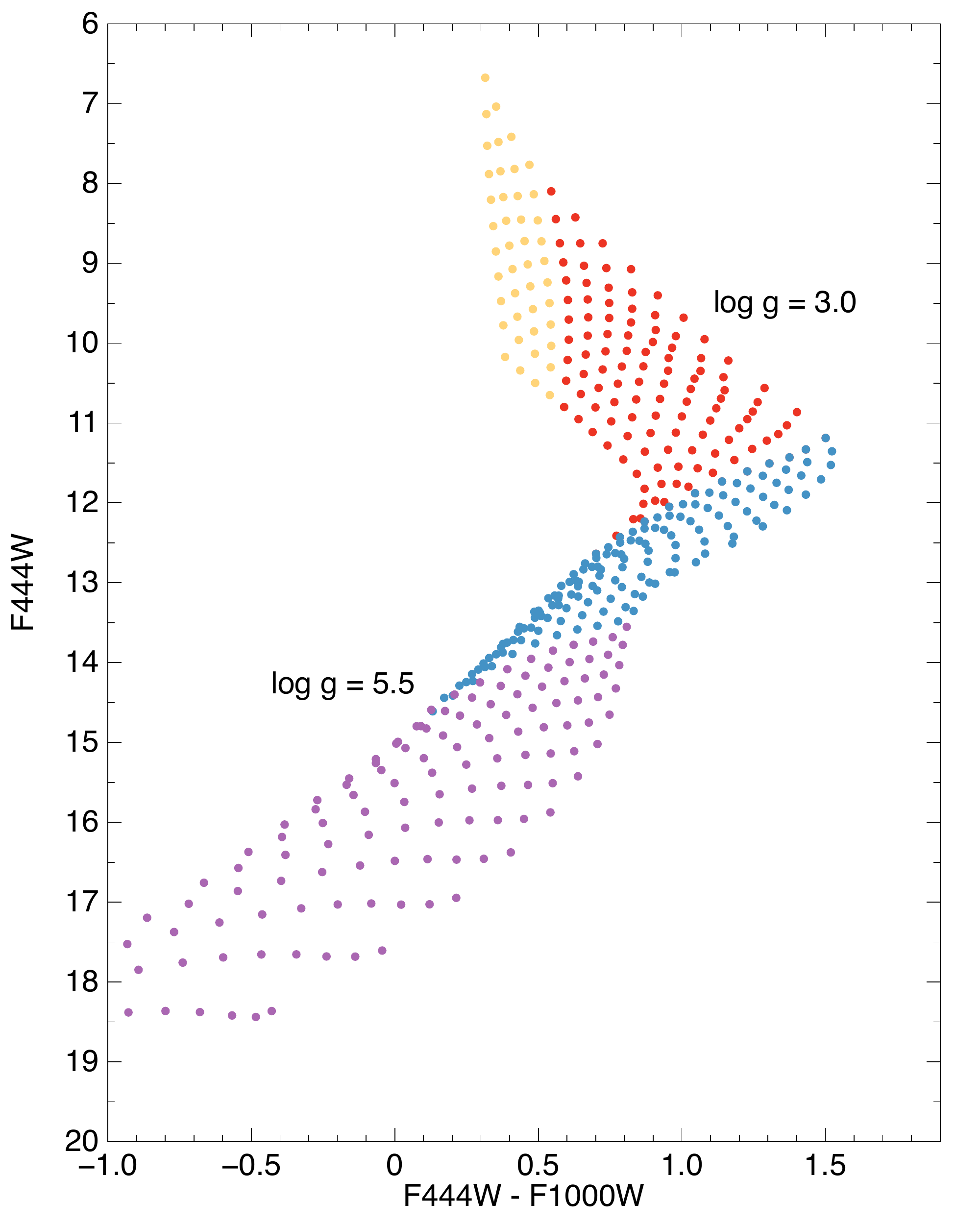}
            
      \caption{Color-magnitude diagrams in a selection of {\em JWST} mid-infrared filters, including two filters, F444W and F540W, commonly considered for searches for low mass companions. Each individual point represents one solar metallicity model. Ranges of effective temperature are denoted by point color from yellow ($> 2000\,\rm K$) to red (2000 to 1000 K) to blue (1000 to 500 K) to purple ($< 500\,\rm K$). Models are shown every 0.25 dex step in $\log g$ to illustrate sensitivity to parameters. Slight clumping among low $g$ models near 1700 K arises from numerical noise in those model temperature-pressure profiles. }

      \label{fig:jwst}
  \end{figure*}

\section{Comparisons to Selected Datasets}

\subsection{Spectra}
Spectra from the Sonora Bobcat model set have already been used in a number of comparisons to various spectral datasets. The most systematic application has been in \citet[][]{Zhang21} and \citet{Zhang21a} who employed the Bayesian inference tool Starfish to  near-infrared spectra ($R\sim 80$ -- 250) of 55 late T dwarfs, including three benchmark (T7.5 and T8) objects. While good spectral fits could be found between model library and observed spectra, there were certain consistent discrepancies across most all objects. Most notably, the peak of {\em J}-band in the best fit spectra was too bright for most objects while the peak of {\em H}-band was too dim. In both cases the discrepancy increases with falling $T_{\rm eff}$ and later spectral types. \citet[][]{Zhang21a} attributed these discrepancies to deep clouds and disequilibrium chemistry, respectively, although further modeling is needed to understand in detail.

Zhang et al. (submitted) also used this model set to quantify the well known degeneracy between metallicity and gravity by considering all of the model spectra. They found that the gravity-metallicity degeneracy can be described with $\Delta \log g \sim 3.42 \Delta \rm [M/H]$. In other words a change of +0.1 in $\rm [M/H]$ has nearly the same effect as a change in $\log g$ of about 0.34.

At higher spectral resolution the models generally reproduce the finer spectral structure of cloudless objects fairly well. Comparing among three model sets, for example, \citet[][]{Tannock21} found Sonora-Bobcat models had the lowest $\chi^2$ in {\em J}, {\em H}, and {\em K-}bands when compared to $R\sim 6,000$ spectra of a T7 dwarf, \added{this is likely due to our choice of recent molecular opacity tables (Table 2) and illustrates the value of considerable effort that underlies the calculation of these tables}.

\subsection{Colors}
The online tables contain model photometry in a selection of photometric passbands, including the MKO, WISE, SDSS, 2MASS, Spitzer/IRAC, and JWST systems. Model absolute magnitudes are computed using radii from the evolution model. Figure \ref{fig:bobcat} shows our predicted Sonora Bobcat model photometry on the familiar ultracool dwarf $J$ vs.\ $J-K$ color magnitude diagram along with a sample of field dwarfs and subdwarfs selected from \citet[][]{ultracoolsheet} for having well constrained photometry and distances. Model colors are shown for three gravities and three metallicities.

The models well reproduce the photometry of the latest M dwarfs and earliest L dwarfs, including the spread in $J-K$ color. This is a substantial improvement over older evolution sets,  which sometimes struggled over the same phase space, likely due to older $\rm H_2O$ opacities that lacked `hot' lines. By $J\sim 12$  the cloudless models turn to the blue instead of continuing to slide to the red as obseved in the field L dwarf population, a consequence of the lack of dust opacity in these models. The $[{\rm M/H}]=+0.5$ models are redder and make the turn about half a magnitude later. Conversely the $[
{\rm M/H}]=-0.5$ are bluer in $J-K$ from even brighter magnitudes, although this models of this metallicity are still not as blue as the majority of the M and L subdwarfs. 

The best agreement with observed photometry is among the mid-T dwarfs, about T3 to T7, which are generally known to be well matched by cloudless model spectra \citep[e.g.,][]{Marley96}. At still later spectral types the observed colors are again redder, likely a consequence of the formation of alkali clouds \citep[][]{Morley12}. 

While there is a fair amount of structure in the individual tracks for each gravity and metallicity, generally speaking lower metallicity models are always bluer than those with higher metallicity. This is a consequence of the $J$ spectral bandpass being very sensitive to the total column gas opacity. At lower metallicities flux from deep atmospheric layers emerges, keeping $J$ magnitudes bright and $J-K$ blue. As metallicity increases the $J$ window closes as $\rm H_2O$ and other gas opacity squeeze in from the sides and the $J-K$ contrast is reduced. 

The dependence on gravity is generally more complex. Gravity affects the thermal profile, including the location and spacing of convection zones. The interaction of the changing thermal profile structure and the atmospheric opacity and chemistry alters the individual tracks for each gravity. Generally speaking lower gravity tracks are redder in $J-K$ at all three metallicities. 

Two additional color-magnitude diagrams in three {\em JWST} filters are shown in Figure \ref{fig:jwst}. In this case individual models are shown to better illustrate sensitivity to model parameters. F444W and F560W are  commonly considered for stellar substellar companion searches as they capture the five micron excess flux seen in Figure \ref{fig:sequence2}. The color difference with F1000W both show a turn as the flux progressively moves redward with falling effective temperature. $\rm F444W - F1000W$ is reddest near 800 K while $\rm F560W - F1000W$ turns from red to blue at just a few hundred Kelvin.

\subsection{Mass-Radius Relationship}

Intensive searches for transiting exoplanets over the past decade have uncovered many transiting brown dwarfs and giant planets, providing valuable determinations of the 
radius of each object. In favorable cases, follow up has led to the determination of the mass of the substellar companion.
In addition, the characterization of the primary star of a system can constrain its age and its metallicity giving a
lower limit on the metal content of the companion.  Combined with the orbital parameters, the degree of insolation in these relatively close binaries 
and any corresponding increases in radii can be estimated.

The mass-radius relation $R(M)$ of very-low mass stars, brown dwarfs and giant planets has been extensively discussed and its main features were recognized early on 
\citep{bhl89, saumon96}. More recently, \cite{chabrier09_MR} discuss the physics that drives its 
characteristic shape in details and \cite{bhn11} explore the role of the helium abundance, metallicity and clouds. Briefly, for ages greater than $\sim 1\,$Gyr, the $R(M)$ relation 
first rises in the Jupiter mass range as the interior consists mostly of 
atomic/molecular fluid in the outer envelope and, deeper, a plasma where the electrons are moderately degenerate. Qualitatively, this corresponds to the 
regime of ``normal matter'' where the volume 
of an object increases with its mass.  The radius peaks at $\sim 4\,M_{\rm J}$ beyond which it declines steadily as 
most of the mass becomes a degenerate plasma and the $R(M)$ relation behaves very much like that of white dwarfs (but with a hydrogen composition). 
When hydrogen fusion starts to contribute to the energy balance of the brown dwarfs, the star is partially supported by thermal pressure and the radius begins 
to rise again. For hydrogen burning stars on the 
main sequence, the radius rises steadily with mass.
Thus, the substellar $R(M)$ relation of gaseous substellar objects has a local maximum around 4$\,M_{\rm J}$ and a minimum at  60-70$\,M_{\rm J}$. It  is
remarkable that the radius of substellar objects remains within $\sim 15$\% of 1$\,R_{\rm J}$ for masses spanning two orders of magnitude (0.5 - 70$\,M_{\rm J}$).

\begin{figure*}[hbtp]
\centering
 \includegraphics[width=0.6\textwidth]{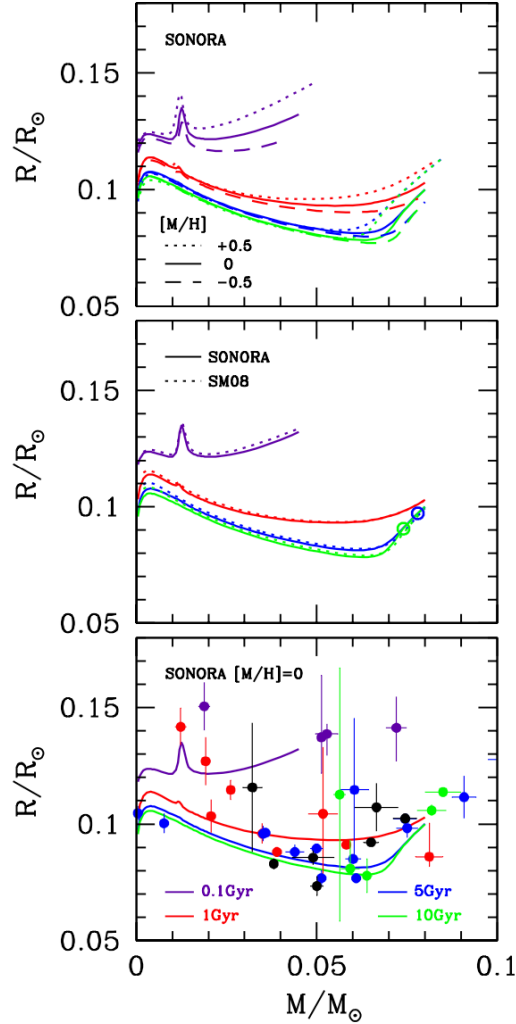}
  \caption{Mass-radius relation of evolution models of ultracool dwarfs. Each panel shows isochrones of the $M$-$R$ relation: 
	  0.1$\,$Gyr (purple), 1$\,$Gyr (red), 5$\,$Gyr (blue) and 10$\,$Gyr (green). Each isochrone is truncated at $T_{\rm eff}=2400\,$K, the limit of our
	   atmosphere model grid. 
	   {\it Top panel:} Dependence of the radius on the metallicity.  Metal-rich objects have systematically larger radii because they cool more slowly. 
	   The sharp peak in the 100$\,$Myr isochrone is caused by deuterium burning. 
	   {\it Middle panel:} Comparison with the cloudless, solar metallicity models of \cite{Saumon08} (SM08). 
           Large open circles indicate the location of the hydrogen burning minimum mass for each isochrone.
	   {\it Bottom panel:} Comparison of solar metallicity models with data. Data points are colored according 
           to their age in bins corresponding to the model isochrones: 0.03 - 0.3$\,$Gyr (purple), 0.3 - 2.2$\,$Gyr (red), 
	   2.2 - 7$\,$Gyr (blue), $>7\,$Gyr (green), unknown (black). Data from \cite{subjak20, david19, gillen17, carmichael20, benni20,
           zhou19, carmichael19, boetticher17, hodzik18, nowak17, winn07, mireles20,  canas18, acton20, littlefair14, parsons17, casewell20},
	   T.G. Beatty, pers. comm., S. L. Casewell, pers. comm.
%           [{\it See the electronic edition of the Journal for a color version of this figure.}]
}
      \label{fig:MR}
\end{figure*}

Figure \ref{fig:MR}  summarizes the mass-radius relation of our new models. Since brown dwarfs and exoplanets cool and contract with time, the $R(M)$ relation is a function of time.
The top panel of Figure \ref{fig:MR} shows four isochrones of the $R(M,t)$ relation for three different metallicities.  While the older isochrones (1 - 10$\,$Gyr) 
display the behavior just discussed, the $R(M)$ relation at 100$\,$Myr is different. In particular, it does not show the inverse relation between radius and mass.
The radius keeps rising with mass because the young, relatively hot objects are not supported by the pressure of degenerate electrons.  There is also a prominent peak
at  $\sim 12\,M_{\rm J}$ due to the short-lived phase of deuterium burning. This peaks has almost vanished after 1$\,$Gyr. The metallicity affects the
radius is several ways. There are two dominant effects. A higher metallicity increases the opacity of the atmosphere and slows down the cooling and the contraction, so
at a given $M$ and age, the radius is larger. This matters primarily in the early evolution where the thermal content of the brown dwarf largely determines its radius.
Another effect of the increased atmospheric opacity is that once nuclear fusion turns on, the flow of this additional energy to 
space is impeded, which is compensated by a larger radius. This is significant for $M \gtrsim 0.06\,M_\odot$ and ages of $\sim 1\,$Gyr or more.  Smaller effects 
include the change on the equation of state (see below), where increasing metallicity decreases the radius, primarily at late times of $\sim 5\,$Gyr or more,  and the 
effect of the condensation of water in the atmosphere which decreases the radius of these cloudless models (see \S \ref{sec:evoln_model}), an effect that grows with the metallicity.

The middle panel of Figure \ref{fig:MR} compares the present models with the cloudless, solar metallicity models of SM08 for the same isochrones. 
The Sonora models are systematically smaller by $\sim 1-3$\%. This is due to the inclusion of metals in the equation of state (Equation \ref{Yeff}).
The more massive objects shown eventually settle on the main sequence. The  HBMM is indicated by the open circles for the 5 and 
10$\,$Gyr isochrones. Here, we define a main sequence star as an object for which $>99.9$\% of the luminosity is provided by nuclear fusion. Lower mass objects 
take a longer time to reach that limit if they are massive enough to reach it at all. Thus the HBMM decreases with time, as seen in the figure. 
Note that the HBMM is well above the location of the 
minimum radius of the isochrones.  Objects that fall between the $R(M,t)$ minimum and the HBMM are only partially supported by nuclear fusion 
($L_{\rm nuclear} < L_{\rm bol}$).

The lower panel of Figure \ref{fig:MR} compares the solar metallicity Sonora models to data. The data points are colored according to the estimated age of each object,
and matched to the plotted isochrones (see caption for details).  Although there are considerable uncertainties for several objects and the scatter is significant,
the agreement is generally quite good. In most cases, outliers  have larger radii than predicted by the models, which can be qualitatively explained by the 
role of stellar insolation as many of these objects are in very small orbits around their primary star, with periods of just a few days. This radius increase can be compensated by
increasing the metallicity.
A detailed comparison with the data, which would have to account for the metallicity and the effect of insolation of each object during its evolution is beyond the scope 
of the present discussion. There remain a few problematic objects that have radii smaller than predicted by the 10$\,$Gyr
isochrone.  The most straightforward way to shrink an old brown dwarf or planet is to increase its metallicity, with the heavy elements dispersed throughout the body or concentrated in 
a core. Our models predict that a 0.05$\,M_\odot$ brown dwarf with 10 times the solar metallicity will see its radius decrease by $\sim 0.008\,$R$_\odot$ at
5 and 10 Gyr, which is sufficient to reach even the smallest object shown in Figure 3. It is challenging to explain how a such a massive object could acquire a metallicity that is well
above that of its parent star.

The characteristics of the HBMM of the solar metallicity Sonora models are nearly identical to those of the solar metallicity cloudless SM08 models. For instance, the HBMM 
mass is 0.074$\,M_\odot$ (Sonora) compared to 0.075$\,M_\odot$ (SM08). 
As in SM08, we define the minimum mass for deuterium burning (DBMM) as the mass of an object that burns 90\% of its initial deuterium content by 
the age of 10$\,$Gyr. Again, we find a DBMM of 12.9$\,M_{\rm J}$, which is nearly identical to the SM08 value of 13.1$\,M_{\rm J}$. 
At the DBMM, deuterium fusion can linger to $T_{\rm eff} \lesssim 800\,$K but most of the deuterium is burned at higher temperatures ($T_{\rm eff} \gtrsim 1200\,$K) 
where clouds largely control the evolution. Deuterium fusion is likely to be affected by the process of cloud clearing at the L/T transition that occurs
around $1200-1400\,$K (e.g. the hybrid model of SM08).  The dependence on metallicity
of the DBMM and HBMM of these cloudless models is of rather academic interests since it is well established that the HBMM occurs at $T_{\rm eff} \sim 1500\,$K  and that
the bulk of deuterium is burned at 
$T_{\rm eff} \gtrsim 1200\,$K with clouds playing an important role in both cases. 

\added{The characteristics of the models near the HBMM and DBMM are quite sensitive to the input physics of the models, mainly because of the extreme dependence of the nuclear fusion rates with temperature at these low temperatures where nuclear fusion reactions are marginal. This is illustrated by the evolution models of \cite{fernandes19} and \cite{Phillips20} who make slightly different choices for the EOS. The former uses EOS tables of carbon and oxygen as surrogates for all metals instead of an increased He mass fraction, and the latter uses new EOS tables for H and He with an increased He mass fraction to represent the metals. In both studies, the models are very close to those of SM08, with small differences noticeable only near the HBMM and DBMM.}

\section{Conclusions}

\added{The forward modeling effort reported here represents
another stepping stone in our efforts to understand the
atmospheres and evolution of substellar objects. As
molecular opacities have improved over the past decade
the limiting step in our understanding of their atmospheres now lies in atmospheric chemistry, mixing, and cloud
processes. The models presented in this paper address
the first of these.  In upcoming papers we will present
new sets models that explicitly explore disequilibrium chemistry and cloud processes.

The Sonora Bobcat models provide a comprehensive set of properties of 
cloudless models, including atmospheric $(T,P)$ profiles, high resolution
spectra tailored to the capabilities of {\it JWST}, absolute magnitudes in all
the bandpasses of key ground-based and space telescopes, and evolutionary
tracks consistent with the model atmospheres. The models cover an expanded set
of parameter space, most notably considering non-solar C/O and down to $T_{\rm
eff}=200\,$K. The new models reach well into the $T_{\rm eff}$ range of spectral class Y and anticipates {\it JWST}
discoveries of new objects nearly as cool as Jupiter. 

The next sets of models, by allowing for clouds an disequilibrium chemistry,
will be better matched to the observed MLT sequence and will provide new
insights into the L/T transition and L and T subdwarfs.  A refined hybrid
evolution model that accounts for the cloud clearing at the L/T transition
will become possible. The number of well-characterized brown dwarfs is now
large enough that comparisons with such a hybrid model will provide fresh
insights into the L/T transition. 

Further progress in the field will hinge on detailed, systematic comparisons of all available model sets to
the wealth of observational data already available and soon to come from {\it JWST}. The combination of retrieval 
and forward model fitting studies will illuminate where the atmosphere and
evolution models excel and still fall short and provide essential clues as to
the yet to be discovered physics that is missing in forward models.}

\section{Acknowledgments}
This work has made use of the VALD database, operated at Uppsala University, 
the Institute of Astronomy RAS in Moscow, and the University of Vienna. This work has benefited from The UltracoolSheet, maintained by Will Best, Trent Dupuy, Michael Liu, Rob Siverd, and Zhoujian Zhang, and developed from compilations by \citet[][]{Dupuy12, Dupuy13, Liu16, Best18}, and Best et al. (in press). We thank Mike Liu for assistance with preparing Figure \ref{fig:bobcat} and general comments and Ivan Hubeny for helpful discussions on the various atmospheres modeling schools.
Part of this work was performed under the auspices of the U.S. Department of Energy under Contract No. 89233218CNA000001.

\bibliographystyle{aasjournal}
\bibliography{references}

\begin{thebibliography}{}
\expandafter\ifx\csname natexlab\endcsname\relax\def\natexlab#1{#1}\fi
\providecommand{\url}[1]{\href{#1}{#1}}
\providecommand{\dodoi}[1]{doi:~\href{http://doi.org/#1}{\nolinkurl{#1}}}
\providecommand{\doeprint}[1]{\href{http://ascl.net/#1}{\nolinkurl{http://ascl.net/#1}}}
\providecommand{\doarXiv}[1]{\href{https://arxiv.org/abs/#1}{\nolinkurl{https://arxiv.org/abs/#1}}}

\bibitem[{{Acton} {et~al.}(2020){Acton}, {Goad}, {Casewell}, {Vines},
  {Burleigh}, {Eigm{\"u}ller}, {Nielsen}, {G{\"a}nsicke}, {Bayliss}, {Bouchy},
  {Bryant}, {Gill}, {Gillen}, {G{\"u}nther}, {Jenkins}, {McCormac}, {Moyano},
  {Raynard}, {Tilbrook}, {Udry}, {Watson}, {West}, \& {Wheatley}}]{acton20}
{Acton}, J.~S., {Goad}, M.~R., {Casewell}, S.~L., {et~al.} 2020, \mnras, 498,
  3115, \dodoi{10.1093/mnras/staa2513}

\bibitem[{{Allard} {et~al.}(2007{\natexlab{a}}){Allard}, {Allard}, {Homeier},
  {Kielkopf}, {McCaughrean}, \& {Spiegelman}}]{Allard07}
{Allard}, F., {Allard}, N.~F., {Homeier}, D., {et~al.} 2007{\natexlab{a}},
  \aap, 474, L21, \dodoi{10.1051/0004-6361:20078362}

\bibitem[{{Allard} {et~al.}(2001){Allard}, {Hauschildt}, {Alexander},
  {Tamanai}, \& {Schweitzer}}]{Allard01}
{Allard}, F., {Hauschildt}, P.~H., {Alexander}, D.~R., {Tamanai}, A., \&
  {Schweitzer}, A. 2001, \apj, 556, 357

\bibitem[{{Allard} {et~al.}(1996){Allard}, {Hauschildt}, {Baraffe}, \&
  {Chabrier}}]{Allard96}
{Allard}, F., {Hauschildt}, P.~H., {Baraffe}, I., \& {Chabrier}, G. 1996,
  \apjl, 465, L123, \dodoi{10.1086/310143}

\bibitem[{{Allard} {et~al.}(2000){Allard}, {Hauschildt}, \&
  {Schwenke}}]{Allard00}
{Allard}, F., {Hauschildt}, P.~H., \& {Schwenke}, D. 2000, \apj, 540, 1005,
  \dodoi{10.1086/309366}

\bibitem[{{Allard} {et~al.}(2014){Allard}, {Homeier}, \& {Freytag}}]{allard14}
{Allard}, F., {Homeier}, D., \& {Freytag}, B. 2014, in Astronomical Society of
  India Conference Series, Vol.~11, Astronomical Society of India Conference
  Series, 33--45

\bibitem[{{Allard} {et~al.}(2007{\natexlab{b}}){Allard}, {Kielkopf}, \&
  {Allard}}]{Allard07a}
{Allard}, N.~F., {Kielkopf}, J.~F., \& {Allard}, F. 2007{\natexlab{b}}, Eur.
  Phys. J. D, 44, 507, \dodoi{10.1140/epjd/e2007-00230-6}

\bibitem[{{Allard} {et~al.}(2016){Allard}, {Spiegelman}, \&
  {Kielkopf}}]{Allard16}
{Allard}, N.~F., {Spiegelman}, F., \& {Kielkopf}, J.~F. 2016, \aap, 589, A21,
  \dodoi{10.1051/0004-6361/201628270}

\bibitem[{{Allard} {et~al.}(2019){Allard}, {Spiegelman}, {Leininger}, \&
  {Molliere}}]{Allard19}
{Allard}, N.~F., {Spiegelman}, F., {Leininger}, T., \& {Molliere}, P. 2019,
  \aap, 628, A120, \dodoi{10.1051/0004-6361/201935593}

\bibitem[{{Amundsen} {et~al.}(2017){Amundsen}, {Tremblin}, {Manners},
  {Baraffe}, \& {Mayne}}]{Amundsen17}
{Amundsen}, D.~S., {Tremblin}, P., {Manners}, J., {Baraffe}, I., \& {Mayne},
  N.~J. 2017, \aap, 598, A97, \dodoi{10.1051/0004-6361/201629322}

\bibitem[{{Arcangeli} {et~al.}(2018){Arcangeli}, {D{\'e}sert}, {Line}, {Bean},
  {Parmentier}, {Stevenson}, {Kreidberg}, {Fortney}, {Mansfield}, \&
  {Showman}}]{Arcangeli18}
{Arcangeli}, J., {D{\'e}sert}, J.-M., {Line}, M.~R., {et~al.} 2018, \apjl, 855,
  L30, \dodoi{10.3847/2041-8213/aab272}

\bibitem[{{Asplund} {et~al.}(2021){Asplund}, {Amarsi}, \&
  {Grevesse}}]{Asplund21}
{Asplund}, M., {Amarsi}, A.~M., \& {Grevesse}, N. 2021, arXiv e-prints,
  arXiv:2105.01661.
\newblock \doarXiv{2105.01661}

\bibitem[{{Asplund} {et~al.}(2009){Asplund}, {Grevesse}, {Sauval}, \&
  {Scott}}]{Asplund09}
{Asplund}, M., {Grevesse}, N., {Sauval}, A.~J., \& {Scott}, P. 2009, \araa, 47,
  481, \dodoi{10.1146/annurev.astro.46.060407.145222}

\bibitem[{Azzam {et~al.}(2015)Azzam, Lodi, Yurchenko, \& Tennyson}]{Azzam15}
Azzam, A.~A., Lodi, L., Yurchenko, S.~N., \& Tennyson, J. 2015, Journal of
  Quantitative Spectroscopy and Radiative Transfer, 161, 41 ,
  \dodoi{https://doi.org/10.1016/j.jqsrt.2015.03.029}

\bibitem[{{Baraffe} {et~al.}(2002){Baraffe}, {Chabrier}, {Allard}, \&
  {Hauschildt}}]{Baraffe02}
{Baraffe}, I., {Chabrier}, G., {Allard}, F., \& {Hauschildt}, P.~H. 2002, A\&A,
  382, 563, \dodoi{10.1051/0004-6361:20011638}

\bibitem[{{Baraffe} {et~al.}(2003){Baraffe}, {Chabrier}, {Barman}, {Allard}, \&
  {Hauschildt}}]{baraffe03}
{Baraffe}, I., {Chabrier}, G., {Barman}, T.~S., {Allard}, F., \& {Hauschildt},
  P.~H. 2003, \aap, 402, 701

\bibitem[{{Baraffe} {et~al.}(2015){Baraffe}, {Homeier}, {Allard}, \&
  {Chabrier}}]{baraffe15}
{Baraffe}, I., {Homeier}, D., {Allard}, F., \& {Chabrier}, G. 2015, \aap, 577,
  A42, \dodoi{10.1051/0004-6361/201425481}

\bibitem[{{Barber} {et~al.}(2006){Barber}, {Tennyson}, {Harris}, \&
  {Tolchenov}}]{Barber06}
{Barber}, R.~J., {Tennyson}, J., {Harris}, G.~J., \& {Tolchenov}, R.~N. 2006,
  \mnras, 368, 1087, \dodoi{10.1111/j.1365-2966.2006.10184.x}

\bibitem[{{Barnes} \& {Fortney}(2003)}]{Barnes03}
{Barnes}, J.~W., \& {Fortney}, J.~J. 2003, \apj, 588, 545

\bibitem[{Barton {et~al.}(2013)Barton, Yurchenko, \& Tennyson}]{Barton13}
Barton, E.~J., Yurchenko, S.~N., \& Tennyson, J. 2013, Monthly Notices of the
  Royal Astronomical Society, 434, 1469, \dodoi{10.1093/mnras/stt1105}

\bibitem[{{Benni} {et~al.}(2020){Benni}, {Burdanov}, {Krushinsky}, {Bonfanti},
  {H{\'e}brard}, {Almenara}, {Dalal}, {Demangeon}, {Tsantaki}, {Pepper},
  {Stassun}, {Vanderburg}, {Belinski}, {Kashaev}, {Barkaoui}, {Kim}, {Kang},
  {Antonyuk}, {Dyachenko}, {Rastegaev}, {Beskakotov}, {Mitrofanova},
  {Pozuelos}, {Popov}, {Kiefer}, {Wilson}, {Ricker}, {Vanderspek}, {Latham},
  {Seager}, {Jenkins}, {Sokov}, {Sokova}, {Marchini}, {Papini}, {Salvaggio},
  {Banfi}, {Ba{\textcommabelow s}t{\"u}rk}, {Torun}, {Yal{\c{c}}{\i}nkaya},
  {Ivanov}, {Valyavin}, {Jehin}, {Gillon}, {Pak{\v{s}}tien{\.{e}}}, {Hentunen},
  {Shadick}, {Bretton}, {W{\"u}nsche}, {Garlitz}, {Jongen}, {Molina},
  {Girardin}, {Grau Horta}, {Naves}, {Benkhaldoun}, {Joner}, {Spencer},
  {Bieryla}, {Stevens}, {Jensen}, {Collins}, {Charbonneau}, {Quintana},
  {Mullally}, \& {Henze}}]{benni20}
{Benni}, P., {Burdanov}, A.~Y., {Krushinsky}, V.~V., {et~al.} 2020, arXiv
  e-prints, arXiv:2009.11899.
\newblock \doarXiv{2009.11899}

\bibitem[{Best {et~al.}(2020)Best, Dupuy, Liu, Siverd, \&
  Zhang}]{ultracoolsheet}
Best, W. M.~J., Dupuy, T.~J., Liu, M.~C., Siverd, R.~J., \& Zhang, Z. 2020,
  \dodoi{10.5281/zenodo.4169085}

\bibitem[{{Best} {et~al.}(2020){Best}, {Liu}, {Magnier}, \& {Dupuy}}]{best20}
{Best}, W. M.~J., {Liu}, M.~C., {Magnier}, E.~A., \& {Dupuy}, T.~J. 2020, \aj,
  159, 257, \dodoi{10.3847/1538-3881/ab84f4}

\bibitem[{{Best} {et~al.}(2018){Best}, {Magnier}, {Liu}, {Aller}, {Zhang},
  {Burgett}, {Chambers}, {Draper}, {Flewelling}, {Kaiser}, {Kudritzki},
  {Metcalfe}, {Tonry}, {Wainscoat}, \& {Waters}}]{Best18}
{Best}, W. M.~J., {Magnier}, E.~A., {Liu}, M.~C., {et~al.} 2018, \apjs, 234, 1,
  \dodoi{10.3847/1538-4365/aa9982}

\bibitem[{{Burcat} \& Ruscic(2005)}]{Burcat05}
{Burcat}, A., \& Ruscic, B. 2005, Third millenium Ideal Gas and Condensed Phase
  Thermochemical Database for Combustion with updates from Active
  Thermochemical Tables, TAE 960, ANL-05/20, Argonne National Laboratory

\bibitem[{{Burningham} {et~al.}(2017){Burningham}, {Marley}, {Line}, {Luiu},
  {Visscher}, {Morley}, {Saumon}, \& {Freedman}}]{Burningham17}
{Burningham}, B., {Marley}, M.~S., {Line}, M.~R., {et~al.} 2017, \mnras, 470,
  1177, \dodoi{10.1093/mnras/stx1246}

\bibitem[{{Burningham} {et~al.}(2021){Burningham}, {Faherty}, {Gonzales},
  {Marley}, {Visscher}, {Lupu}, {Gaarn}, {Bieger}, {Freedman}, \&
  {Saumon}}]{burningham21}
{Burningham}, B., {Faherty}, J.~K., {Gonzales}, E.~C., {et~al.} 2021, \mnras,
  \dodoi{10.1093/mnras/stab1361}

\bibitem[{{Burrows} {et~al.}(2002){Burrows}, {Burgasser}, {Kirkpatrick},
  {Liebert}, {Milsom}, {Sudarsky}, \& {Hubeny}}]{Burrows02}
{Burrows}, A., {Burgasser}, A.~J., {Kirkpatrick}, J.~D., {et~al.} 2002, \apj,
  573, 394, \dodoi{10.1086/340584}

\bibitem[{{Burrows} {et~al.}(2011){Burrows}, {Heng}, \& {Nampaisarn}}]{bhn11}
{Burrows}, A., {Heng}, K., \& {Nampaisarn}, T. 2011, \apj, 736, 47,
  \dodoi{10.1088/0004-637X/736/1/47}

\bibitem[{{Burrows} {et~al.}(1989){Burrows}, {Hubbard}, \& {Lunine}}]{bhl89}
{Burrows}, A., {Hubbard}, W.~B., \& {Lunine}, J.~I. 1989, \apj, 345, 939,
  \dodoi{10.1086/167964}

\bibitem[{{Burrows} {et~al.}(2001){Burrows}, {Hubbard}, {Lunine}, \&
  {Liebert}}]{Burrows01}
{Burrows}, A., {Hubbard}, W.~B., {Lunine}, J.~I., \& {Liebert}, J. 2001,
  Reviews of Modern Physics, 73, 719

\bibitem[{{Burrows} {et~al.}(2004){Burrows}, {Hubeny}, {Hubbard}, {Sudarsky},
  \& {Fortney}}]{Burrows04}
{Burrows}, A., {Hubeny}, I., {Hubbard}, W.~B., {Sudarsky}, D., \& {Fortney},
  J.~J. 2004, \apjl, 610, L53, \dodoi{10.1086/423173}

\bibitem[{{Burrows} {et~al.}(2000{\natexlab{a}}){Burrows}, {Marley}, \&
  {Sharp}}]{Burrows00b}
{Burrows}, A., {Marley}, M.~S., \& {Sharp}, C.~M. 2000{\natexlab{a}}, \apj,
  531, 438, \dodoi{10.1086/308462}

\bibitem[{{Burrows} {et~al.}(2000{\natexlab{b}}){Burrows}, {Marley}, \&
  {Sharp}}]{BMS}
---. 2000{\natexlab{b}}, \apj, 531, 438, \dodoi{10.1086/308462}

\bibitem[{{Burrows} {et~al.}(2006){Burrows}, {Sudarsky}, \&
  {Hubeny}}]{Burrows06}
{Burrows}, A., {Sudarsky}, D., \& {Hubeny}, I. 2006, \apj, 650, 1140,
  \dodoi{10.1086/507269}

\bibitem[{{Burrows} {et~al.}(1997){Burrows}, {Marley}, {Hubbard}, {Lunine},
  {Guillot}, {Saumon}, {Freedman}, {Sudarsky}, \& {Sharp}}]{Burrows97}
{Burrows}, A., {Marley}, M., {Hubbard}, W.~B., {et~al.} 1997, \apj, 491, 856

\bibitem[{{Ca{\~n}as} {et~al.}(2018){Ca{\~n}as}, {Bender}, {Mahadevan},
  {Fleming}, {Beatty}, {Covey}, {De Lee}, {Hearty},
  {Garc{\'\i}a-Hern{\'a}ndez}, {Majewski}, {Schneider}, {Stassun}, \&
  {Wilson}}]{canas18}
{Ca{\~n}as}, C.~I., {Bender}, C.~F., {Mahadevan}, S., {et~al.} 2018, \apjl,
  861, L4, \dodoi{10.3847/2041-8213/aacbc5}

\bibitem[{{Caffau} {et~al.}(2011){Caffau}, {Ludwig}, {Steffen}, {Freytag}, \&
  {Bonifacio}}]{Caffau11}
{Caffau}, E., {Ludwig}, H.-G., {Steffen}, M., {Freytag}, B., \& {Bonifacio}, P.
  2011, \solphys, 268, 255, \dodoi{10.1007/s11207-010-9541-4}

\bibitem[{{Carmichael} {et~al.}(2019){Carmichael}, {Latham}, \&
  {Vanderburg}}]{carmichael19}
{Carmichael}, T.~W., {Latham}, D.~W., \& {Vanderburg}, A.~M. 2019, \aj, 158,
  38, \dodoi{10.3847/1538-3881/ab245e}

\bibitem[{{Carmichael} {et~al.}(2020){Carmichael}, {Quinn}, {Zhou}, {Grieves},
  {Bouchy}, {Collins}, {Kielkopf}, {Schwarz}, {Vanderburg}, {Irwin},
  {Charbonneau}, {Ziegler}, {Briceno}, {Law}, {Mann}, {Huang}, {Shporer},
  {Rodriguez}, {Stassun}, \& {Latham}}]{carmichael20}
{Carmichael}, T.~W., {Quinn}, S.~N., {Zhou}, G., {et~al.} 2020, arXiv e-prints,
  arXiv:2009.13515.
\newblock \doarXiv{2009.13515}

\bibitem[{{Casewell} {et~al.}(2020){Casewell}, {Belardi}, {Parsons},
  {Littlefair}, {Braker}, {Hermes}, {Debes}, {Vanderbosch}, {Burleigh},
  {G{\"a}nsicke}, {Dhillon}, {Marsh}, {Winget}, \& {Winget}}]{casewell20}
{Casewell}, S.~L., {Belardi}, C., {Parsons}, S.~G., {et~al.} 2020, \mnras, 497,
  3571, \dodoi{10.1093/mnras/staa1608}

\bibitem[{{Chabrier} \& {Baraffe}(1997)}]{cb97}
{Chabrier}, G., \& {Baraffe}, I. 1997, \aap, 327, 1039.
\newblock \doarXiv{astro-ph/9704118}

\bibitem[{{Chandrasekhar} \& {M{\"u}nch}(1946)}]{chandra46}
{Chandrasekhar}, S., \& {M{\"u}nch}, G. 1946, \apj, 104, 446,
  \dodoi{10.1086/144875}

\bibitem[{{Chase}(1998)}]{Chase98}
{Chase}, M.~W. 1998, Journal of Physical and Chemical Reference Data, 28,
  Monograph No.~9, New York: AIP

\bibitem[{{Cushing} {et~al.}(2008){Cushing}, {Marley}, {Saumon}, {Kelly},
  {Vacca}, {Rayner}, {Freedman}, {Lodders}, \& {Roellig}}]{Cushing08}
{Cushing}, M.~C., {Marley}, M.~S., {Saumon}, D., {et~al.} 2008, \apj, 678,
  1372, \dodoi{10.1086/526489}

\bibitem[{{David} {et~al.}(2019){David}, {Hillenbrand}, {Gillen}, {Cody},
  {Howell}, {Isaacson}, \& {Livingston}}]{david19}
{David}, T.~J., {Hillenbrand}, L.~A., {Gillen}, E., {et~al.} 2019, \apj, 872,
  161, \dodoi{10.3847/1538-4357/aafe09}

\bibitem[{Dulick {et~al.}(2003)Dulick, C.~W.~Bauschlicher, Burrows, Sharp, Ram,
  \& Bernath}]{Dulick03}
Dulick, M., C.~W.~Bauschlicher, J., Burrows, A., {et~al.} 2003, The
  Astrophysical Journal, 594, 651, \dodoi{10.1086/376791}

\bibitem[{{Dupuy} \& {Kraus}(2013)}]{Dupuy13}
{Dupuy}, T.~J., \& {Kraus}, A.~L. 2013, Science, 341, 1492,
  \dodoi{10.1126/science.1241917}

\bibitem[{{Dupuy} \& {Liu}(2012)}]{Dupuy12}
{Dupuy}, T.~J., \& {Liu}, M.~C. 2012, \apjs, 201, 19,
  \dodoi{10.1088/0067-0049/201/2/19}

\bibitem[{{Dupuy} \& {Liu}(2017)}]{Dupuy17}
---. 2017, The Astrophysical Journal Supplement Series, 231, 15,
  \dodoi{10.3847/1538-4365/aa5e4c}

\bibitem[{{Fegley} \& {Lodders}(1994)}]{Fegley94}
{Fegley}, B.~J., \& {Lodders}, K. 1994, Icarus, 110, 117

\bibitem[{{Fegley} \& {Lodders}(1996)}]{Fegley96}
---. 1996, \apjl, 472, L37

\bibitem[{{Fernandes} {et~al.}(2019){Fernandes}, {Van Grootel}, {Salmon},
  {Aringer}, {Burgasser}, {Scuflaire}, {Brassard}, \& {Fontaine}}]{fernandes19}
{Fernandes}, C.~S., {Van Grootel}, V., {Salmon}, S. J.~A.~J., {et~al.} 2019,
  \apj, 879, 94, \dodoi{10.3847/1538-4357/ab2333}

\bibitem[{{Fortney} {et~al.}(2011){Fortney}, {Ikoma}, {Nettelmann}, {Guillot},
  \& {Marley}}]{Fortney11}
{Fortney}, J.~J., {Ikoma}, M., {Nettelmann}, N., {Guillot}, T., \& {Marley},
  M.~S. 2011, \apj, 729, 32, \dodoi{10.1088/0004-637X/729/1/32}

\bibitem[{{Fortney} {et~al.}(2005){Fortney}, {Marley}, {Lodders}, {Saumon}, \&
  {Freedman}}]{Fortney05}
{Fortney}, J.~J., {Marley}, M.~S., {Lodders}, K., {Saumon}, D., \& {Freedman},
  R. 2005, \apjl, 627, L69, \dodoi{10.1086/431952}

\bibitem[{{Fortney} {et~al.}(2008){Fortney}, {Marley}, {Saumon}, \&
  {Lodders}}]{Fortney08b}
{Fortney}, J.~J., {Marley}, M.~S., {Saumon}, D., \& {Lodders}, K. 2008, \apj,
  683, 1104, \dodoi{10.1086/589942}

\bibitem[{{Fortney} {et~al.}(2013){Fortney}, {Mordasini}, {Nettelmann},
  {Kempton}, {Greene}, \& {Zahnle}}]{Fortney13}
{Fortney}, J.~J., {Mordasini}, C., {Nettelmann}, N., {et~al.} 2013, \apj, 775,
  80, \dodoi{10.1088/0004-637X/775/1/80}

\bibitem[{{Freedman} {et~al.}(2014){Freedman}, {Lustig-Yaeger}, {Fortney},
  {Lupu}, {Marley}, \& {Lodders}}]{Freedman14}
{Freedman}, R.~S., {Lustig-Yaeger}, J., {Fortney}, J.~J., {et~al.} 2014, \apjs,
  214, 25, \dodoi{10.1088/0067-0049/214/2/25}

\bibitem[{{Freedman} {et~al.}(2008){Freedman}, {Marley}, \&
  {Lodders}}]{Freedman08}
{Freedman}, R.~S., {Marley}, M.~S., \& {Lodders}, K. 2008, \apjs, 174, 504

\bibitem[{{Garland} \& {Irwin}(2019)}]{garland19}
{Garland}, R., \& {Irwin}, P.~G.~J. 2019, arXiv e-prints, arXiv:1903.03997.
\newblock \doarXiv{1903.03997}

\bibitem[{{Gillen} {et~al.}(2017){Gillen}, {Hillenbrand}, {David}, {Aigrain},
  {Rebull}, {Stauffer}, {Cody}, \& {Queloz}}]{gillen17}
{Gillen}, E., {Hillenbrand}, L.~A., {David}, T.~J., {et~al.} 2017, \apj, 849,
  11, \dodoi{10.3847/1538-4357/aa84b3}

\bibitem[{{Gordon} \& {McBride}(1994)}]{Gordon94}
{Gordon}, S., \& {McBride}, B.~J. 1994, Computer Program for Calculation of
  Complex Chemical Equilibrium Compositions and Applications, Reference
  Publication 1311, NASA

\bibitem[{{Gurvich} {et~al.}(1989){Gurvich}, {Veyts}, \& {Alcock}}]{Gurvich89}
{Gurvich}, L.~V., {Veyts}, I.~V., \& {Alcock}, C.~B. 1989, Thermodynamic
  Properties of Individual Substances, 4th edn., Vol.~1 (New York: Hemisphere
  Publishing)

\bibitem[{{Gurvich} {et~al.}(1991){Gurvich}, {Veyts}, \& {Alcock}}]{Gurvich91}
---. 1991, Thermodynamic Properties of Individual Substances, 4th edn., Vol.~2
  (New York: Hemisphere Publishing)

\bibitem[{{Gurvich} {et~al.}(1994){Gurvich}, {Veyts}, \& {Alcock}}]{Gurvich94}
---. 1994, Thermodynamic Properties of Individual Substances, 4th edn., Vol.~3
  (Boca Raton, FL: CRC Press)

\bibitem[{Hargreaves {et~al.}(2010)Hargreaves, Hinkle, Bauschlicher, Wende,
  Seifahrt, \& Bernath}]{Hargreaves10}
Hargreaves, R.~J., Hinkle, K.~H., Bauschlicher, C.~W., {et~al.} 2010, The
  Astronomical Journal, 140, 919, \dodoi{10.1088/0004-6256/140/4/919}

\bibitem[{{Harris} {et~al.}(2008){Harris}, {Larner}, {Tennyson}, {Kaminsky},
  {Pavlenko}, \& {Jones}}]{Harris08}
{Harris}, G.~J., {Larner}, F.~C., {Tennyson}, J., {et~al.} 2008, \mnras, 390,
  143, \dodoi{10.1111/j.1365-2966.2008.13642.x}

\bibitem[{{Harris} {et~al.}(2006){Harris}, {Tennyson}, {Kaminsky}, {Pavlenko},
  \& {Jones}}]{Harris06}
{Harris}, G.~J., {Tennyson}, J., {Kaminsky}, B.~M., {Pavlenko}, Y.~V., \&
  {Jones}, H.~R.~A. 2006, \mnras, 367, 400,
  \dodoi{10.1111/j.1365-2966.2005.09960.x}

\bibitem[{{Heng} \& {Kitzmann}(2017)}]{Heng17}
{Heng}, K., \& {Kitzmann}, D. 2017, \apjs, 232, 20,
  \dodoi{10.3847/1538-4365/aa8907}

\bibitem[{{Heng} {et~al.}(2016){Heng}, {Lyons}, \& {Tsai}}]{Heng16}
{Heng}, K., {Lyons}, J.~R., \& {Tsai}, S.-M. 2016, \apj, 816, 96,
  \dodoi{10.3847/0004-637X/816/2/96}

\bibitem[{{Hod{\v{z}}i{\'c}} {et~al.}(2018){Hod{\v{z}}i{\'c}}, {Triaud},
  {Anderson}, {Bouchy}, {Collier Cameron}, {Delrez}, {Gillon}, {Hellier},
  {Jehin}, {Lendl}, {Maxted}, {Pepe}, {Pollacco}, {Queloz}, {S{\'e}gransan},
  {Smalley}, {Udry}, \& {West}}]{hodzik18}
{Hod{\v{z}}i{\'c}}, V., {Triaud}, A. H.~M.~J., {Anderson}, D.~R., {et~al.}
  2018, \mnras, 481, 5091, \dodoi{10.1093/mnras/sty2512}

\bibitem[{{Huang} {et~al.}(2014){Huang}, {Gamache}, {Freedman}, {Schwenke}, \&
  {Lee}}]{Huang14}
{Huang}, X., {Gamache}, R.~R., {Freedman}, R.~S., {Schwenke}, D.~W., \& {Lee},
  T.~J. 2014, \jqsrt, 147, 134, \dodoi{10.1016/j.jqsrt.2014.05.015}

\bibitem[{{Hubeny}(1988)}]{Hubeny88}
{Hubeny}, I. 1988, Computer Physics Communications, 52, 103,
  \dodoi{10.1016/0010-4655(88)90177-4}

\bibitem[{{Hubeny}(2017)}]{Hubeny17}
---. 2017, \mnras, 469, 841, \dodoi{10.1093/mnras/stx758}

\bibitem[{{Hubeny} \& {Burrows}(2007)}]{Hubeny07}
{Hubeny}, I., \& {Burrows}, A. 2007, \apj, 669, 1248

\bibitem[{{Hubeny} \& {Lanz}(1995)}]{Hubeny95}
{Hubeny}, I., \& {Lanz}, T. 1995, \apj, 439, 875, \dodoi{10.1086/175226}

\bibitem[{{Hubeny} \& {Mihalas}(2014)}]{Mihalas14}
{Hubeny}, I., \& {Mihalas}, D. 2014, {Theory of Stellar Atmospheres}

\bibitem[{{Jensen-Clem} {et~al.}(2020){Jensen-Clem}, {Millar-Blanchaer}, {van
  Holstein}, {Mawet}, {Graham}, {Sengupta}, {Marley}, {Snik}, {Vigan},
  {Hinkley}, {de Boer}, {Girard}, {De Rosa}, {Bowler}, {Wiktorowicz}, {Perrin},
  {Crepp}, \& {Macintosh}}]{Clem20}
{Jensen-Clem}, R., {Millar-Blanchaer}, M.~A., {van Holstein}, R.~G., {et~al.}
  2020, \aj, 160, 286, \dodoi{10.3847/1538-3881/abc33d}

\bibitem[{{Joergens}(2014)}]{joergens2014}
{Joergens}, V. 2014, {50 Years of Brown Dwarfs: From Prediction to Discovery to
  Forefront of Research}, Vol. 401, \dodoi{10.1007/978-3-319-01162-2}

\bibitem[{{Kataria} {et~al.}(2016){Kataria}, {Sing}, {Lewis}, {Visscher},
  {Showman}, {Fortney}, \& {Marley}}]{Kataria16}
{Kataria}, T., {Sing}, D.~K., {Lewis}, N.~K., {et~al.} 2016, \apj, 821, 9,
  \dodoi{10.3847/0004-637X/821/1/9}

\bibitem[{{Kirkpatrick}(2005)}]{Kirkpatrick05}
{Kirkpatrick}, J.~D. 2005, \araa, 43, 195

\bibitem[{{Kirkpatrick} {et~al.}(2020){Kirkpatrick}, {Gelino}, {Faherty},
  {Meisner}, {Caselden}, {Schneider}, {Marocco}, {Cayago}, {Smart},
  {Eisenhardt}, {Kuchner}, {Wright}, {Cushing}, {Allers}, {Bardalez Gagliuffi},
  {Burgasser}, {Gagne}, {Logsdon}, {Martin}, {Ingalls}, {Lowrance}, {Abrahams},
  {Aganze}, {Gerasimov}, {Gonzales}, {Hsu}, {Kamraj}, {Kiman}, {Rees},
  {Theissen}, {Ammar}, {Stevnbak Andersen}, {Beaulieu}, {Colin}, {Elachi},
  {Goodman}, {Gramaize}, {Hamlet}, {Hong}, {Jonkeren}, {Khalil}, {Martin},
  {Pendrill}, {Pumphrey}, {Rothermich}, {Sainio}, {Stenner}, {Tanner},
  {Thevenot}, {Voloshin}, {Walla}, \& {Wedracki}}]{kirkpatrick20}
{Kirkpatrick}, J.~D., {Gelino}, C.~R., {Faherty}, J.~K., {et~al.} 2020, arXiv
  e-prints, arXiv:2011.11616.
\newblock \doarXiv{2011.11616}

\bibitem[{Kissel {et~al.}(2002)Kissel, Sumpf, Kronfeldt, Tikhomirov, \&
  Ponomarev}]{Kissel02}
Kissel, A., Sumpf, B., Kronfeldt, H.-D., Tikhomirov, B., \& Ponomarev, Y. 2002,
  Journal of Molecular Spectroscopy, 216, 345 ,
  \dodoi{https://doi.org/10.1006/jmsp.2002.8630}

\bibitem[{{Kitzmann} {et~al.}(2020){Kitzmann}, {Heng}, {Oreshenko}, {Grimm},
  {Apai}, {Bowler}, {Burgasser}, \& {Marley}}]{kitzmann20}
{Kitzmann}, D., {Heng}, K., {Oreshenko}, M., {et~al.} 2020, \apj, 890, 174,
  \dodoi{10.3847/1538-4357/ab6d71}

\bibitem[{{Kurucz}(2011)}]{Kurucz11}
{Kurucz}, R.~L. 2011, Canadian Journal of Physics, 89, 417,
  \dodoi{10.1139/p10-104}

\bibitem[{{Kurucz} \& {Avrett}(1981)}]{Kurucz1981}
{Kurucz}, R.~L., \& {Avrett}, E.~H. 1981, SAO Special Report, 391

\bibitem[{{Lacis} \& {Oinas}(1991)}]{Lacis91}
{Lacis}, A.~A., \& {Oinas}, V. 1991, \jgr, 96, 9027, \dodoi{10.1029/90JD01945}

\bibitem[{Li {et~al.}(2015)Li, Gordon, Rothman, Tan, Hu, Kassi, Campargue, \&
  Medvedev}]{Li15}
Li, G., Gordon, I.~E., Rothman, L.~S., {et~al.} 2015, The Astrophysical Journal
  Supplement Series, 216, 15, \dodoi{10.1088/0067-0049/216/1/15}

\bibitem[{{Liebert} {et~al.}(2000){Liebert}, {Reid}, {Burrows}, {Burgasser},
  {Kirkpatrick}, \& {Gizis}}]{Liebert00}
{Liebert}, J., {Reid}, I.~N., {Burrows}, A., {et~al.} 2000, \apjl, 533, L155,
  \dodoi{10.1086/312619}

\bibitem[{{Line} {et~al.}(2015){Line}, {Teske}, {Burningham}, {Fortney}, \&
  {Marley}}]{Line15}
{Line}, M.~R., {Teske}, J., {Burningham}, B., {Fortney}, J.~J., \& {Marley},
  M.~S. 2015, \apj, 807, 183, \dodoi{10.1088/0004-637X/807/2/183}

\bibitem[{{Line} {et~al.}(2017){Line}, {Marley}, {Liu}, {Burningham}, {Morley},
  {Hinkel}, {Teske}, {Fortney}, {Freedman}, \& {Lupu}}]{Line17}
{Line}, M.~R., {Marley}, M.~S., {Liu}, M.~C., {et~al.} 2017, \apj, 848, 83,
  \dodoi{10.3847/1538-4357/aa7ff0}

\bibitem[{{Littlefair} {et~al.}(2014){Littlefair}, {Casewell}, {Parsons},
  {Dhillon}, {Marsh}, {G{\"a}nsicke}, {Bloemen}, {Catalan}, {Irawati}, {Hardy},
  {Mcallister}, {Bours}, {Richichi}, {Burleigh}, {Burningham}, {Breedt}, \&
  {Kerry}}]{littlefair14}
{Littlefair}, S.~P., {Casewell}, S.~L., {Parsons}, S.~G., {et~al.} 2014,
  \mnras, 445, 2106, \dodoi{10.1093/mnras/stu1895}

\bibitem[{{Liu} {et~al.}(2016){Liu}, {Dupuy}, \& {Allers}}]{Liu16}
{Liu}, M.~C., {Dupuy}, T.~J., \& {Allers}, K.~N. 2016, \apj, 833, 96,
  \dodoi{10.3847/1538-4357/833/1/96}

\bibitem[{{Lodders}(1999)}]{Lodders99}
{Lodders}, K. 1999, \apj, 519, 793

\bibitem[{{Lodders}(2002)}]{Lodders02b}
---. 2002, \apj, 577, 974, \dodoi{10.1086/342241}

\bibitem[{{Lodders}(2010)}]{Lodders10}
---. 2010, {in Formation and Evolution of Exoplanets, ed. R. Barnes} (Wiley),
  157--186, arXiv:0910.0811

\bibitem[{{Lodders}(2020)}]{Lodders20}
---. 2020, Oxford Research Encyclopedia of Planetary Science,
  \dodoi{10.1093/acrefore/9780190647926.013.145}

\bibitem[{{Lodders} \& {Fegley}(2002)}]{Lodders02}
{Lodders}, K., \& {Fegley}, B. 2002, Icarus, 155, 393

\bibitem[{{Lodders} \& {Fegley}(2006)}]{Lodders06}
---. 2006, {Astrophysics Update 2} (Springer Praxis Books, Berlin: Springer,
  2006)

\bibitem[{{Lupu} {et~al.}(2014){Lupu}, {Zahnle}, {Marley}, {Schaefer},
  {Fegley}, {Morley}, {Cahoy}, {Freedman}, \& {Fortney}}]{Lupu14}
{Lupu}, R.~E., {Zahnle}, K., {Marley}, M.~S., {et~al.} 2014, \apj, 784, 27,
  \dodoi{10.1088/0004-637X/784/1/27}

\bibitem[{{Malik} {et~al.}(2017){Malik}, {Grosheintz}, {Mendon{\c{c}}a},
  {Grimm}, {Lavie}, {Kitzmann}, {Tsai}, {Burrows}, {Kreidberg}, {Bedell},
  {Bean}, {Stevenson}, \& {Heng}}]{Malik17}
{Malik}, M., {Grosheintz}, L., {Mendon{\c{c}}a}, J.~M., {et~al.} 2017, \aj,
  153, 56, \dodoi{10.3847/1538-3881/153/2/56}

\bibitem[{{Marley} {et~al.}(2007){Marley}, {Fortney}, {Hubickyj},
  {Bodenheimer}, \& {Lissauer}}]{Marley07}
{Marley}, M.~S., {Fortney}, J.~J., {Hubickyj}, O., {Bodenheimer}, P., \&
  {Lissauer}, J.~J. 2007, \apj, 655, 541, \dodoi{10.1086/509759}

\bibitem[{{Marley} {et~al.}(1999){Marley}, {Gelino}, {Stephens}, {Lunine}, \&
  {Freedman}}]{Marley99}
{Marley}, M.~S., {Gelino}, C., {Stephens}, D., {Lunine}, J.~I., \& {Freedman},
  R. 1999, \apj, 513, 879

\bibitem[{{Marley} \& {Robinson}(2015)}]{MarlRob15}
{Marley}, M.~S., \& {Robinson}, T.~D. 2015, \araa, 53, 279,
  \dodoi{10.1146/annurev-astro-082214-122522}

\bibitem[{{Marley} {et~al.}(2012){Marley}, {Saumon}, {Cushing}, {Ackerman},
  {Fortney}, \& {Freedman}}]{Marley12}
{Marley}, M.~S., {Saumon}, D., {Cushing}, M., {et~al.} 2012, \apj, 754, 135,
  \dodoi{10.1088/0004-637X/754/2/135}

\bibitem[{{Marley} {et~al.}(1996){Marley}, {Saumon}, {Guillot}, {Freedman},
  {Hubbard}, {Burrows}, \& {Lunine}}]{Marley96}
{Marley}, M.~S., {Saumon}, D., {Guillot}, T., {et~al.} 1996, Science, 272, 1919

\bibitem[{{Marley} {et~al.}(2002){Marley}, {Seager}, {Saumon}, {Lodders},
  {Ackerman}, {Freedman}, \& {Fan}}]{Marley02}
{Marley}, M.~S., {Seager}, S., {Saumon}, D., {et~al.} 2002, \apj, 568, 335

\bibitem[{{Marois} {et~al.}(2008){Marois}, {Macintosh}, {Barman}, {Zuckerman},
  {Song}, {Patience}, {Lafreni{\`e}re}, \& {Doyon}}]{Marois08}
{Marois}, C., {Macintosh}, B., {Barman}, T., {et~al.} 2008, Science, 322, 1348,
  \dodoi{10.1126/science.1166585}

\bibitem[{{McKay} {et~al.}(1989){McKay}, {Pollack}, \& {Courtin}}]{Mckay89}
{McKay}, C.~P., {Pollack}, J.~B., \& {Courtin}, R. 1989, Icarus, 80, 23

\bibitem[{{McKemmish} {et~al.}(2016){McKemmish}, {Yurchenko}, \&
  {Tennyson}}]{McKemmish16}
{McKemmish}, L.~K., {Yurchenko}, S.~N., \& {Tennyson}, J. 2016, \mnras, 463,
  771, \dodoi{10.1093/mnras/stw1969}

\bibitem[{{Miles} {et~al.}(2020){Miles}, {Skemer}, {Morley}, {Marley},
  {Fortney}, {Allers}, {Faherty}, {Geballe}, {Visscher}, {Schneider}, {Lupu},
  {Freedman}, \& {Bjoraker}}]{miles20}
{Miles}, B.~E., {Skemer}, A. J.~I., {Morley}, C.~V., {et~al.} 2020, \aj, 160,
  63, \dodoi{10.3847/1538-3881/ab9114}

\bibitem[{{Mireles} {et~al.}(2020){Mireles}, {Shporer}, {Grieves}, {Zhou},
  {G{\"u}nther}, {Brahm}, {Ziegler}, {Stassun}, {Huang}, {Nielsen}, {dos
  Santos}, {Udry}, {Bouchy}, {Ireland}, {Wallace}, {Sarkis}, {Henning},
  {Jord{\'a}n}, {Law}, {Mann}, {Paredes}, {James}, {Jao}, {Henry}, {Butler},
  {Rodriguez}, {Yu}, {Flowers}, {Ricker}, {Latham}, {Vanderspek}, {Seager},
  {Winn}, {Jenkins}, {Furesz}, {Hesse}, {Quintana}, {Rose}, {Smith},
  {Tenenbaum}, {Vezie}, {Yahalomi}, \& {Zhan}}]{mireles20}
{Mireles}, I., {Shporer}, A., {Grieves}, N., {et~al.} 2020, \aj, 160, 133,
  \dodoi{10.3847/1538-3881/aba526}

\bibitem[{{Morley} {et~al.}(2013){Morley}, {Fortney}, {Kempton}, {Marley},
  {Visscher}, \& {Zahnle}}]{Morley13}
{Morley}, C.~V., {Fortney}, J.~J., {Kempton}, E.~M.-R., {et~al.} 2013, \apj,
  775, 33, \dodoi{10.1088/0004-637X/775/1/33}

\bibitem[{{Morley} {et~al.}(2012){Morley}, {Fortney}, {Marley}, {Visscher},
  {Saumon}, \& {Leggett}}]{Morley12}
{Morley}, C.~V., {Fortney}, J.~J., {Marley}, M.~S., {et~al.} 2012, \apj, 756,
  172, \dodoi{10.1088/0004-637X/756/2/172}

\bibitem[{{Morley} {et~al.}(2014{\natexlab{a}}){Morley}, {Marley}, {Fortney},
  \& {Lupu}}]{Morley14b}
{Morley}, C.~V., {Marley}, M.~S., {Fortney}, J.~J., \& {Lupu}, R.
  2014{\natexlab{a}}, \apjl, 789, L14, \dodoi{10.1088/2041-8205/789/1/L14}

\bibitem[{{Morley} {et~al.}(2014{\natexlab{b}}){Morley}, {Marley}, {Fortney},
  {Lupu}, {Saumon}, {Greene}, \& {Lodders}}]{Morley14a}
{Morley}, C.~V., {Marley}, M.~S., {Fortney}, J.~J., {et~al.}
  2014{\natexlab{b}}, \apj, 787, 78, \dodoi{10.1088/0004-637X/787/1/78}

\bibitem[{{Morley} {et~al.}(2019){Morley}, {Skemer}, {Miles}, {Line}, {Lopez},
  {Brogi}, {Freedman}, \& {Marley}}]{Morley19}
{Morley}, C.~V., {Skemer}, A.~J., {Miles}, B.~E., {et~al.} 2019, \apjl, 882,
  L29, \dodoi{10.3847/2041-8213/ab3c65}

\bibitem[{{Moses} {et~al.}(2013){Moses}, {Line}, {Visscher}, {Richardson},
  {Nettelmann}, {Fortney}, {Stevenson}, \& {Madhusudhan}}]{Moses13}
{Moses}, J.~I., {Line}, M.~R., {Visscher}, C., {et~al.} 2013, ArXiv e-prints.
\newblock \doarXiv{1306.5178}

\bibitem[{{Nowak} {et~al.}(2017){Nowak}, {Palle}, {Gandolfi}, {Dai}, {Lanza},
  {Hirano}, {Barrag{\'a}n}, {Fukui}, {Bruntt}, {Endl}, {Cochran}, {Prada
  Moroni}, {Prieto-Arranz}, {Kiilerich}, {Nespral}, {Hatzes}, {Albrecht},
  {Deeg}, {Winn}, {Yu}, {Kuzuhara}, {Grziwa}, {Smith}, {Guenther}, {Van Eylen},
  {Csizmadia}, {Fridlund}, {Cabrera}, {Eigm{\"u}ller}, {Erikson}, {Korth},
  {Narita}, {P{\"a}tzold}, {Rauer}, \& {Ribas}}]{nowak17}
{Nowak}, G., {Palle}, E., {Gandolfi}, D., {et~al.} 2017, \aj, 153, 131,
  \dodoi{10.3847/1538-3881/aa5cb6}

\bibitem[{{Oppenheimer} {et~al.}(1995){Oppenheimer}, {Kulkarni}, {Matthews}, \&
  {Nakajima}}]{Oppenheimer95}
{Oppenheimer}, B.~R., {Kulkarni}, S.~R., {Matthews}, K., \& {Nakajima}, T.
  1995, Science, 270, 1478

\bibitem[{{Orton} {et~al.}(1996){Orton}, {Ortiz}, {Baines}, {Bjoraker},
  {Carsenty}, {Colas}, {Dayal}, {Deming}, {Drossart}, {Frappa}, {Friedson},
  {Goguen}, {Golisch}, {Griep}, {Hernandez}, {Hoffmann}, {Jennings},
  {Kaminski}, {Kuhn}, {Laques}, {Limaye}, {Lin}, {Lecacheux}, {Martin},
  {McCabe}, {Momary}, {Parker}, {Puetter}, {Ressler}, {Reyes}, {Sada},
  {Spencer}, {Spitale}, {Stewart}, {Varsik}, {Warell}, {Wild},
  {Yanamandra-Fisher}, {Fazio}, {Hora}, \& {Deutsch}}]{Orton96}
{Orton}, G., {Ortiz}, J.~L., {Baines}, K., {et~al.} 1996, Science, 272, 839,
  \dodoi{10.1126/science.272.5263.839}

\bibitem[{{Parsons} {et~al.}(2017){Parsons}, {Hermes}, {Marsh}, {G{\"a}nsicke},
  {Tremblay}, {Littlefair}, {Sahman}, {Ashley}, {Green}, {Rattanasoon},
  {Dhillon}, {Burleigh}, {Casewell}, {Buckley}, {Braker}, {Irawati}, {Dennihy},
  {Rodr{\'\i}guez-Gil}, {Winget}, {Winget}, {Bell}, \& {Kilic}}]{parsons17}
{Parsons}, S.~G., {Hermes}, J.~J., {Marsh}, T.~R., {et~al.} 2017, \mnras, 471,
  976, \dodoi{10.1093/mnras/stx1610}

\bibitem[{{Patience} {et~al.}(2012){Patience}, {King}, {De Rosa}, {Vigan},
  {Witte}, {Rice}, {Helling}, \& {Hauschildt}}]{patience12}
{Patience}, J., {King}, R.~R., {De Rosa}, R.~J., {et~al.} 2012, \aap, 540, A85,
  \dodoi{10.1051/0004-6361/201118058}

\bibitem[{{Phillips} {et~al.}(2020){Phillips}, {Tremblin, P.}, {Baraffe, I.},
  {Chabrier, G.}, {Allard, N. F.}, {Spiegelman, F.}, {Goyal, J. M.}, {Drummond,
  B.}, \& {H\'ebrard, E.}}]{Phillips20}
{Phillips}, M.~W., {Tremblin, P.}, {Baraffe, I.}, {et~al.} 2020, A\&A, 637,
  A38, \dodoi{10.1051/0004-6361/201937381}

\bibitem[{{Piette} \& {Madhusudhan}(2020)}]{Piette20}
{Piette}, A. A.~A., \& {Madhusudhan}, N. 2020, \mnras, 497, 5136,
  \dodoi{10.1093/mnras/staa2289}

\bibitem[{Pine(1992)}]{Pine92}
Pine, A.~S. 1992, The Journal of Chemical Physics, 97, 773,
  \dodoi{10.1063/1.463943}

\bibitem[{{Potekhin} \& {Chabrier}(2012)}]{pc12}
{Potekhin}, A.~Y., \& {Chabrier}, G. 2012, \aap, 538, A115,
  \dodoi{10.1051/0004-6361/201117938}

\bibitem[{{Robie} \& {Hemingway}(1995)}]{Robie95}
{Robie}, R.~A., \& {Hemingway}, B.~S. 1995, Thermodynamic Properties of
  Minerals and Related Substances at 298.15 K and 1 Bar (10$^{5}$ Pascals)
  Pressure and at Higher Temperatures (USGS Bulletin 2131)

\bibitem[{{Robinson} \& {Catling}(2014)}]{robinson14c}
{Robinson}, T.~D., \& {Catling}, D.~C. 2014, Nature Geoscience, 7, 12,
  \dodoi{10.1038/ngeo2020}

\bibitem[{Rossi \& Pascale(1985)}]{Rossi85}
Rossi, F., \& Pascale, J. 1985, Phys. Rev. A, 32, 2657,
  \dodoi{10.1103/PhysRevA.32.2657}

\bibitem[{Rothman {et~al.}(2010)Rothman, Gordon, Barber, Dothe, Gamache,
  Goldman, Perevalov, Tashkun, \& Tennyson}]{Rothman10}
Rothman, L., Gordon, I., Barber, R., {et~al.} 2010, Journal of Quantitative
  Spectroscopy and Radiative Transfer, 111, 2139 ,
  \dodoi{https://doi.org/10.1016/j.jqsrt.2010.05.001}

\bibitem[{{Rothman} {et~al.}(2013){Rothman}, {Gordon}, {Babikov}, {Barbe},
  {Chris Benner}, {Bernath}, {Birk}, {Bizzocchi}, {Boudon}, {Brown},
  {Campargue}, {Chance}, {Cohen}, {Coudert}, {Devi}, {Drouin}, {Fayt}, {Flaud},
  {Gamache}, {Harrison}, {Hartmann}, {Hill}, {Hodges}, {Jacquemart}, {Jolly},
  {Lamouroux}, {Le Roy}, {Li}, {Long}, {Lyulin}, {Mackie}, {Massie},
  {Mikhailenko}, {M{\"u}ller}, {Naumenko}, {Nikitin}, {Orphal}, {Perevalov},
  {Perrin}, {Polovtseva}, {Richard}, {Smith}, {Starikova}, {Sung}, {Tashkun},
  {Tennyson}, {Toon}, {Tyuterev}, \& {Wagner}}]{Rothman13}
{Rothman}, L.~S., {Gordon}, I.~E., {Babikov}, Y., {et~al.} 2013, \jqsrt, 130,
  4, \dodoi{10.1016/j.jqsrt.2013.07.002}

\bibitem[{{Ryabchikova} {et~al.}(2015){Ryabchikova}, {Piskunov}, {Kurucz},
  {Stempels}, {Heiter}, {Pakhomov}, \& {Barklem}}]{Ryabchikova15}
{Ryabchikova}, T., {Piskunov}, N., {Kurucz}, R., {et~al.} 2015, Physica
  Scripta, 90, 054005, \dodoi{10.1088/0031-8949/90/5/054005}

\bibitem[{{Salem} {et~al.}(2004){Salem}, {Bouanich}, {Walrand}, {Aroui}, \&
  {Blanquet}}]{Salem04}
{Salem}, J., {Bouanich}, J.-P., {Walrand}, J., {Aroui}, H., \& {Blanquet}, G.
  2004, Journal of Molecular Spectroscopy, 228, 23,
  \dodoi{10.1016/j.jms.2004.06.015}

\bibitem[{{Saumon} {et~al.}(1995){Saumon}, {Chabrier}, \& {van Horn}}]{SCVH}
{Saumon}, D., {Chabrier}, G., \& {van Horn}, H.~M. 1995, \apjs, 99, 713

\bibitem[{{Saumon} {et~al.}(1996){Saumon}, {Hubbard}, {Burrows}, {Guillot},
  {Lunine}, \& {Chabrier}}]{saumon96}
{Saumon}, D., {Hubbard}, W.~B., {Burrows}, A., {et~al.} 1996, \apj, 460, 993,
  \dodoi{10.1086/177027}

\bibitem[{{Saumon} \& {Marley}(2008)}]{Saumon08}
{Saumon}, D., \& {Marley}, M.~S. 2008, \apj, 689, 1327, \dodoi{10.1086/592734}

\bibitem[{{Saumon} {et~al.}(2012){Saumon}, {Marley}, {Abel}, {Frommhold}, \&
  {Freedman}}]{Saumon12}
{Saumon}, D., {Marley}, M.~S., {Abel}, M., {Frommhold}, L., \& {Freedman},
  R.~S. 2012, \apj, 750, 74, \dodoi{10.1088/0004-637X/750/1/74}

\bibitem[{{Schwenke}(1998)}]{Schwenke98}
{Schwenke}, D.~W. 1998, Faraday Discussions, 109, 321, \dodoi{10.1039/a800070k}

\bibitem[{{Sengupta} \& {Marley}(2010)}]{Sengupta2010}
{Sengupta}, S., \& {Marley}, M.~S. 2010, \apjl, 722, L142,
  \dodoi{10.1088/2041-8205/722/2/L142}

\bibitem[{{Skemer} {et~al.}(2016){Skemer}, {Morley}, {Allers}, {Geballe},
  {Marley}, {Fortney}, {Faherty}, {Bjoraker}, \& {Lupu}}]{Skemer16}
{Skemer}, A.~J., {Morley}, C.~V., {Allers}, K.~N., {et~al.} 2016, \apjl, 826,
  L17, \dodoi{10.3847/2041-8205/826/2/L17}

\bibitem[{{Sousa-Silva} {et~al.}(2015){Sousa-Silva}, {Al-Refaie}, {Tennyson},
  \& {Yurchenko}}]{SousaSilva15}
{Sousa-Silva}, C., {Al-Refaie}, A.~F., {Tennyson}, J., \& {Yurchenko}, S.~N.
  2015, \mnras, 446, 2337, \dodoi{10.1093/mnras/stu2246}

\bibitem[{{Stempels}(2009)}]{chabrier09_MR}
{Stempels}, E., ed. 2009, American Institute of Physics Conference Series, Vol.
  1094, {The mass-radius relationship from solar-type stars to terrestrial
  planets: a review}, ed. E.~{Stempels}, 102--111

\bibitem[{{Stephens} {et~al.}(2009){Stephens}, {Leggett}, {Cushing}, {Marley},
  {Saumon}, {Geballe}, {Golimowski}, {Fan}, \& {Noll}}]{Stephens09}
{Stephens}, D.~C., {Leggett}, S.~K., {Cushing}, M.~C., {et~al.} 2009, \apj,
  702, 154, \dodoi{10.1088/0004-637X/702/1/154}

\bibitem[{{Tannock} {et~al.}(2021){Tannock}, {Metchev}, {Heinze},
  {Miles-P{\'a}ez}, {Gagn{\'e}}, {Burgasser}, {Marley}, {Apai}, {Su{\'a}rez},
  \& {Plavchan}}]{Tannock21}
{Tannock}, M.~E., {Metchev}, S., {Heinze}, A., {et~al.} 2021, arXiv e-prints,
  arXiv:2103.01990.
\newblock \doarXiv{2103.01990}

\bibitem[{{Tennyson} \& {Yurchenko}(2018)}]{Tennyson18}
{Tennyson}, J., \& {Yurchenko}, S. 2018, Atoms, 6, 26,
  \dodoi{10.3390/atoms6020026}

\bibitem[{{Tennyson} \& {Yurchenko}(2012)}]{Tennyson12}
{Tennyson}, J., \& {Yurchenko}, S.~N. 2012, \mnras, 425, 21,
  \dodoi{10.1111/j.1365-2966.2012.21440.x}

\bibitem[{{Toon} {et~al.}(1989){Toon}, {McKay}, {Ackerman}, \&
  {Santhanam}}]{Toon89}
{Toon}, O.~B., {McKay}, C.~P., {Ackerman}, T.~P., \& {Santhanam}, K. 1989,
  Journal of Geophysical Research, 94, 16287

\bibitem[{{Visscher}(2012)}]{Visscher12}
{Visscher}, C. 2012, \apj, 757, 5, \dodoi{10.1088/0004-637X/757/1/5}

\bibitem[{{Visscher} {et~al.}(2010){Visscher}, {Lodders}, \&
  {Fegley}}]{Visscher10}
{Visscher}, C., {Lodders}, K., \& {Fegley}, Jr., B. 2010, \apj, 716, 1060,
  \dodoi{10.1088/0004-637X/716/2/1060}

\bibitem[{{Visscher} {et~al.}(2006){Visscher}, {Lodders}, \&
  {Fegley}}]{Visscher06}
{Visscher}, C., {Lodders}, K., \& {Fegley}, B.~J. 2006, \apj, 648, 1181,
  \dodoi{10.1086/506245}

\bibitem[{{von Boetticher} {et~al.}(2017){von Boetticher}, {Triaud}, {Queloz},
  {Gill}, {Lendl}, {Delrez}, {Anderson}, {Collier Cameron}, {Faedi}, {Gillon},
  {G{\'o}mez Maqueo Chew}, {Hebb}, {Hellier}, {Jehin}, {Maxted}, {Martin},
  {Pepe}, {Pollacco}, {S{\'e}gransan}, {Smalley}, {Udry}, \&
  {West}}]{boetticher17}
{von Boetticher}, A., {Triaud}, A. H.~M.~J., {Queloz}, D., {et~al.} 2017, \aap,
  604, L6, \dodoi{10.1051/0004-6361/201731107}

\bibitem[{{{\v{S}}ubjak} {et~al.}(2020){{\v{S}}ubjak}, {Sharma}, {Carmichael},
  {Johnson}, {Gonzales}, {Matthews}, {Boffin}, {Brahm}, {Chaturvedi},
  {Chakraborty}, {Ciardi}, {Collins}, {Esposito}, {Fridlund}, {Gan},
  {Gandolfi}, {Garc{\'\i}a}, {Guenther}, {Hatzes}, {Latham}, {Mathis},
  {Mathur}, {Persson}, {Relles}, {Schlieder}, {Barclay}, {Dressing},
  {Crossfield}, {Howard}, {Rodler}, {Zhou}, {Quinn}, {Esquerdo}, {Calkins},
  {Berlind}, {Stassun}, {Bla{\v{z}}ek}, {Skarka}, {{\v{S}}pokov{\'a}},
  {{\v{Z}}{\'a}k}, {Albrecht}, {Sobrino}, {Beck}, {Cabrera}, {Carleo},
  {Cochran}, {Csizmadia}, {Dai}, {Deeg}, {de Leon}, {Eigm{\"u}ller}, {Endl},
  {Erikson}, {Fukui}, {Georgieva}, {Gonz{\'a}lez-Cuesta}, {Grziwa}, {Hidalgo},
  {Hirano}, {Hjorth}, {Knudstrup}, {Korth}, {Lam}, {Livingston}, {Lund},
  {Luque}, {Rodr{\'\i}guez}, {Murgas}, {Narita}, {Nespral}, {Niraula}, {Nowak},
  {Pall{\'e}}, {P{\"a}tzold}, {Prieto-Arranz}, {Rauer}, {Redfield}, {Ribas},
  {Smith}, {Eylen}, \& {Kab{\'a}th}}]{subjak20}
{{\v{S}}ubjak}, J., {Sharma}, R., {Carmichael}, T.~W., {et~al.} 2020, \aj, 159,
  151, \dodoi{10.3847/1538-3881/ab7245}

\bibitem[{{Wakeford} {et~al.}(2017){Wakeford}, {Visscher}, {Lewis}, {Kataria},
  {Marley}, {Fortney}, \& {Mand ell}}]{Wakeford17}
{Wakeford}, H.~R., {Visscher}, C., {Lewis}, N.~K., {et~al.} 2017, \mnras, 464,
  4247, \dodoi{10.1093/mnras/stw2639}

\bibitem[{{Weck} {et~al.}(2004){Weck}, {Schweitzer}, {Kirby}, {Hauschildt}, \&
  {Stancil}}]{Weck04}
{Weck}, P.~F., {Schweitzer}, A., {Kirby}, K., {Hauschildt}, P.~H., \&
  {Stancil}, P.~C. 2004, \apj, 613, 567, \dodoi{10.1086/423029}

\bibitem[{{Weck} {et~al.}(2003){Weck}, {Schweitzer}, {Stancil}, {Hauschildt},
  \& {Kirby}}]{Weck03b}
{Weck}, P.~F., {Schweitzer}, A., {Stancil}, P.~C., {Hauschildt}, P.~H., \&
  {Kirby}, K. 2003, \apj, 582, 1059, \dodoi{10.1086/344722}

\bibitem[{Wilzewski {et~al.}(2016)Wilzewski, Gordon, Kochanov, Hill, \&
  Rothman}]{Wilzewski16}
Wilzewski, J., Gordon, I., Kochanov, R., Hill, C., \& Rothman, L. 2016, Journal
  of Quantitative Spectroscopy and Radiative Transfer, 168, 193,
  \dodoi{10.1016/j.jqsrt.2015.09.003}

\bibitem[{{Winn} {et~al.}(2007){Winn}, {Holman}, {Henry}, {Roussanova}, {Enya},
  {Yoshii}, {Shporer}, {Mazeh}, {Johnson}, {Narita}, \& {Suto}}]{winn07}
{Winn}, J.~N., {Holman}, M.~J., {Henry}, G.~W., {et~al.} 2007, \aj, 133, 1828,
  \dodoi{10.1086/512159}

\bibitem[{{Yurchenko} {et~al.}(2011){Yurchenko}, {Barber}, \&
  {Tennyson}}]{Yurchenko11}
{Yurchenko}, S.~N., {Barber}, R.~J., \& {Tennyson}, J. 2011, \mnras, 413, 1828,
  \dodoi{10.1111/j.1365-2966.2011.18261.x}

\bibitem[{{Yurchenko} \& {Tennyson}(2014)}]{Yurchenko14}
{Yurchenko}, S.~N., \& {Tennyson}, J. 2014, \mnras, 440, 1649,
  \dodoi{10.1093/mnras/stu326}

\bibitem[{{Yurchenko} {et~al.}(2013){Yurchenko}, {Tennyson}, {Barber}, \&
  {Thiel}}]{Yurchenko13}
{Yurchenko}, S.~N., {Tennyson}, J., {Barber}, R.~J., \& {Thiel}, W. 2013,
  Journal of Molecular Spectroscopy, 291, 69, \dodoi{10.1016/j.jms.2013.05.014}

\bibitem[{{Zalesky} {et~al.}(2019){Zalesky}, {Line}, {Schneider}, \&
  {Patience}}]{Zalesky19}
{Zalesky}, J.~A., {Line}, M.~R., {Schneider}, A.~C., \& {Patience}, J. 2019,
  \apj, 877, 24, \dodoi{10.3847/1538-4357/ab16db}

\bibitem[{{Zhang} {et~al.}(2020){Zhang}, {Liu}, {Marley}, {Line}, \&
  {Best}}]{Zhang21}
{Zhang}, Z., {Liu}, M.~C., {Marley}, M.~S., {Line}, M.~R., \& {Best}, W. M.~J.
  2020, arXiv e-prints, arXiv:2011.12294.
\newblock \doarXiv{2011.12294}

\bibitem[{{Zhang} {et~al.}(2021){Zhang}, {Liu}, {Marley}, {Line}, \&
  {Best}}]{Zhang21a}
---. 2021, arXiv e-prints, arXiv:2105.05256.
\newblock \doarXiv{2105.05256}

\bibitem[{{Zhou} {et~al.}(2019){Zhou}, {Bakos}, {Bayliss}, {Bento}, {Bhatti},
  {Brahm}, {Csubry}, {Espinoza}, {Hartman}, {Henning}, {Jord{\'a}n}, {Mancini},
  {Penev}, {Rabus}, {Sarkis}, {Suc}, {de Val-Borro}, {Rodriguez}, {Osip},
  {Kedziora-Chudczer}, {Bailey}, {Tinney}, {Durkan}, {L{\'a}z{\'a}r}, {Papp},
  \& {S{\'a}ri}}]{zhou19}
{Zhou}, G., {Bakos}, G.~{\'A}., {Bayliss}, D., {et~al.} 2019, \aj, 157, 31,
  \dodoi{10.3847/1538-3881/aaf1bb}

\end{thebibliography}

\end{document}